\documentclass[12pt,oneside,reqno,english]{amsbook}
\usepackage[T1]{fontenc}
\usepackage[latin9]{inputenc}
\usepackage{geometry}
\usepackage{babel}
\usepackage{amstext}
\usepackage{amsthm}
\usepackage{amssymb}
\makeindex
\usepackage{graphicx}
\usepackage{setspace}
\usepackage{wasysym}
\usepackage[unicode=true,pdfusetitle,
 bookmarks=true,bookmarksnumbered=false,bookmarksopen=false,
 breaklinks=false,pdfborder={0 0 0},pdfborderstyle={},backref=false,colorlinks=false]
 {hyperref}

\makeatletter
\numberwithin{section}{chapter}
\numberwithin{equation}{section}
\theoremstyle{plain}
\ifx\thechapter\undefined
	\newtheorem{thm}{\protect\theoremname}
\else
	\newtheorem{thm}{\protect\theoremname}[chapter]
\fi
\theoremstyle{definition}
\newtheorem{example}[thm]{\protect\examplename}
\theoremstyle{definition}
\newtheorem{defn}[thm]{\protect\definitionname}
\theoremstyle{plain}
\newtheorem{lem}[thm]{\protect\lemmaname}
\theoremstyle{remark}
\newtheorem{rem}[thm]{\protect\remarkname}
\theoremstyle{plain}
\newtheorem{prop}[thm]{\protect\propositionname}

\usepackage{chngcntr}
\usepackage{xstring}
\usepackage{caption}

\renewcommand{\tocappendix}[3]{%
  \indentlabel{\IfStrEq{#3}{Bibliography}{}{#1}\@ifnotempty{#2}{ #2.\quad}}#3}
\usepackage{xstring}
\renewcommand{\tocappendix}[3]{%
  \indentlabel{\IfStrEq{#3}{Index}{}{#1}\@ifnotempty{#2}{ #2.\quad}}#3}
\counterwithin{figure}{chapter}

\numberwithin{figure}{chapter}

\hypersetup{
	pdfcreator={Mengzhen Zhang},   
    pdfproducer={},  
}

\makeatother

\providecommand{\definitionname}{Definition}
\providecommand{\examplename}{Example}
\providecommand{\lemmaname}{Lemma}
\providecommand{\propositionname}{Proposition}
\providecommand{\remarkname}{Remark}
\providecommand{\theoremname}{Theorem}

\begin{document}
\thispagestyle{empty}

\pagestyle{empty}
\begin{center}
\textbf{\large{}Abstract}{\large\par}
\par\end{center}

\smallskip{}

\begin{center}
\textbf{\large{}PROPERTIES AND APPLICATIONS OF GAUSSIAN QUANTUM PROCESSES}{\large\par}
\par\end{center}

\begin{center}
Mengzhen Zhang
\par\end{center}

\begin{center}
December 2020
\par\end{center}

\medskip{}

Gaussian states, operations, and measurements are central building
blocks for continuous-variable quantum information processing which
paves the way for abundant applications, especially including network-based
quantum computation and communication. To make the most use of the
Gaussian processes, it is required to understand and utilize suitable
mathematical tools such as the symplectic space, symplectic algebra,
and Wigner representation. Applying these mathematical tools to practical
quantum scenarios, we developed various schemes for faithful quantum
transduction, interference-based bosonic mode permutation and ultra-sensitive
bosonic sensing. Quantum transduction, proposed for transferring quantum
information between different bosonic platforms, will enable the construction
of large-scale hybrid quantum networks. We demonstrated that generic
coupler characterized by Gaussian unitary process can be transformed
into a high-fidelity transducer, assuming the availability of infinite
squeezing and high-precision adaptive feedforward with homodyne measurements,
all of which are Gaussian operations, measurements or classical communication
channels and can be easily analyzed using symplectic algebra. To address
the practical limitation of finite squeezing, we also explored the
potential of interference-based protocols. It turns out that these
protocols can let us freely permute bosonic modes only assuming the
access to single-mode Gaussian unitary operations and multiple uses
of a given generic multi-mode Gaussian process. Thus, such a scheme
not only enables universal decoupling for multi-mode bosonic systems,
which can be useful for suppressing undesired coupling between the
system and the environment, but also efficient and faithful bidirectional
single-mode quantum transduction. Moreover, noticing that the Gaussian
processes are appropriate theoretical models for optical sensors,
we studied the quantum noise theory for optical parameter sensing
and its potential in providing great measurement precision enhancement.
We also extended the Gaussian theories to discrete variable systems,
with several examples such as quantum (gate) teleportation. All the
analyses and conclusions originated from the fundamental quantum commutation
relations, and therefore are widely applicable.

\frontmatter

\thispagestyle{empty}

\setcounter{page}{1}
\begin{center}
\textbf{\Large{}PROPERTIES AND APPLICATION OF GAUSSIAN QUANTUM PROCESSES}{\Large\par}
\par\end{center}

\vfill{}

\begin{center}
A Dissertation
\par\end{center}

\begin{center}
Presented to the Faculty of the Graduate School
\par\end{center}

\begin{center}
of
\par\end{center}

\begin{center}
Yale University
\par\end{center}

\begin{center}
in Cnadidacy for the Degree of Doctor of Philosophy
\par\end{center}

\vfill{}

\begin{center}
by
\par\end{center}

\begin{center}
Mengzhen Zhang
\par\end{center}

\smallskip{}

\begin{center}
Dissertation Director: A. Douglas Stone
\par\end{center}

\begin{center}
December 2020
\par\end{center}

\pagestyle{headings}

\hphantom{}

\vfill{}

\begin{center}
\copyright 2020 by Mengzhen Zhang
\par\end{center}

\begin{center}
All rights reserved.
\par\end{center}

\vfill{}

\hphantom{}

\tableofcontents{}

\listoffigures

\chapter*{List of Abbreviations and Symbols}

\begin{tabbing} 

$\hat{q},\hat{p},\hat{x}$\=~~~~~~~~ \= Canonical quadrature
operatos. \\

$M^{t}$ \>~~~~~~~~\>Transpose of a matrix $M$. \\

$\text{Id}$ \>~~~~~~~~\>Identity of groups and matrix algebras;
the identity matrix. \\ 

$\text{Tr}\left[\cdot\right]$ \>~~~~~~~~\>Trace of operators
or matrices. \\

$\sigma\left(\cdot,\cdot\right),\Omega$ \>~~~~~~~~\>Symplectic
form. \\

$\hat{U}_{S}$ \>~~~~~~~~\> Unitary operator associated
with a symplectic transformation $S$. \\

$\hat{D}_{u}$ \>~~~~~~~~\> Displacement operation associated
with a vector $u$ in the\\ 

~\>~~~~~~~~\> symplectic space. \\

$W_{\hat{A}}$ \>~~~~~~~~\>Winger function associated with
an operator $\hat{A}$. \\

$\hat{\Pi}_{l}\left(\eta\right)$\>~~~~~~~~\>Infinitely
squeezed state associated with Lagrangian plane $l$ and \\

~\>~~~~~~~~\>position vector $\eta$. \\

$\hat{\rho}\left(\bar{x},V\right)$ \>~~~~~~~~\>Gaussian
state associated with first moments $\bar{x}$ and \\

~\>~~~~~~~~\>covariance matrix $V$. \\

$\mathcal{G}_{T,N}$ \>~~~~~~~~\>Gaussian channel associated
with matrix data $T$ and $N$.

\end{tabbing}

\chapter*{Acknowledgement}

It has been quite a pleasure and honor to work in Yale Quantum Institute
(YQI). I have received a lot of help from many people from YQI, to
whom I wan to express my great gratitude.

Firstly, I cannot be more grateful to Liang Jiang for being my advisor.
Liang showed me the way to the fantastic field of quantum information.
He can understand and accept novel theoretical tools effortlessly,
and incorporate them into practical applications, which shaped my
research patterns. In particular, his emphasis on experiments always
keeps me from wasting my life on playing meaningless mathematical
games. Liang has also created a relaxing working environment so that
all the people in his can freely share their wildest ideas. My collaboration
with A. Douglas Stone is absolutely one of my most valuable experiences
at Yale. Doug's wonderful physical intuition and ability of clarifying
complicated concepts constantly amaze me, not to mention his well-known
talents for writing. Among all the courses I took at Yale, Michel
Devoret's introductory lectures on quantum noise theory is the most
helpful to me. His patient and detailed explanation of the elementary
concepts helped me quickly master this powerful tool, which has played
a central role in my doctoral research. Although I did not have many
conversations with Hong Tang, his unreserved and constructive comments
on my early projects and our later collaboration taught me a great
deal.

Next, I must thank Changling Zou and Chao Shen for their unselfish
help and fruitful discussions during their stay at YQI. Without them,
the beginning stage of me being a researcher would not have been so
smooth and painless. Their scientific and emotional support will always
be cherished by me. I thank Chia Wei Hsu and William Sweeney for their
countless valuable scientific inputs and friendship. I am grateful
to Shoumik Chowdhury, still an undergraduate student at Yale, for
our pleasant collaboration due to his patience and diligence. We also
owe the nice presentation of one of our most complicated ideas to
him. Also, I cannot forget to mention Aashish Clerk. His comments
provide me with new perspectives of my work and are influencing some
of my ongoing projects.

I have enjoyed my life so much at Yale, thanks to each and every one
of my friends and my fellow group members. For example, Stephan Krastanov's
recognized achievement in teaching and organizing outreaching activities
has not only deeply touched me and brought me a lot of fun, but also
improve my communication skills.

I will forever be indebted to Yanhong Xiao, who kindly encouraged
and supported me as an undergraduate to pursue the fun of physics.
I was very lucky to have learnt the basics of scientific research
and some still essential physical concepts from her. 

Last but absolutely not the least, I would like to express my gratitude
to my family. My mother and father, Yunfang Han and Lijun Zhang, have
supplied me with the most needed support as always, otherwise this
thesis could not have been possible.

\mainmatter

\chapter{Introduction}

In classical mechanics, states of a physical system are completely
determined by the values of its position and momentum variables. Thus,
each state of a classical physical system corresponds to a point in
a multi-dimensional space, the phase space, which enables the geometrization
of the time evolution of classical systems. The momenta and positions
are conjugate pairs of variables, such that their Poisson brackets
must be preserved under any physical dynamics. Therefore, phase space
must be regarded as a symplectic manifold, and physical dynamics should
be covariant under symplectic diffeomorphisms \cite{DeGosson2006,Arnold2013}.
Thanks to the fundamental correspondence between Poisson brackets
and commutators, phase space is also compatible with quantum mechanics,
along with its symplectic-geometrical structure. Furthermore, this
allows us to analyze the quantum states using the Wigner function,
a quasi-probabilistic function and hence the qauntum counterpart of
the statistical ensemble in statistical mechanics \cite{Wigner1997}.

These concepts--phase space, symplectic geometry and Wigner function--are
extensively applied to studying problems related to Gaussian quantum
processes \cite{Weedbrook2012}. In such a process, dynamics of a
quantum physical system is tracked by the linear change of the expectation
values of the momentum and position operators, and therefore most
closely resembles its classical counterparts. As the consequence,
the symplectic geometry methods greatly reduce the complexity of representation
and analysis of Gaussian quantum processes, leading to many successful
applications in bosonic quantum information and computation. However,
the potential of these tools have not been fully exploited for solving
practical bosonic quantum engineering problems, despite the ubiquitous
involvement of Gaussian processes in such problems. In view of the
rapid development and increasing demand of bosonic quantum engineering,
I will demonstrate in this thesis how these powerful symplectic-geometry-related
mathematical tools can be utilized to derive promising systematic
solutions to some intriguing problems.

Firstly, applications are found in quantum transduction. Successful
construction of hybrid quantum networks \cite{Kimble2008,Duan2010},
which often involves bosonic modes with different physical characteristics
such as frequencies, polarizations or even mode carriers, relies on
the performance of quantum transducers. They are devices that can
efficiently and faithfully convert quantum signals between different
bosonic platforms. For example, with a high-quality microwave-to-optical-frequency
quantum transducer, one can transfer information between superconducting
circuits and optical waveguides with 100\% fidelity, which will enable
the construction of large scale quantum computers with an enormous
number of qubits free of size-dependent penalties.

Many theoretical and experimental efforts have been made to realize
the quantum transducers, including frequency conversion between the
microwave and optical modes \cite{Hafezi2012,Bochmann2013,Andrews2014,Tian2015,Hisatomi2016,Rueda2016,Vainsencher2016,Higginbotham2018,Han2020,Mirhosseini2020},
between the microwave and mechanical/spin-wave modes \cite{Palomaki2013,Zhang2014,Tabuchi2015},
and between quantum processors and quantum memory \cite{Julsgaard2004,Sherson2006,Rabl2006,Stannigel2010,Grezes2014}.
These attempts are usually based on tuning the linear scattering process
to the working point where the transmittance between two wanted modes
is maximized and hence the reflectance simultaneously vanishes. However,
because of the limited parameter controllability and more importantly
the inevitable presence of sideband modes \cite{Rueda2016,Soltani2017},
often ignored in theoretical proposals, such working points in the
experimental environments either do not exist or cannot be achieved
and thus high-fidelity quantum transducers are still far from ready. 

We note that in the existing attempts using linear scattering processes,
the reflected signal is usually discarded although it often carries
a significant part of the information of the input signal. Then it
is natural to ask: Can we collect the information from the reflected
signal and restore the quantum state of the input signal? Based on
this conception and by applying the symplectic geometry tools, we
find that by infinitely squeezing the auxiliary modes, measuring the
reflected signal homodyne measurements and displacing the output signal
according to the measurement outcomes, the quantum information can
be fully restored, which is universally applicable to general Gaussian
processes \cite{Zhang2018}. In addition, all of the key components--the
squeezed auxiliary modes, the homodyne measurement and the adaptive
displacements--also appear in the continuous variable teleportation
scheme. Actually, we will see in this thesis that our scheme for quantum
transduction is a generalization of the teleportation scheme, where
the special Gaussian process, the balanced beam splitters, is replaced
with a general Gaussian process.

However, there is a major drawback of this scheme: large squeezing
is too hard to achieve in laboratories. In the search of a solution
this problem, a promising interference-based scheme came into our
sight. The idea was originally proposed for transducers involving
two bosonic modes \cite{Lau2019}: We consider a imperfect quantum
transducer corresponding to a unitary Gaussian process and assume
all the single-mode Gaussian unitary processes for any mode are available.
Then we let the input signal pass a sequence of copies of this unitary
process with single-mode Gaussian unitary processes interspersed them.
It turns out that in general, the quantum signal can be faithfully
transduced if the interspersed single-mode Gaussian unitary processes
are designed appropriately. However, the two bosonic modes constraint
of this proposal is impractical in most practical situations as we
have explained above. However, we find that this proposal enjoys a
hidden underlying mathematical mechanism that can be concisely explained
by some most fundamental facts in symplectic geometry and hence is
extendable to the multi-mode cases \cite{Zhang2020}. Moreover, the
flexibility of the choice of the single-mode Gaussian processes enable
us to decouple arbitrary modes without the need to change the overall
structure of the scheme, which may benefit noise and error reduction
in quantum information \cite{Ofek2016,Hu2019,CampagneIbarcq2019,Heinze2019}.
As the result, these observations together lead to the interference-based
bosonic permutation scheme that will be presented in this thesis.

Next, symplectic geometry is also helpful for understanding exotic
properties of optical systems with special spectrum structure, known
as exceptional points. Exceptional points have attracted a lot of
attention recently. Specifically, we consider a non-Hermitian optical
Hamiltonian depending on some parameter, e.g. the coupling strength
between different modes. When the parameter is tuned to a certain
value which is usually called the exceptional point, the originally
distinct eigenmodes will coalesce into a single eigenmode, which is
mathematically equivalent to the appearance of a non-trivial Jordan
normal form in the matrix representation of the Hamiltonian \cite{Kato2013,Heiss1999,Heiss2000,Berry2004}.
This phenomenon inspires many novel application-oriented proposals
\cite{Lin2011,Chang2014,Peng2014,Xu2016,Doppler2016,Makris2008,Zhen2015,Rueter2010,Regensburger2012,Feng2014,Zhang2016},
among which one of the most promising is enhancing sensing precision
of optical sensors \cite{Wiersig2014,Wiersig2016,Liu2016,Chen2017,Hodaei2017}.
People found the split of eigen-frequencies near the exceptional point
varies non-linearly with respect to the parameter change: The smaller
the parameter change, the larger the amplification frequency-split.
Therefore, considering the frequency split to be measurable signal,
one may expect to see a great improvement in sensitivity when the
parameter is small. However, as presented in this thesis, this optimistic
outlook is challenged when the process is carefully analyzed using
quantum noise theory \cite{gardiner_quantum_2004}. The quantum noise
will increase proportionately as the signal increases. Nevertheless,
such a seemingly pessimistic conclusion does not lead to a complete
failure of this attempt, since we find that by properly changing the
measurement scheme, an even larger sensitivity enhancement can be
genuinely achieved \cite{Zhang2019}. Moreover, with results derived
from symplectic geometry, this exceptional-point sensing scheme along
with other sensing schemes based on Gaussian processes can be dilated
to a unitary Gaussian process and understood as a phenomenon related
to squeezing, details of which will be demonstrated in this thesis.

Apart from the applications in the continuous variable systems, the
concepts related to symplectic geometry can also be generalized to
discrete variable systems \cite{Gross2013}. In the last chapter of
this thesis, I will briefly show by examples how some of the results
derived for the continuous variable cases can be applied effortlessly
to the discrete variable systems. Last but not the least, in the same
chapter, I systematically generalized most of the symplectic geometry
toolbox to arbitrary discrete variable systems, compatible with the
conventional definition of generalized Pauli operators, which has
not been fully revealed before.

Here is an overview of the thesis: The necessary mathematical concepts
of symplectic geometry is introduced in Ch.~\ref{chap:Symplectic-geometry-for},
while Gaussian states and Gaussian processes are introduced in the
following Ch.~\ref{chap:Gaussian-states-and}. Ch.~\ref{chap:Quantum-Transduction}
and Ch.~\ref{chap:Interference-based-Gaussian-cont} provide details
of our investigation of quantum transduction using symplectic geometry.
In Ch.~\ref{chap:Applications-to-bosonic}, we explore the application
of Gaussian processes to analyzing bosonic sensing schemes. The extension
of symplectic geometry to the discrete variable system is demonstrated
in the last chapter, Ch.~\ref{chap:Beyond-the-continuous}.

\chapter{Symplectic geometry for bosonic systems\label{chap:Symplectic-geometry-for}}

\section{Introduction}

In classical mechanics, a physical state of a system is completely
determined by the momenta $p_{j}$ and positions $q_{j}$ of its components.
Momenta and positions are the canonical coordinates of the phase space.
Therefore, each point in the phase space represents a unique physical
state and the evolution of the system geometrically corresponds to
a trajectory. Poisson brackets--which plays a central role in classical
mechanics--of the canonical coordinates satisfy the following conditions
\[
\left\{ q_{j},q_{k}\right\} =\left\{ p_{j},p_{k}\right\} =0,
\]
and
\[
\left\{ q_{j},p_{k}\right\} =\delta_{jk},
\]
where $\delta_{jk}$ is the Kronecker delta. The Poisson brackets
impose a special geometrical structure on the phase space: the coordinate
transformations of the canonical coordinates should keep the Poisson
brackets invariant. Intuitively, these coordinate transformations
preserve the area enclosed by a closed trajectory in the phase space.
Such Poisson-brackets-preserving coordinate transformations in the
phase space are called the symplectic transformations, and the induced
geometrical structure is known as the symplectic geometry.

\begin{figure}[h]
\includegraphics[width=0.9\textwidth]{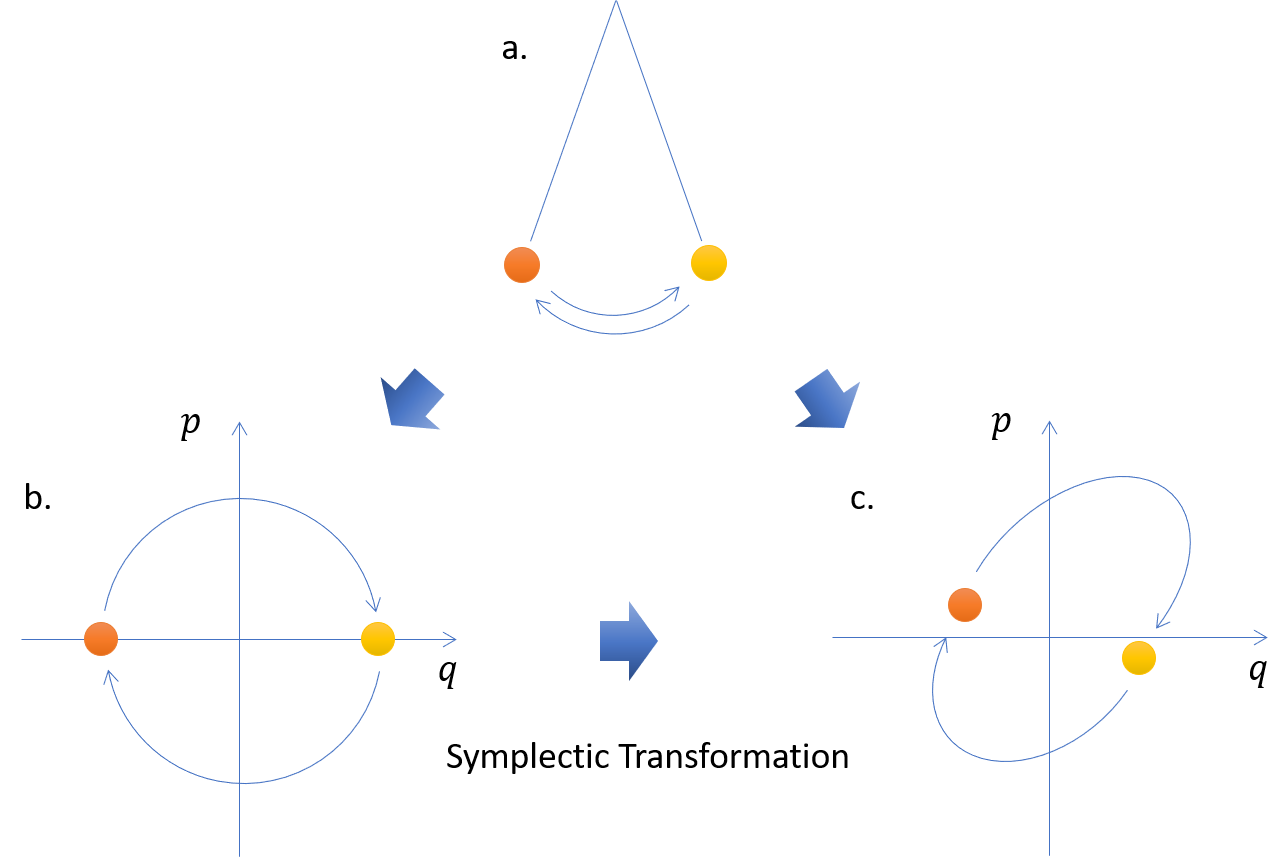}
\begin{doublespace}
\centering{}\caption{\emph{Evolution of a pendulum in the phase space.} }
\begin{minipage}[t]{0.9\textwidth}%
\begin{doublespace}
\begin{flushleft}
The figure shows the phase space representation of a pendulum. As
shown in \textbf{b}, the classical physical state of a pendulum (as
in \textbf{a}) at each instant can be represented by a point in the
phase space. The evolution of a pendulum, as a closed physical system,
is thus represented by a closed trajectory, i.e. the circle in \textbf{b}.
Under symplectic transformations (i.e. any physical evolution or transformation
preserving the Poisson brackets), the area encircled by the closed
trajectory should be preserved. Therefore, the enclosed area in \textbf{c
}is equal to the enclosed area in \textbf{b}. Moreover, as for the
case in \textbf{c}, the physical state of the pendulum is not directly
represented by its position and momentum, but another pair of canonical
variables transformed from the position and the momentum via the corresponding
symplectic transformation.
\par\end{flushleft}
\end{doublespace}
\end{minipage}
\end{doublespace}
\end{figure}

What are the symplectic transformations? The first candidates are
the length-preserving coordinate transformations, which are obviously
area-preserving. However, not all length-preserving transformations
are symplectic transformations. For example, if the canonical coordinates
are transformed as
\[
q_{1}\rightarrow p_{2},\,q_{2}\rightarrow q_{1},\,p_{1}\rightarrow q_{2},\,p_{2}\rightarrow p_{1},
\]
then the Poisson-brackets-relations will not be broken while the area
of a closed trajectory is still preserved. For the same reason, we
must also exclude the mirror operations, such as
\[
q_{1}\rightarrow q_{1},\,p_{1}\rightarrow-p_{1}.
\]
In addition to the length-preserving coordinates, the enclosed area
of a closed trajectory can also be preserved by the squeezing operations,
for instance
\[
q_{1}\rightarrow2q_{1},\,p_{1}\rightarrow(1/2)p_{1}.
\]
Later, we will see that any symplectic transformation is product of
feasible rotation-like and squeezing-like coordinate transformations
in the phase space.

We have seen how the Poisson brackets geometrize the classical mechanics.
It turns out the quantum mechanics can also be geometrized similarly.
Thanks to the canonical correspondence principles
\[
\left\{ f,g\right\} \rightarrow\frac{1}{i\hbar}\left[\hat{f},\hat{g}\right]
\]
with $\hat{f},\hat{g}$ being the quantizations of the classical observables
$f$ and $g$, the Poisson brackets serve as a bridge connecting classical
and quantum physics. Specifically, the Poisson-bracket relations of
the canonical coordinates translate to canonical quantization relations
(CCRs):
\[
\left[\hat{q}_{j},\hat{q}_{k}\right]=\left[\hat{p}_{j},\hat{p}_{k}\right]=0,
\]
and
\[
\left[\hat{q}_{j},\hat{p}_{k}\right]=\delta_{jk}.
\]
Therefore, phase space along with its symplectic geometry structure
(the Poisson-bracket-preserving requirement is replaced with the commutator-preserving
requirement) can be naturally established in the quantum mechanics:
Each point in the phase with the canonical coordinates being $u$
(representing the positions) and $v$ (representing the momenta) now
corresponds uniquely to a quadrature operator
\[
\sum\left(u_{j}\hat{q}_{j}+v_{j}\hat{p}_{j}\right)
\]
which is a linear combination of the canonical $\hat{q}$ and $\hat{p}$
quadratures.

So far, the extension from classical mechanics and quantum mechanics
has been quite straightforward. However, due to the Heisenberg's uncertainty
principles derived from the CCRs, states of a quantum system is always
probabilistic. It turns out that we should no longer represent a physical
state by its canonical coordinates, but by a quasi-probabilistic distribution,
such as the Wigner function and its equivalent variants. Intuitively,
this means instead of being by points, the evolution of a physical
system should be described by the a trajectory of smeared blobs (the
image of the Wigner function) in the phase space.

The Wigner function is covariant with the symplectic transformations
insofar as these coordinate transformations in phase space will also
influence the shape of the quantum states. For example, this means
a rotation-like symplectic transformation can transform a coherent
state to some other coherent state and a squeezing-like symplectic
transformation can turn a vacuum state into a squeezed vacuum state
in bosonic systems. Therefore, symplectic geometry is not only essential
to capture the kinetics of a physical system but also useful for manipulating
the quantum states.

To see that symplectic transformations are purely fictitious mathematical
concepts, we use beamsplitter in quantum optics as an example to show
they can be embodied physically. A beamsplitter can mix the two input
modes through the following unitary transformation\cite{Weedbrook2012}:
\[
\hat{q}_{1}\rightarrow\cos\theta\hat{q}_{1}+\sin\theta\hat{q}_{2},
\]
\[
\hat{p}_{1}\rightarrow\cos\theta\hat{p}_{1}+\sin\theta\hat{p}_{2},
\]
\[
\hat{q}_{2}\rightarrow-\sin\theta\hat{q}_{1}+\cos\theta\hat{q}_{2},
\]
\[
\hat{p}_{2}\rightarrow-\sin\theta\hat{p}_{1}+\cos\theta\hat{p}_{2},
\]
which is apparently area-preserving in the phase space and hence symplectically
transform the canonical coordinates in the phase space. Actually,
all rotation-like symplectic transformations can be realized as beamsplitters
or phase shifters in quantum optics. Similarly, the squeezing-like
symplectic transformations can be realized by (multi-mode) squeezing
operations. Later on, we will define this connection rigorously using
the concept metaplectic operators, and introduce other physical ways
to implement symplectic transformations. 

Moreover, as coordinate transformations, the symplectic transformations
can be conveniently described by matrices. For instance, the symplectic
transformation corresponding to a beamsplitter as shown above can
be described by the $4\times4$ real matrix
\[
\begin{pmatrix}\cos\theta & 0 & -\sin\theta & 0\\
0 & \cos\theta & 0 & -\sin\theta\\
\sin\theta & 0 & \cos\theta & 0\\
0 & \sin\theta & 0 & \cos\theta
\end{pmatrix}.
\]
It follows that the composition of the unitary operations corresponding
to the symplectic transformations can be efficiently calculated by
matrix algebra. In the following sections, we introduce the notations,
definitions and properties to provide a precise description of symplectic
geometry in the bosonic systems, which will be necessary for proving
the results in the later chapters of this thesis. We will start with
the symplectic space (the mathematical term for the phase space),
and then define symplectic transformations as structure-preserving
linear transformations on the symplectic space. The basis properties
and physical constructions of symplectic transformations will be introduced
and discussed \footnote{Throughout this chapter, we follow the conventions in \cite{DeGosson2006}.
The detailed derivations of the statements in this chapter can also
be found in the same reference.}.

\section{Symplectic space\label{section:Quadrature_operator_and_symplectic_algebra}}

Consider an $N$-mode bosonic system, of which the annihilation operators
are $\hat{a}_{1},\hat{a}_{2},\dots,\hat{a}_{N}$ (correspondingly
the $N$ creation operators $\hat{a}_{1}^{\dagger},\hat{a}_{2}^{\dagger},\dots,\hat{a}_{N}^{\dagger}$).
The annihilation and creation operators satisfy 
\[
[\hat{a}_{j},\hat{a}_{k}]=[\hat{a}_{j}^{\dagger},\hat{a}_{k}^{\dagger}]=0,
\]
and
\[
[\hat{a}_{j},\hat{a}_{k}^{\dagger}]=\delta_{j,k},
\]
for any $j,k\in\{1,2,\dots,N\}$. These relations amount to the CCRs
if the canonical quadrature operators are defined as follows:
\[
\hat{q}_{j}:=\frac{\hat{a}_{j}+\hat{a}_{j}^{\dagger}}{\sqrt{2}}\text{ and \ensuremath{\hat{p}_{j}}=\ensuremath{\frac{\hat{a}_{j}-\hat{a}_{j}^{\dagger}}{\sqrt{2}i}}}.
\]
For convenience, we deonte
\[
\hat{x}_{2j-1}=\hat{q}_{j},\ \forall j\in\{1,2,\dots,N\},
\]
and
\[
\hat{x}_{2j}=\hat{p}_{j},\ \forall j\in\{1,2,\dots,N\}.
\]
Then the CCRs can be expressed in a more compact form:
\[
[\hat{x}_{j},\hat{x}_{k}]=i\Omega_{j,k},\ \forall j,k\in\{1,2,\dots,N\}
\]
where the matrix $\Omega$ on the right hand side is defined as
\[
\Omega=\oplus_{j=1}^{N}\begin{pmatrix}0 & -1\\
1 & 0
\end{pmatrix}=\begin{pmatrix}0 & -1\\
1 & 0\\
 &  & 0 & -1\\
 &  & 1 & 0\\
 &  &  &  & \ddots\\
 &  &  &  &  & 0 & -1\\
 &  &  &  &  & 1 & 0
\end{pmatrix}
\]
and will be called the symplectic form\index{Symplectic!form}.

The name, symplectic form, is obtained from the mathematical definition
of symplectic space\index{Symplectic!space}. A $2N$-dimensional
real vector space $E$ is called a symplectic space if it is equipped
with a non-degenerate skew-symmetric bilinear binary map $\sigma:E\times E\rightarrow E$\footnote{This means that given any $u,v\in E$, we have $\sigma\left(u,v\right)=-\sigma\left(v,u\right)$
and $\sigma\left(\lambda u,\mu v\right)=\lambda\mu\sigma\left(u,v\right)$for
any $\lambda,\mu\in\mathbb{R}$ and $\sigma\left(u,v\right)=0$ for
every $v\in E$ if and only if $u=0$.}. In such a symplectic space, we can always find $2N$ linearly independent
vectors $e^{(1)},f^{(1)},e^{(2)},f^{(2)},\dots,e^{(N)},f^{(N)}$,
such that
\[
\sigma(e^{(j)},e^{(k)})=0,
\]
\[
\sigma(f^{(j)},f^{(k)})=0,
\]
and
\[
\sigma(e^{(j)},f^{(k)})=-\sigma(f^{(j)},e^{(k)})=\delta_{jk},
\]
for all $j,k\in\{1,2,\dots,N\}.$ This basis is called the symplectic
canoncial basis\index{Symplectic!canonical basis}, and the map $\sigma$
is called the symplectic form. The canonical basis can be chosen to
be the orthonormal basis of $E$ viewed as an Euclidean space\footnote{This perspective allows us to represent linear transformations by
real matrices.}. Then it is easy to check that

\[
\sigma(u,v)=u^{t}\Omega v,\ \forall u,v\in E,
\]
where $\Omega$ is the aforementioned matrix defining the CCRs.

Now we relate the mathematical terms to the physical concepts. First,
symplectic space is the mathematical counterpart of the phase space:
Each $e^{\left(j\right)}$corresponds to a canonical quadrature operator\index{Canonical!quadrature operator}
$\hat{q}_{j}$ while each $f^{\left(j\right)}$ corresponds to a canonical
quadrature operator $\hat{p}_{j}$, and each element $u$ in $E$
corresponds to a linear combination of the canonical operators (which
is itself a genera quadrature operator)
\[
\hat{u}=\sum_{j=1}^{2N}u_{j}\hat{x}_{j}.
\]
Then for all $u,v\in E$, we can quickly verify that
\[
\left[\hat{u},\hat{v}\right]=i\sigma\left(u,v\right)=iu^{t}\Omega v,
\]
which naturally leads to the CCRs. In short, now we can conveniently
treat each quadrature operator as a vector in the symplectic space
and the commutativity between quadrature operators can be checked
easily a linear algebra operation, the symplectic form.

Before moving on to the symplectic transformations, it is helpful
to understand the structure of a symplectic space in more detail.
Especially, certain subspaces of a symplectic space play a crucial
role in deriving the results in later chapters, and thus should be
introduced here. A subspace of $E$ is spanned by a set of linearly
independent vectors. This set of vectors can be grouped as a matrix.
Specifically, let $u^{(1)},u^{(2)},\dots,u^{(M)}$ be $M$ linearly
independent vectors in the $2N$-dimensional symplectic space $E$
spanning a subspace. Then these $M$ vectors can be grouped as the
following $2N\times M$ real matrix
\[
F=\begin{pmatrix}u_{1}^{(1)} & u_{1}^{(2)} & \cdots & u_{1}^{(M)}\\
u_{2}^{(1)} & u_{2}^{(2)} & \cdots & u_{2}^{(M)}\\
\vdots & \vdots & \vdots & \vdots\\
u_{2N}^{(1)} & u_{2N}^{(2)} & \cdots & u_{2N}^{(M)}
\end{pmatrix},
\]
where the vectors $u^{(1)},u^{(2)},\dots,u^{(M)}$ are the $M$ corresponding
columns. We say the subspace spanned the $M$ vectors is determined
by a such a matrix $F$ as above. In this language, there are two
kinds of subspaces of particular interest: The subspaces determined
by a $2N\times M$ matrix $F$ such that
\[
F^{t}\Omega F=0,
\]
and the subspaces determined by a $2N\times2M$ matrix $F$ such that
\[
F^{t}\Omega F=A
\]
with $A$ being a skew-symmetric matrix ($A=-A^{t}$) and $\text{rank}A=2M$.
The former are called the isotropic subspaces\index{Isotropic subspace}
and the latter are called the symplectic subspaces\index{Symplectic!subspace}.

For the isotropic subspaces, specifically, the columns of $F$, i.e.
the $M$ linearly independent vectors spanning the this subspace,
are symplectically orthogonal\index{Symplectically orthogonal}, which
means 
\[
\sigma(u^{(j)},u^{(k)})=\left(u^{(j)}\right)^{t}\Omega u^{(k)}=0,\ \forall j,k\in\{1,2,\dots,M\}.
\]
The number of the linearly independent vectors spanning an isotropic
space does not exceed the number of modes $N$. In particular, when
$M=N$, an isotropic subspace is called a Lagrangian plane\index{Lagrangian plane}.
Let $l$ be a subspace determined by the matrix $F$ as introduced
above. (We will sometimes denote $l$ as $l_{F}$.) The symplectic
complement\index{Symplectic!complement} of $l$ in $E$, denoted
by $l^{\perp}$, will be the following subspace containing vectors
that are symplectically orthogonal to every vector in $l$, i.e.
\[
l^{\perp}=\{u\in E\ |\ \sigma(u,v)=0,\ \forall v\in l\}.
\]
Then a Lagrangian plane $l$ can be alternatively defined as a subspace
satisfying the condition $l=l^{\perp}$. Moreover, we call a subspace
$l'$ the symplectic conjugate\index{Symplectic!conjugate} of $l$
if $l'\cap l=\{0\}$ and
\begin{equation}
E\cong l\oplus l'.
\end{equation}
In other words, $l'$ is isomorphic the quotient space $E/l$. It
is easy to check that the symplectic conjugate of a Lagrangian plane
is also a Lagrangian plane. Intuitively, we demonstrate their physical
concepts in the following example:
\begin{example}
Consider the $3$-mode bosonic system with quadrature operators $\hat{q}_{1},\hat{q}_{2},\hat{q}_{3},\hat{p}_{1},\hat{p}_{2},\hat{p}_{3}$.
The subspace spanned by $\hat{q}_{1},\hat{q}_{2},\hat{q}_{3}$ with
$M\le N$ is determined by the $6\times3$ matrix
\[
F=\begin{pmatrix}1 & 0 & 0\\
0 & 1 & 0\\
0 & 0 & 1\\
0 & 0 & 0\\
0 & 0 & 0\\
0 & 0 & 0
\end{pmatrix}
\]
satisfying $F^{t}\Omega F=0$, and is therefore an isotropic subspace.
It is also a Lagrangian plane since the number of the basis of this
subspace equals the number of modes. And its symplectic conjugate
is spanned by all the $\hat{p}$-quadratures.
\end{example}

For the symplectic subspaces, the $2M$ linearly independent vectors
$g^{\left(1\right)},$ $g^{\left(2\right)}$, $\dots$, $g^{\left(M\right)}$,
$h^{\left(1\right)}$, $h^{\left(2\right)}$,$\dots$ $h^{(M)}$,
which are the $2M$ columns of the determinant matrix $F$, satisfy
\[
\sigma(g^{(j)},g^{(k)})=\sigma(h^{(j)},h^{(k)}),
\]
and
\[
\sigma(g^{(j)},h^{(k)})=\delta_{jk},
\]
for all $j,k\in\{1,2,\dots,M\}$. Therefore, such a symplectic subspace
is also a $2M$-dimensional symplectic space. If $\{g^{(1)},g^{(2)},\dots,g^{(M)},h^{(1)},h^{(2)},\dots,h^{(M)}\}$
happens to be a subset of the canonical basis $\{e^{(1)},f^{(1)},e^{(2)},f^{(2)},\dots,e^{(N)},f^{(N)}\}$
of $E$, they will span a canonical symplectic subspace\index{Canonical!symplectic subspace}.
The symplectic conjugate of a symplectic subspace is also a symplectic
subspace. For convenience, we also denote the subspace spanned by
members of the canonical basis, $e^{\left(j\right)},f^{\left(j\right)}$,
for $j\in\{1,2,\dots,N\}$, by $E_{j}$. Thus, we have
\begin{equation}
E\cong E_{1}\oplus E_{2}\oplus\cdots\oplus E_{N}.
\end{equation}
The symplectic form of a canonical symplectic subspace $E_{j}$ of
$E$ is naturally
\[
\Omega_{j}=\begin{pmatrix}0 & -1\\
1 & 0
\end{pmatrix}.
\]

\begin{example}
Consider the $3$-mode bosonic system with quadrature operators $\hat{q}_{1},\hat{q}_{2},\hat{q}_{3},\hat{p}_{1},\hat{p}_{2},\hat{p}_{3}$.
The canonical quadratures $\hat{q}_{2},\hat{p}_{2}$ span a canonical
symplectic subspace which can be denoted by $E_{2}$.
\end{example}

Moreover, we can prove that for $l$ being an isotropic space (but
strictly not a Lagrangian plane), there is always a symplectic subspace
$H$, such that
\begin{equation}
E\cong H\oplus l\oplus l^{H},
\end{equation}
where $l^{H}$ is the symplectic conjugate of $l$ in the symplectic
conjugate $H'$ of $H$ as follows:
\begin{equation}
H'\cong l\oplus l^{H}.
\end{equation}
The isotropic subspace $l$ is thus a Lagrangian plane of $H'$.
\begin{example}
The symplectic space of the 3-mode bosonic system can be decomposed
as the direct sum of a canonical subspace spanned by $\hat{q}_{1},\hat{p}_{1}$,
an isotropic subspace spanned by $\hat{q}_{2},\hat{q}_{3}$ and $\hat{p}_{2},\hat{p}_{3}$.
The latter two subspaces are symplectic conjugate in the symplectic
conjugate, spanned by $\hat{q}_{1},\hat{p}_{1},\hat{q}_{2},\hat{p}_{2}$,
of the canonical symplectic subspace $\hat{q}_{1},\hat{p}_{1}$.
\end{example}

Each subspace of a symplectic space is associated with a projection
map. Let $l$ be a subspace of $E$. We denote the corresponding projection
by $\pi_{l}:E\rightarrow l$. Since this operation will be frequently
used later, for any linear transformation $S:E\rightarrow E$and any
subspaces $l,l'$ of $E$, we will often denote
\[
S_{l,l'}=\pi_{l}S\pi_{l'}.
\]

\begin{example}
Let $E_{j}$ be a canonical symplectic subspace for $j\in\{1,2,\dots,N\}$,
then 
\[
\Omega_{j}=\pi_{j}\Omega\pi_{j},
\]
just as expected.
\end{example}

\begin{example}
Consider a beamsplitter, which corresponds to a linear transformation
\[
S=\begin{pmatrix}\cos\theta & 0 & -\sin\theta & 0\\
0 & \cos\theta & 0 & -\sin\theta\\
\sin\theta & 0 & \cos\theta & 0\\
0 & \sin\theta & 0 & \cos\theta
\end{pmatrix}
\]
in the phase space. Let $E_{1}$ be the canonical symplectic subspace
spanned by $\hat{q}_{1},\hat{p}_{1}$ and $l$ be the Lagrangian plane
spanned by $\hat{q}_{1},\hat{q}_{2}$. Then we have the submatrix
\[
S_{E_{1},l}=\begin{pmatrix}\cos\theta & -\sin\theta\\
0 & 0
\end{pmatrix}.
\]
\end{example}

Last but not the least, changing the ordering of the canonical basis
of a symplectic basis will change the presentation of the linear transformations,
which can often improve our understanding of the derivations. Aside
from the 
\[
\hat{q}_{1},\hat{q}_{2},\dots,\hat{q}_{N},\hat{p}_{1},\hat{p}_{2},\dots,\hat{p}_{N}
\]
 ordering we have been used so far, we can also order the basis as
\[
\hat{q}_{1},\hat{p}_{1},\hat{q}_{2},\hat{p}_{2},\dots,\hat{q}_{N},\hat{p}_{N}.
\]
These are the two major ordering that are best suited to the later
development of our theory. In consideration of their frequent use,
we will refer to the former as the default ordering\index{Default ordering}
and the latter the alternative ordering\index{Alternative ordering}.
Without notice, the default ordering is always assumed.
\begin{example}
Consider the linear transformation corresponding to the beamsplitter.
With the default ordering, we have
\[
S=\begin{pmatrix}\cos\theta & 0 & -\sin\theta & 0\\
0 & \cos\theta & 0 & -\sin\theta\\
\sin\theta & 0 & \cos\theta & 0\\
0 & \sin\theta & 0 & \cos\theta
\end{pmatrix}.
\]
If we rearrange the canonical basis according to the alternative ordering,
the above matrix will be of the following form:
\[
S=\begin{pmatrix}\cos\theta & -\sin\theta & 0 & 0\\
\sin\theta & \cos\theta & 0 & 0\\
0 & 0 & \cos\theta & -\sin\theta\\
0 & 0 & \sin\theta & \cos\theta
\end{pmatrix}.
\]
\end{example}

\section{Symplectic transformation\label{symplectic_transformation}}

As we mentioned in the first section of this chapter, certain coordinate
transformations of the phase space (i.e linear transformations in
the symplectic space) preserve the CCRs and therefore the symplectic
geometrical structure. Such coordinate transformations are symplectic
transformations, which are defined mathematically as follows:
\begin{defn}
Let $E$ be a symplectic space equipped with the symplectic form $\sigma:E\times E\rightarrow E$.
A linear transformation $S$ on $E$ is a\index{Symplectic!transformation}
\emph{symplectic transformation }if
\[
\sigma\left(Su,Sv\right)=\sigma\left(u,v\right)
\]
for any $u,v\in E$.
\end{defn}

Using the matrix representation of the symplectic form, the above
defining condition is equivalent to

\begin{equation}
S^{t}\Omega S=\Omega,\label{eq:symplectic matrix}
\end{equation}
where $\Omega$ is the matrix representation of the symplectic form.
\begin{example}
As we mentioned before, a beamsplitter can be described by a linear
transformation
\[
S=\begin{pmatrix}\cos\theta & 0 & -\sin\theta & 0\\
0 & \cos\theta & 0 & -\sin\theta\\
\sin\theta & 0 & \cos\theta & 0\\
0 & \sin\theta & 0 & \cos\theta
\end{pmatrix},
\]
 in the symplectic space. It is a symplectic transformation since
it satisfies the condition $S^{t}\Omega S=\Omega$.
\end{example}

Just as orthogonal transformations in an Euclidean space form the
orthogonal group. The set of all symplectic transformations in a symplectic
space also form a group--the symplectic group\index{Symplectic!group}:
Let $E$ be a $2N$-dimensional symplectic space; The symplectic group,
denoted by $\text{Sp}\left(2N,\mathbb{R}\right)$, is the set of symplectic
transformations on $E$ equipped with the normal matrix product. The
identity\index{Identity} of the symplectic group, denoted by $\text{Id}$,
is the identity matrix\footnote{As identity matrix always serves as the identity of the algebraic
structures throughout this thesis, we will always denote identity
matrices by $\text{Id}$. We will not clarify the dimensions of an
identity matrix every time it appears unless necessary, since they
can usually be deduced from the context.}. The inverse of a symplectic transformations can be directly derived
from its defintion as below
\[
S^{-1}=-\Omega S^{t}\Omega.
\]
The above equation, in combination with the fact that the symplectic
form $\Omega$ is itself a symplectic matrix, leads to the fact that
the transpose of a symplectic matrix is also symplectic. Moreover,
a symplectic transformation $S$ has the following notable property:
\[
\text{det}\left(S\right)=1.
\]

The symplectic group $\text{Sp}\left(2N,\mathbb{R}\right)$ is a
non-compact Lie group. Its Lie algebra\index{Symplectic!Lie algebra}
$\text{sp}\left(2N,\mathbb{R}\right)$ consists of the $2N\times2N$
matrices $M$ satisfying the condition that
\begin{equation}
M\Omega+\Omega M^{t}=0.\label{eq:Lie algebra}
\end{equation}
The elements in $\text{sp}\left(2N,\mathbb{R}\right)$ are one-to-one
corresponding to the real symmetric matrices: Given any $2N\times2N$
real symmetric matrix $H$, we have $\Omega H\in\text{sp}\left(2N,\mathbb{R}\right)$,
and for any$M\in\text{sp}\left(2N,\mathbb{R}\right)$, $\Omega M$
is a real symmetric matrix. However, since the symplectic group is
non-compact, exponential map $\text{Exp}:\text{sp}\left(2N,\mathbb{R}\right)\rightarrow\text{Sp}\left(2N,\mathbb{R}\right)$,
defined by
\[
\text{Exp}\left(M\right)=e^{M},\ \forall M\in\text{sp}\left(2N,\mathbb{R}\right),
\]
is not surjective\footnote{A theorem in \cite{Cornwell1986} tells us the exponential map is
surjective if and only if the Lie group is compact, connected and
linear.}. Simply speaking, there exist symplectic transformations that cannot
be generated by elements in $\text{sp}\left(2N,\mathbb{R}\right)$
through the exponential map\footnote{This feature marks the significant difference between symplectic groups
and special orthogonal groups (so that the determinant is positive
and equal to 1). Special orthogonal groups are compact Lie groups,
so that they can be faithfully generated by their Lie algebras with
no exception.}.

Aside form the exponential map, we can also generate $\text{Sp}\left(2N,\mathbb{R}\right)$
with $\text{sp}\left(2N,\mathbb{R}\right)$ using symplectic Cayley
transform\index{Symplectic!Cayley transform}. The symplectic Cayley
transform works as follows: Let $M$ be an element of $\text{sp}\left(2N,\mathbb{R}\right)$;
Then
\begin{align}
S & =\left(M-\frac{1}{2}\text{Id}\right)\left(M+\frac{1}{2}\text{Id}\right)^{-1}\nonumber \\
 & =\text{Id}-\left(M+\frac{1}{2}\text{Id}\right)^{-1}\label{eq:Cayley_Transform}
\end{align}
is a symplectic transformation\footnote{Cayley transform is originally invented for special orthogonal groups,
and later on generated to other Lie groups with proper adaptations.}. Cayley transform is not surjective either, since it cannot generate
a symplectic transformation satisfying $\text{det}\left(\text{Id}-S\right)=0$
(e.g., $S=\text{Id}$). However, any symplectic transformation $S$
with $\text{det}\left(\text{Id}-S\right)=0$ is a product of two symplectic
transformations, both of which can be generated from elements in $\text{sp}\left(2N,\mathbb{R}\right)$
using the Cayley transform.

These properties of symplectic groups are not only mathematical concepts
but also closely related to physical processes. Actually, the exponential
map and symplectic Cayley transform provides for us convenient tools
to systematically construct symplectic transformations using physical
processes. To realize a symplectic transformation $S$, we must find
a unitary operation (since the CCRs have to be preserved), denoted
by $\hat{U}_{S}$, such that
\begin{equation}
\hat{U}_{S}\hat{x}_{j}\hat{U}_{S}^{\dagger}=\sum_{k=1}^{2N}S_{j,k}\hat{x}_{k},
\end{equation}
for any $j,k\in\left\{ 1,2,\dots,N\right\} $, so that the quadratures
can be transformed as expected in the symplectic space\footnote{In fact, any unitary operation that can transform the canonical quadrature
operators linearly as above corresponds to a symplectic transformation
in the symplectic space: Let $\hat{U}$ be a unitary operation with
$\hat{U}\hat{x}_{j}\hat{U}^{\dagger}=\sum_{k=1}^{2N}S_{j,k}\hat{x}_{k}$.
Then we have 
\begin{align*}
\hat{U}\left[\hat{x}_{j},\hat{x}_{k}\right]\hat{U}^{\dagger} & =\left[\hat{x}_{j},\hat{x}_{k}\right],\\
\Leftrightarrow[\sum_{l=1}^{2N}S_{j,l}\hat{x}_{l},\sum_{m=1}^{2N}S_{k,m}\hat{x}_{m}] & =\left[\hat{x}_{j},\hat{x}_{k}\right],\\
\Leftrightarrow\sum_{l=1}^{2N}\sum_{m=1}^{2N}S_{j,l}S_{k,m}[\hat{x}_{l},\hat{x}_{m}] & =\left[\hat{x}_{j},\hat{x}_{k}\right],\\
\Leftrightarrow\left(S^{T}\Omega S\right)_{j,k} & =\Omega_{j,k},\\
\Leftrightarrow S^{T}\Omega S & =\Omega.
\end{align*}
Therefore, the linear transformation $S$ induced by the unitary operation
$\hat{U}$ is a symplectic transformation. Such unitary operations
form a Lie group, the metaplectic group, which is denoted by $\text{Mp}\left(2N,\mathbb{R}\right)$.
Note that there are always more than one metaplecitc operations associated
to a symplecitc transormation, since a adding a global phase to a
metaplectic operation will not change the corresponding symplectic
transformation. For convenience, we will refer to the metaplectic
operators directly as the symplectic transformations.}. Here, the $\hat{x}$-operators are the aforementioned universal
notation for all canonical quadrature operators.  Now consider
the following quadratic Hamiltonian\index{Quadratic Hamiltonian}
for a bosonic system
\[
\hat{H}=\sum_{j,k=1}^{2N}H_{j,k}\hat{x}_{j}\hat{x}_{k},
\]
where the matrix $H$ on the right hand side is a $2N\times2N$ real
matrix\footnote{The set of all $i\hat{H}$'s form the metaplectic Lie algebra $\text{mp}\left(2N,\mathbb{R}\right)$,
which is the Lie algebra of the metaplectic group $\text{Mp}\left(2N,\mathbb{R}\right)$.}. Then the exponential map between $\text{sp}\left(2N,\mathbb{R}\right)$
and $\text{Sp}\left(2N,\mathbb{R}\right)$ implies that
\[
\hat{U}_{S}=e^{-i\hat{H}t}
\]
induces the symplectic transformation
\[
S=e^{\Omega Ht}.
\]
Note that the product of $\hat{U}_{S}$, $\hat{U}_{S'}$ corresponding
to two different symplectic transformations $S$ and $S'$ is a unitary
operation $\hat{U}_{S'S}$ corresponds to the symplectic transformation
$S'S$, i.e.
\[
\hat{U}_{S'S}=\hat{U}_{S}\hat{U}_{S'}.
\]

\begin{example}
\label{exa:single-mode-squeezing}Consider a single mode bosonic system
with Hamiltonian
\[
\hat{H}=\hat{q}\hat{p}+\hat{p}\hat{q}=i\left(\hat{a}^{\dagger2}-\hat{a}^{2}\right),
\]
which can be described by a real matrix
\[
H=\begin{pmatrix}0 & 1\\
1 & 0
\end{pmatrix}.
\]
Then the symplectic transformation corresponding to the unitary operation
$e^{-i\hat{H}t}$, which is the single-mode squeezing operation, is
\[
Z=e^{\Omega Ht}=\begin{pmatrix}e^{t} & 0\\
0 & e^{-t}
\end{pmatrix}.
\]
\end{example}

We have related the exponential map to the Hamiltonain evolution of
a bosonic quantum system. It turns out that the Cayley transform can
also be related to a physical process--the noiseless scattering process
of a bosonic system. This relation is based on the observation that
the symplectic Cayley transform (Eq.~\ref{eq:Cayley_Transform})
shares a similar form with the formulas for scattering amplitudes
in optics and quantum mechanics. In fact, a noiseless bosonic scattering
process is governed by a dimensionless quadratic Hamiltonian, which
is described by a real symmetric matrix $H$. Then the noiseless scattering
process, yielding a unitary operation $\hat{U}_{S}$, induces a symplectic
transformation
\[
S=\text{Id}-\left(\Omega H+\frac{1}{2}\text{Id}\right)^{-1},
\]
where we have used property that $M=\Omega H$ belongs to the symplectic
algebra $\text{sp}\left(2N,\mathbb{R}\right)$. More details about
bosonic scattering processes can be found in Appx.~\ref{chap:Scattering-process-and}.

 Sometimes the symplectic transformations obtained from a physical
process can be rather complicated. The best way to intuitively understand
the function of a complicated symplectic transformation is to decompose
it into simple fundamental mental building blocks that we are familiar
with. As we have promised, any symplectic transformation $S$ is a
composition of rotation-like symplectic transformations and squeezing-like
transformaions:
\[
S=RZR'
\]
where $R$ and $R'$ are rotation-like transformations (e.g. combinations
of phase-shifting and beam-splitting operations in quantum optics)
and $Z$ is a squeezing-like operation (for instance, the operation
in Example.~\ref{exa:single-mode-squeezing}). Specifically, $R$
and $R'$ are special orthogonal matrices satisfying
\[
RR^{t}=R'R'^{t}=\text{Id},\,\text{det}\left(R\right)=\text{det}\left(R'\right)=1,
\]
and $Z$ is a diagonal symplectic transformation. This decomposition
is known as the polar (or Euler) decomposition\index{Polar decomposition}
of symplectic transformations.

The Euler decomposition is not the only decomposition. In general,
$R,R',Z$ may not be uniquely determined (e.g. there are infinitely
many ways to decompose the identity according to the Euler decomposition).
Therefore, although it will not be applied in the rest of this thesis,
we would like to introduce the pre-Iwasawa decomposition\index{Pre-Iwasawa decomposition},
which can decompose a given symplectic transformations uniquely into
three families of building blocks. The pre-Iwasawa decomposition states
the following: Let us shift the ordering of the canonical basis from
the default ordering to the alternative ordering $\hat{q}_{1},\hat{q}_{2},\dots,\hat{p}_{N},\hat{p}_{1},\hat{p}_{2},\dots,\hat{p}_{N}$;
Then a $2N$-dimensional symplectic transformation $S$ can be decomposed
as
\begin{equation}
S=\begin{pmatrix}\text{Id} & 0\\
P & \text{Id}
\end{pmatrix}\begin{pmatrix}L^{t} & 0\\
0 & L^{-1}
\end{pmatrix}\begin{pmatrix}V & W\\
-W & V
\end{pmatrix},
\end{equation}
where $P$ is an $N$-order symmetric matrix, $L$ is a invertible
$N$-order square matrix, and $V,W$ are $N$-order square matrices
satisfying that $U=V+iW$ is a unitary matrix. The three factors of
this decomposition
\[
\begin{pmatrix}\text{Id} & 0\\
P & \text{Id}
\end{pmatrix},\ \begin{pmatrix}L^{t} & 0\\
0 & L^{-1}
\end{pmatrix},\ \begin{pmatrix}V & W\\
-W & V
\end{pmatrix}
\]
are all symplectic transformations. Physically, the first factor may
be referred to as a quantum-non-demolition operation (e.g. a symplectic
transformation generated by the Hamiltonian $\hat{H}=\hat{q}_{1}\hat{q}_{2}$);
the second factor corresponds to a multi-mode squeezing operation;
the third factor is an orthogonal matrix (a combination of phase-shifting
and beam-splitting). In fact, any orthogonal and symplectic transformation
is of the form the third factor of the pre-Iwasawa decomposition as
shown above.

\chapter{Gaussian states and Gaussian processes\label{chap:Gaussian-states-and}}

\section{Introduction}

In the previous chapter, we established symplectic geometrical concepts
for bosonic quantum systems. We mentioned that due to the probabilistic
feature of quantum mechanics, symplectic transformations are not only
tools for investigating kinetics but also important operations for
preparing or manipulating quantum states. We have also seen how the
linear algebraic essence of symplectic transformations allow us to
calculate, operate, and understand symplectic transformations. Therefore,
if quantum states can also be represented by linear algebraic objects
in symplectic space such as vectors or simple real matrices, we can
imagine how conveniently and efficiently the calculation of the evolution
of a bosonic quantum system can be executed on a classical computer.
However, as we know quantum computation is superior to classical computation
and quantum circuits in general cannot be simulated efficiently on
a classical computer, we must restrict our search of quantum states
with such good properties to a subset of all quantum states. These
states exist and known as the Gaussian states. They are the quantum
states of which the Wigner functions are Gaussian functions. Since
Gaussian functions are completely determined by the mean value (first
moments) and the variance (the covariance matrix or the second moments),
the evolution of a Gaussian state under a symplectic transformation
(the state will always be Gaussian during the evolution) can be easily
captured by the corresponding linear algebraic calculations. Gaussian
states are thus the best quantum analogues of the classical states
and will be used to simulate practical quantum noise in the following
chapters.

Symplectic transformations belong to a family of quantum processes,
the members of which transform Gaussian states to Gaussian states.
We call such quantum processes the Gaussian processes\index{Gaussian!process}.
It turns out that if a Gaussian process is unitary, it can be either
a symplectic transformation or a displacement operation in the phase
space. The latter will be introduced and discussed in this chapter
along with its covariance with symplectic geometry which can be used
as an equivalent definition of symplectic transformations. We will
also briefly discuss the noisy Gaussian processes--the Gaussian channels.
They are convenient theoretical models of quantum noise, which can
facilitate some analyses in the later chapters.

More importantly, Gaussian states and Gaussian processes are ubiquitous
in quantum physics. For example, in quantum optics, the well-known
vacuum state, coherent states, squeezed states, thermal states, etc.
are all Gaussian states; common operations such as beam-splitting,
squeezing, phase-shifting are unitary Gaussian processes (since they
are symplectic transformations); more generally, bosonics scattering
processes generated by quadratic Hamiltonians are also Gaussian processes.
Gaussian process is also a widely used theoretical tool in Gaussian
quantum information\cite{Weedbrook2012}. In this thesis, we hope
to demonstrate that Gaussian process, especially its symplectic geometrical
features, can be applied to solving more practical quantum engineering
problems. This chapter will lay the ground for some of the key steps
of our results.

We will start with displacement operations, and then introduce Wigner
functions, Gaussian states and Gaussian channels. In the last section
of this chapter, we show our systematic method of embedding a noisy
Gaussian process in a unitary Gaussian process.

\section{Displacement operation}

Physical laws are independent of the choice of the coordinate systems,
including the origin. We have seen the CCRs are preserved under symplectic
transformations, which are coordinate transformations preserving the
position of the origin. Then it is natural to wonder what are the
CCR-preserving transformations that change the origin of the phase
space in the meanwhile. It turns out that the simplest operations
satisfying these requirements are the displacement operations\index{Displacement operation}
transforming the canonical quadrature operators as follows
\[
\hat{x}_{j}\rightarrow\hat{x}_{j}+u_{j},
\]
where $u$ is a real vector. We denote this operation by a map $\mathcal{D}_{u}:E\rightarrow E$,
where $E$ is a $2N$-dimensional symplectic space. It follows that
these maps form an Abelian group: For any pair of displacement operations
$\mathcal{D}_{u}$ and $\mathcal{D}_{u'}$, we have
\[
\mathcal{D}_{u}\mathcal{D}_{v}=\mathcal{D}_{u+v},
\]
and the identity of this group is obviously the operation $\mathcal{D}_{u=0}$.

Physically, the displacement operations are realized by the well-known
displacement operators. The physical embodiment of the displacement
operations can be easily constructed using the quadrature operators.
That is, for all $u\in\mathbb{R}^{2N}$, we have a displacement operator
of the following form\footnote{Usually in quantum optics, the displacement vector $u$ is a complex
$N$-dimensional vector. Here we take its real and imaginary parts
separately to form a $2N$-dimensional real vector}:
\[
\hat{D}_{u}=e^{2i\sum_{j=1}^{2N}(\Omega u)_{j}\hat{x}_{j}}.
\]
It follows that any quadrature operator $\hat{v}=\sum_{j}v_{j}\hat{x}_{j}\in E$
is transformed under the displacement operation to 
\[
\hat{D}_{u}\hat{v}\hat{D}_{u}^{\dagger}=\hat{v}+u.
\]
Also, for any $2N$-dimensional real vectors $u,v\in\mathbb{R}^{2N}$,
we have
\begin{align*}
\hat{D}_{u+v}\hat{w}\hat{D}_{u+v}^{\dagger} & =\hat{D}_{u}\hat{D}_{v}\hat{w}\hat{D}_{v}^{\dagger}\hat{D}_{u}^{\dagger}\\
 & =\hat{w}+u+v,
\end{align*}
for any quadrature operator $\hat{w}=\sum_{j=1}^{2N}w_{j}\hat{x}_{j}\in E$.
In particular, we have$\hat{D}_{u}^{\dagger}=\hat{D}_{-u}$. These
relations confirm that the displacement operators are the right physical
realization of $\mathcal{D}_{u}$'s.

Moreover, the displacement operations are covariant with the symplectic
transformations: Let $S$ be a symplectic transformation in the symplectic
space $E$; then we have
\begin{align*}
\hat{U}_{S}\hat{D}_{u}\hat{U}_{S}^{\dagger} & =e^{i\sum_{j,k=1}^{2N}(\Omega u)_{j}S_{jk}\hat{x}_{k}}\\
 & =e^{i\sum_{j=1}^{2N}(S^{t}\Omega u)_{j}\hat{x}_{k}}\\
 & =e^{i\sum_{j=1}^{2N}(\Omega S^{-1}u)_{j}\hat{x}_{k}}\\
 & =\hat{D}_{S^{-1}u}.
\end{align*}
In other words, symplectic transformations preserves the groups
of the displacement operations. Actually, the set of all the unitary
operations that preserve the group of displacement operations also
form a group, containing the unitary operator that may be best addressed
as the Gaussian unitary operators. In fact, a Gaussian unitary operator
$\hat{U}$ is always the product of a symplectic transformation and
a displacement operation, that is
\[
\hat{U}=\hat{D}_{u}\hat{U}_{S}
\]
for some $u\in\mathbb{R}^{2N}$ and $S\in\text{Sp}\left(2N,\mathbb{R}\right)$.
So to speak, symplectic transformations can be defined equivalently
as the quotient group of the group of Gaussian unitary operators modulo
the group of displacement operations.

Last but not the least, exchanging the orders of two displacement
operations in their product will yield an extra phase:
\begin{equation}
\hat{D}_{u}\hat{D}_{v}=e^{-i\sigma\left(u,v\right)}\hat{D}_{v}\hat{D}_{u},\label{eq:phase}
\end{equation}
where $\sigma\left(u,v\right)$ is the symplectic form. Then the covariance
of the displacement operation with the symplectic transformations
amounts to the preservation of this phase:
\begin{equation}
\hat{D}_{S^{-1}u}\hat{D}_{S^{-1}v}=e^{-i\sigma\left(u,v\right)}\hat{D}_{S^{-1}v}\hat{D}_{S^{-1}u}.
\end{equation}
Moreover, we can view it as an equivalent definition of the symplectic
transformations, which will be useful later in this thesis.

The displacement operators are crucial to define Wigner functions
in the next section. Now we can proceed to discussing this important
quasi-probabilistic representation of quantum states. 

\section{Wigner function}

The probabilistic nature of quantum states complicates their representation
in the phase space. Due to the Heisenberg's uncertainty principle,
the momenta and positions of a physical system cannot be precisely
measured at the same time. Therefore, a point in the phase space can
no longer faithfully describe a physical state. To best reflect this
probabilistic nature, a faithful representation of a quantum state
in phase space should assign to each point a ``probability'' showing
how possible the system is of the corresponding momenta and positions.
It turns out these requirements give rise to the concepts of quasi-probabilistic
distributions--functions mapping each point in the phase space to
a real/complex value and sharing important properties with a normal
probability distribution. Among possible constructions of quasi-probabilistic
distributions, we are particularly interested in the Wigner function:
It is real-valued distribution that is equivalent to the usual probabilistic
distribution\footnote{For example, the phase representation of a statistical ensemble in
classical mechanics. } for certain quantum states. In this and the succeeding sections,
we discuss the general properties and useful instances of the Wigner
function, which will be crucial to the rest of this thesis\footnote{We will stick to the conventions in \cite{Weedbrook2012}. Rigorous
proofs of the statements in this section can be found in \cite{DeGosson2006}.}.

Consider an $N$-mode bosonic system with a $2N$-dimensional symplectic
space $E$. The set of the displacement operators is a complete basis
of quantum operators, i.e., any quantum operator $\hat{A}$\footnote{Aside from some mathematical exceptions, we assume no constraint on
the form of operator $\hat{A}$. The operator $\hat{A}$ can be Hermitian,
unitary, non-Hermitian, non-unitary,etc.} can be represented by a linear combination of displacement operators.
Specifically, given an operator $\hat{A}$, the complex coefficient
of this linear combination can be obtained from the following formula:
\begin{equation}
\chi_{\hat{A}}(v)=\text{Tr}[\hat{A}\hat{D}_{v}],
\end{equation}
which is justified by the property
\[
\text{Tr}[\hat{D}_{v}]\neq0,\ \text{if and only if }v=0,\ \forall v\in E
\]
of the displacement operators \footnote{This property can be consider ed to be a generalization of Fourier
transform, which originated from the resemblance between the displacement
operators and the $e^{ikx}$'s in the Fourier transform.}. Therefore, when applied to a quantum state $\hat{\rho}$, this
defines a function mapping each point in the symplectic space to a
complex value. We call $\chi_{\hat{A}}$ the characteristic function\index{Characteristic function}
of $\hat{A}$. It follows that based on the characteristic function
can be further converted into a function, the Wigner function\index{Wigner function}
$W_{\hat{A}}$, through the symplectic Fourier transform\index{Symplectic!Fourier transform},
as follows:
\begin{equation}
W_{\hat{A}}\left(u\right)=\int_{E}\frac{d^{2N}v}{(2\pi)^{2N}}e^{-iu^{t}\Omega v}\chi_{\hat{A}}(v),
\end{equation}
where the integration is carried out over the whole symplectic space.
In particular, when $\hat{A}$ is Hermitian (for example, $\hat{A}$
being the density operator of a quantum state), the Wigner function
$W_{\hat{A}}$ is a real-valued function.
\begin{example}
The Winger functions of the canonical quadrature operators $\hat{q}_{1}$,
$\hat{p}_{1}$, $\hat{q}_{2}$, $\hat{p}_{2}$, $\dots$, $\hat{q}_{N}$,
$\hat{p}_{N}$ are:
\[
W_{\hat{q}_{1}}\left(u\right)=u_{1},\,W_{\hat{p}_{1}}\left(u\right)=u_{2},\,\dots
\]
i.e., the classical canonical coordinates in the phase space.
\end{example}

When applied to a density operator, the Wigner function is a faithful
quasi-probabilistic representation of the quantum state. This can
be seen from the following properties of the Wigner function:

Let $\hat{\rho}$ be the density operator of a quantum state (no matter
it is pure or mixed):
\begin{enumerate}
\item The Wigner functions are normalized. For a Wigner function $W_{\hat{\rho}}$
of any density operator $\hat{\rho}$, we have
\begin{equation}
\int_{E}W_{\hat{\rho}}(v)dv=1
\end{equation}
\item Let $\hat{O}$ be an observable. Then its expectation value can be
calculated by
\end{enumerate}
\begin{equation}
\langle\hat{O}\rangle_{\hat{\rho}}=\text{Tr}\left[\hat{\rho}\hat{O}\right]=\int_{E}W_{\hat{O}}(v)W_{\hat{\rho}}(v)dv.
\end{equation}
Although the above properties seem to be the same as those of a usual
probabilistic distribution, the Wigner function is still quasi-probabilistic
insofar as the Wigner functions of certain quantum states (e.g., the
cat states) can take negative values on some points in the symplectic
space, in contrast to the always-positive property of a probabilistic
distribution.

Moreover, Wigner functions are covariant with the displacement operations
and symplectic transformations:

Let $\hat{A}$ be an operator:
\begin{enumerate}
\item Action of displacement operations: 
\begin{equation}
W_{\hat{D}_{v}\hat{A})\hat{D}_{v}^{\dagger}}(u)=W_{\hat{A}}(u-v)\label{action_of_displacement}
\end{equation}
\item Action of symplectic transformations: 
\begin{equation}
W_{\hat{U}_{S}\hat{A}\hat{U}_{S}^{\dagger}}(u)=W_{\hat{A}}(S^{-1}u)\label{action_of_symplectic_transformations}
\end{equation}
\end{enumerate}
These properties allow us to conveniently manipulate the Wigner functions
of a quantum state using symplectic transformations and displacement
operations. In addition, the Wigner functions are compatible with
more quantum operations that are commonly used in analyzing quantum
processes:

Let $\hat{O}_{1}$, $\hat{O}_{2}$ be Hermitian operators:
\begin{enumerate}
\item Trace and partial trace: Let $\mathcal{H}$ denote the $N$-bosonic-mode
Hilbert space. Let $\mathcal{H}_{\alpha}$ be an $M$-bosonic-mode
Hilbert space (thus a subspace of $\mathcal{H}$). Then 
\begin{equation}
W_{\text{tr}_{\mathcal{H}_{\alpha}}[\hat{O}_{1}\hat{O}_{2}]}(u)=\int_{E_{\alpha}}W_{\hat{A}_{1}}(v)W_{\hat{A}_{2}}(v)dv_{\alpha}
\end{equation}
with the right-hand-side integrated over the phase space $E_{\alpha}$
associated to the Hilbert subspace $\mathcal{H}_{\alpha}$, which
is a symplectic subspace of the symplectic space $E$. 
\item Tensor product: Let $\hat{O}_{1}$ and $\hat{O}_{2}$ be Hermitian
operators on bosonic Hilbert spaces $\mathcal{H}_{1}$ and $\mathcal{H}_{2}$.
Then
\begin{equation}
W_{\hat{O}_{1}\otimes\hat{O}_{2}}(u_{1}\oplus u_{2})=W_{\hat{O}_{1}}(u_{1})W_{\hat{O}_{2}}(u_{2}).
\end{equation}
The phase space associated to the tensor-product Hilbert space $\mathcal{H}_{1}\otimes\mathcal{H}_{2}$
is the direct sum of the phase spaces $\mathcal{H}_{1}$ and $\mathcal{H}_{2}$
associated with $\mathcal{H}_{1}$ and $\mathcal{H}_{2}$ respectively.
Intuitively, this means the tensor product of Hilbert spaces can be
converted into direct sums of symplectic spaces, along with the corresponding
operations.
\end{enumerate}
Before proceeding to the succeeding section where the important instances,
the Gaussian states, are discussed, we conclude the section here with
a special family of quantum states, the infintely squeezed states\index{Infinitely squeezed state}
(or equivalently the eigenstates of quadrature operators, for example
the eigenstate of $\hat{p}$ with wave function being a propagating
wave $\psi\left(q\right)\propto e^{ipq}$ and the eigenstate of $\hat{q}$
with wave function being a delta function $\psi\left(q\right)\propto\delta\left(q-q'\right)$).
Physically speaking, such states do not exist in practice since their
wave functions cannot be normalized. However, these states turn out
to be useful theoretical tools because of the convenient mathematical
properties of delta functions. Consider the infintely squeezed state
which is an eigenstate of $\hat{q}$. Then the deviation of the $\hat{q}$-quadrature
$\left\langle \left(\hat{q}-\left\langle \hat{q}\right\rangle ^{2}\right)\right\rangle $
is vanishing while the deviation of the $\hat{p}$-quadrature $\left\langle \left(\hat{p}-\left\langle \hat{p}\right\rangle ^{2}\right)\right\rangle $
blows up. This means, in the phase space, such a state should be strictly
represented by a straight line (as in mathematics), parallel to the
$\hat{p}$-axis. Therefore, the Wigner function of this state should
be of the form a delta function. Specifically, in a $N$-mode bosonic
system, let $l_{z}$ be a Lagrangian plane (as defined in the preceding chapter) of the symplectic space, which is, simply speaking, spanned
by $N$ quadratures of a symplectic basis of the space (so the other
half of the basis are conjugate with them in respect of the CCRs).
Then this Lagrangian plane determines a infintely squeezed (a common
eigenstate of the $N$ quadrature operators), of which the Wigner
function is given by\footnote{We pretend this state can be normalized by forcing $\int_{E}W_{\hat{\Pi}_{l_{z}}\left(\eta\right)}\left(u\right)=1$.}
\[
W_{\hat{\Pi}_{l_{z}}\left(\eta\right)}\left(u\right)\propto\delta\left(u-\eta\right).
\]
Here $\hat{\Pi}_{l_{z}}\left(\eta\right)$ denotes the ``density
operator'' of the corresponding infinitely squeezed state, with $l_{z}$
being the determinant Lagrangian plane and $\eta$ an $N$-dimensional
vector consisting of the eigenvalues of the $N$ determinant quadrature
operators.
\begin{example}
Consider a Lagrangian plane spanned by $\hat{q}_{1},\hat{q}_{2},\dots,\hat{q}_{N}$.
Then the eigenstate $\hat{\Pi}_{l_{z}}\left(\eta\right)$ denotes
an infinitely squeezed state satisfies that
\[
\text{Tr}\left[\hat{q}_{j}\hat{\Pi}_{l_{z}}\left(\eta\right)\right]=\eta_{j},\,\forall j\in\left\{ 1,2,\dots,N\right\} .
\]
\end{example}

Although the infinitely squeezed states are fictitious states introduced
for theoretical convenience, they can be approximated by practical
quantum states, such as the Gaussian states. In addition, Gaussian
states are best analogues of classical states and are handy models
of quantum noise. In the succeeding section, we introduce and discuss
the definition and properties of the Gaussian states.

\section{Gaussian states}

Statistical ensemble of classical states can be described by a probabilistic
distribution, assigning to each point in the phase space a (non-negative)
probability. Therefore, we could expect the Winger function of a ``classical''
quantum state to be non-negative over the whole symplectic space.
According to the findings in \cite{Hudson1974}, a pure quantum state
is ``classical'' in this sense if and only if its Wigner function
is a Gaussian function. Therefore, we consider the quantum states,
of which the Wigner functions are Gaussian functions, to be the best
analogues of classical states and refer to them as the Gaussian states\footnote{We adopt the conventions of \cite{Weedbrook2012} in this section.}.

Specifically, the Wigner function of a Gaussian state\index{Gaussian!state}
$\hat{\rho}$ is of the following form 
\begin{equation}
W_{\hat{\rho}}(u)=\frac{e^{-\frac{1}{2}(u-\bar{x})^{t}V^{-1}(u-\bar{x})}}{(2\pi)^{N/2}\sqrt{\det V}}
\end{equation}
where the vector
\begin{equation}
\bar{x}=\langle\hat{x}\rangle
\end{equation}
consists of the expectation values of the canonical quadrature operators,
called the\emph{ }\textit{\emph{first moments}}\index{First moments@\textit{\emph{First moments}}},
and the covariance matrix\index{Covariance matrix}
\begin{equation}
V_{j,k}=\langle(\hat{x}_{j}-\bar{x}_{j})(\hat{x}_{k}-\bar{x}_{k})\rangle,\ \forall j,k\in\{1,2,\dots,2N\},
\end{equation}
also known as the \textit{\emph{second moments}}, consists of the
second order correlation between quadrature operators. Intuitively,
we can imagine that a Gaussian state to be a smeared blob in the phase
space, with the position of its center given by the vector $\bar{x}$
and its size as well as shape given by the covariance matrix\footnote{In classical mechanics, the inertia tensor, which is simply a real
symmetric matrix, assigns to each rigid body an ellipsoid. The three
principal components of the inertia tensor determines the stretches
of the ellipsoid along the three principal axes. Similarly, the covariance
matrix assigns to each Gaussian state a multi-dimensional ellipsoid.
It can be diagonalized by rotating the the coordinate system, and
the diagonal elements of the diagonalized covariance matrix thus determines
the size and shape of the multi-dimensional ellipsoid.}.

It turns out under the displacement operations and symplectic transforamtions,
a Gaussian state will still be transformed to a Gaussian state. Moreover,
since the a Gaussian state is completely determined by the first moments
$\bar{x}$ and the covariance matrix $V$, the covariance of its Wigner
function with the these operations can be reflected by the change
of the first moments and the covariance matrix, which can be easily
calculated using simple linear algebra. Specifically, for a Gaussian
state $\hat{\rho}$ with first moments being $\bar{x}$ and covariance
matrix being $V$, according to Eqs.~\ref{action_of_displacement}
\& \ref{action_of_symplectic_transformations}, we have
\begin{align*}
W_{\hat{U}_{S}\hat{D}_{v}\hat{\rho}\hat{D}_{v}^{\dagger}\hat{U}_{S}^{\dagger}}(u) & =\frac{e^{-\frac{1}{2}(S^{-1}(u-v)-\bar{x})^{t}V^{-1}(S^{-1}\left(u-v\right)-\bar{x})}}{(2\pi)^{N/2}\sqrt{\det V}}\\
 & =\frac{e^{-\frac{1}{2}(u-v-S\bar{x})^{t}\left(SV\left(S^{t}\right)\right)^{-1}(u-v-S\bar{x})}}{(2\pi)^{N/2}\sqrt{\det\left(SVS^{t}\right)}}
\end{align*}
for any symplectic transformation $S$ and displacement operation
$\hat{D}_{v}$, which represents a Gaussian state $\hat{\rho}'$ with
first moments
\begin{equation}
\bar{x}'=S\bar{x}+v
\end{equation}
and covariance matrix
\begin{equation}
V'=SVS^{t}.
\end{equation}
In consideration of this nice property, we denote a Gaussian state
using its first moments $\bar{x}$ and covariance matrix $V$ by $\hat{\rho}\left(\bar{x},V\right)$.
\begin{example}
The above calculation describes the following transformation between
Gaussian states:
\[
\hat{\rho}\left(\bar{x},V\right)\rightarrow\hat{\rho}\left(S\bar{x}+v,SVS^{t}\right).
\]
\end{example}

We now demonstrate some examples of Gaussian states, which are important
quantum states in quantum optics.
\begin{example}
(Vacuum state) The simplest example of a Gaussian state is the vacuum
state $|0\rangle\langle0|$, i.e. the pure bosonic quantum state that
can be annihilated by every bosonic annihilation operator of the system.
The first moments of $|0\rangle\langle0|$ is
\[
\bar{x}=0,
\]
and the covariance matrix is
\[
V=\text{Id}.
\]
Therefore, we can also represent the vacuum state by $\hat{\rho}\left(0,\text{Id}\right)$.
\end{example}

\begin{example}
(Coherent state) A \textit{\emph{coherent states}} $|\alpha\rangle$,
which are defined as a state satisfying $\hat{a}_{j}|\alpha\rangle=\alpha_{j}|\alpha\rangle$
with $\alpha_{j}$ being complex number for all $j\in\{1,2,\dots,N\}$,
can be created out of the vacuum state by applying the displacement
operation:
\begin{equation}
\hat{D}_{u}|0\rangle=|\alpha\rangle.
\end{equation}
Here we let $u_{2j-1}=\Re[\alpha_{j}]$ (the real part of $\alpha_{j}$)
and $u_{2j}=\Im[\alpha]$ (the imaginary part of $\alpha_{j}$) for
all $j=\{1,2,\dots,N\}$. The first moments and the covaraince matrix
of such a coherent state are respectively
\[
\bar{x}=u,
\]
and
\[
V=\text{Id}.
\]
We can represent such a coherent state by $\hat{\rho}\left(u,\text{Id}\right)$.
\end{example}

\begin{example}
(Thermal state) Gaussian states can also be mixed states. Let $\hat{\rho}_{th}$
be a thermal state satsifying $\text{Tr}[\hat{\rho}\hat{n}_{j}]=\bar{n}_{j}$
with $\hat{n}_{j}=\hat{a}_{j}^{\dagger}\hat{a}_{j}$ for $j\in\{1,2,\dots,N\}$
(i.e. different modes are surrounded by different thermal baths).
The thermal state $\hat{\rho}_{th}$ is a Gaussian state, with the
first moments being
\[
\bar{x}=0,
\]
and the covariance matrix being
\[
V=\begin{pmatrix}2\bar{n}_{1}+1 & 0 & \cdots & 0 & 0\\
0 & 2\bar{n}_{1}+1 & \cdots & 0 & 0\\
\vdots & \vdots & \ddots & 0 & 0\\
0 & 0 & 0 & 2\bar{n}_{N}+1 & 0\\
0 & 0 & 0 & 0 & 2\bar{n}_{N}+1
\end{pmatrix}.
\]
\end{example}

We have seen the covaraince matrix plays a crucial role in describing
Gaussian states. The following theorem allows to reveal the structure
the covaraince matrix in great detail:
\begin{thm}
(Williamson's theorem for strictly positve symmetric matrices\footnote{The proof of this theorem can be found in \cite{DeGosson2006}})
Let $V$ be a $2n$-dimensional real symmetric matrix, with strictly
positive eigenvalues. There exists a symplectic matrix $S\in\mathrm{Sp}(2n,\mathbb{R})$,
such that $SVS^{t}$ is a diagonal matrix with positive eigenvalues
$\{\bar{n}_{1},\bar{n}_{1},\bar{n}_{2},\bar{n}_{2},\dots,\bar{n}_{N},\bar{n}_{N}\}$.
\end{thm}

\noindent Simply speaking, this theorem tells us that given a Gaussian
state $\hat{\rho}\left(\bar{x},V\right)$can be transformed into a
thermal state by displacement operations and symplectic transformations:

\begin{equation}
\hat{\rho}_{th}=\hat{U}_{S}\hat{D}_{-\bar{x}}\hat{\rho}_{th}\hat{D}_{\bar{x}}^{\dagger}\hat{U}_{S}^{\dagger}.
\end{equation}
Here, we let $S$ be a symplectic transformation that can diagonalize
the covariance matrix $V$ using the above theorem. And the thermal
excitation numbers, i.e. $n_{j}=\text{Tr}\left[\hat{\rho}_{th}\hat{a}_{j}^{\dagger}\hat{a}_{j}\right]$
, for each mode are can be extracted from the diagonalized covariance
matrix of the resulting thermal state $\hat{\rho}_{th}$.

\begin{figure}[h]
\includegraphics[width=0.9\textwidth]{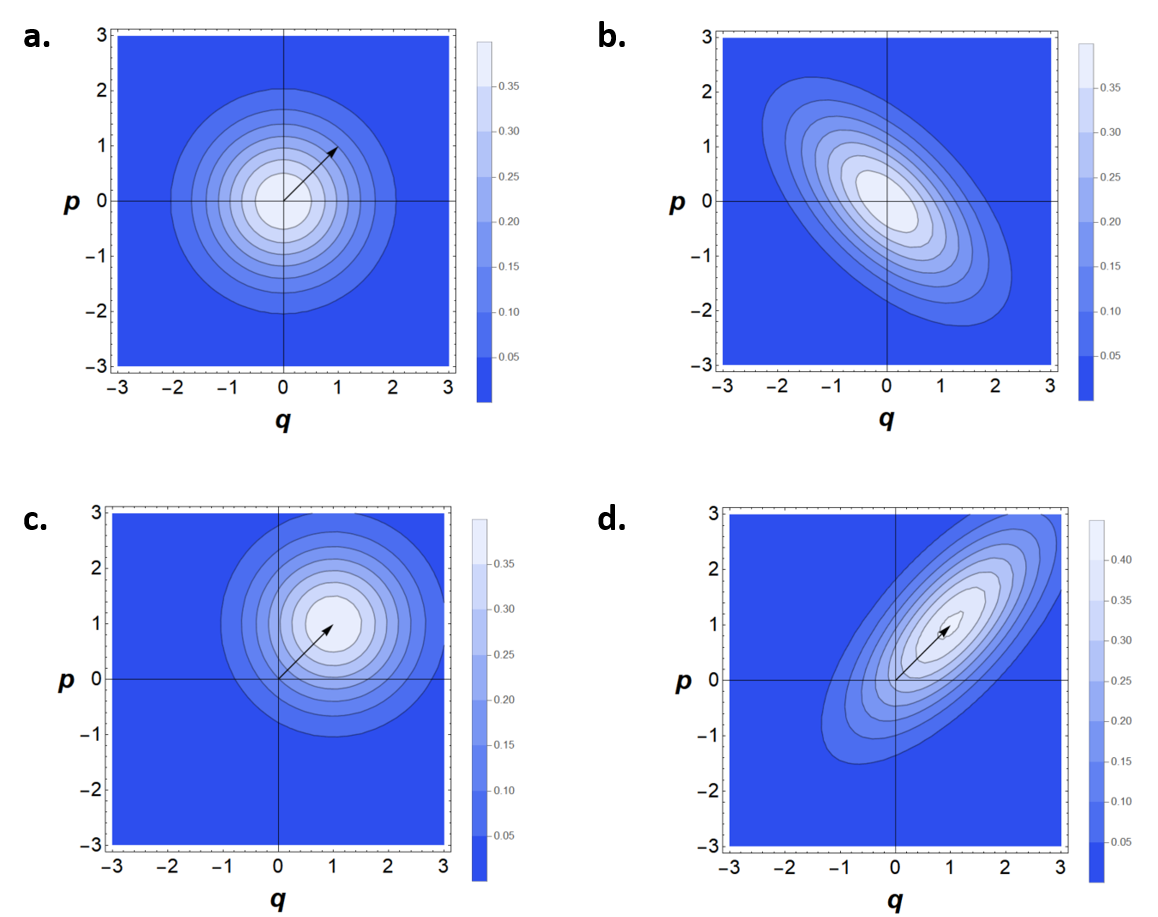}
\begin{doublespace}
\centering{}\caption{\emph{Gaussian states.}}
\begin{minipage}[t]{0.9\textwidth}%
\begin{doublespace}
These are the illustrations of Wigner functions of some Gaussian states
in the symplectic space: (a) A vacuum state. (b) A squeezed state.
(c) A coherent state with first moments $(1,1)^{t}$. (d) A randomly
generated Gaussian state.
\end{doublespace}
\end{minipage}
\end{doublespace}
\end{figure}

In the end of the preceding section, we mentioned that the fictitious
infinitely squeezed states can be approximated by Gaussian states.
To demonstrate it, we consider the following squeezing-like symplectic
transformation
\[
Z(\zeta)=\begin{pmatrix}\zeta & 0 & \cdots & 0 & 0\\
0 & \zeta^{-1} & \cdots & 0 & 0\\
\vdots & \vdots & \ddots & 0 & 0\\
0 & 0 & 0 & \zeta & 0\\
0 & 0 & 0 & 0 & \zeta^{-1}
\end{pmatrix},
\]
with $\zeta$ being a nonzero real number. Then given any Gaussian
state $\hat{\rho}\left(\bar{x},V\right)$, we can generate an infinitely
squeezed state
\[
\hat{\Pi}_{l_{z}}\left(\eta\right)=\lim_{\zeta\rightarrow0}\hat{U}_{Z\left(\zeta\right)}\hat{\rho}\left(\bar{x},V\right)\hat{U}_{Z(\zeta)}^{\dagger},
\]
where $l_{z}$ is a Lagrangian plane spanned by all the $\hat{q}$-quadratures
(seen from the form of $Z\left(\zeta\right)$), and $\eta$ is a $N$-dimensional
vector defined by
\[
\eta_{j}=\bar{x}_{2j-1},\,\forall j\in\left\{ 1,2,\dots,N\right\} .
\]
Therefore, when $\zeta$ is small enough (i.e. $\zeta^{-1}$ is large
enough), the Gaussian state $\hat{U}_{Z\left(\zeta\right)}\hat{\rho}\left(\bar{x},V\right)\hat{U}_{Z(\zeta)}^{\dagger}$
is a good approximate of the infinitely squeezed state $\hat{\Pi}_{l_{z}}\left(\eta\right)$.
It is not difficult to realize that every infinitely squeezed state
can be generated in a similar way. 

\begin{figure}[h]
\includegraphics[width=0.9\textwidth]{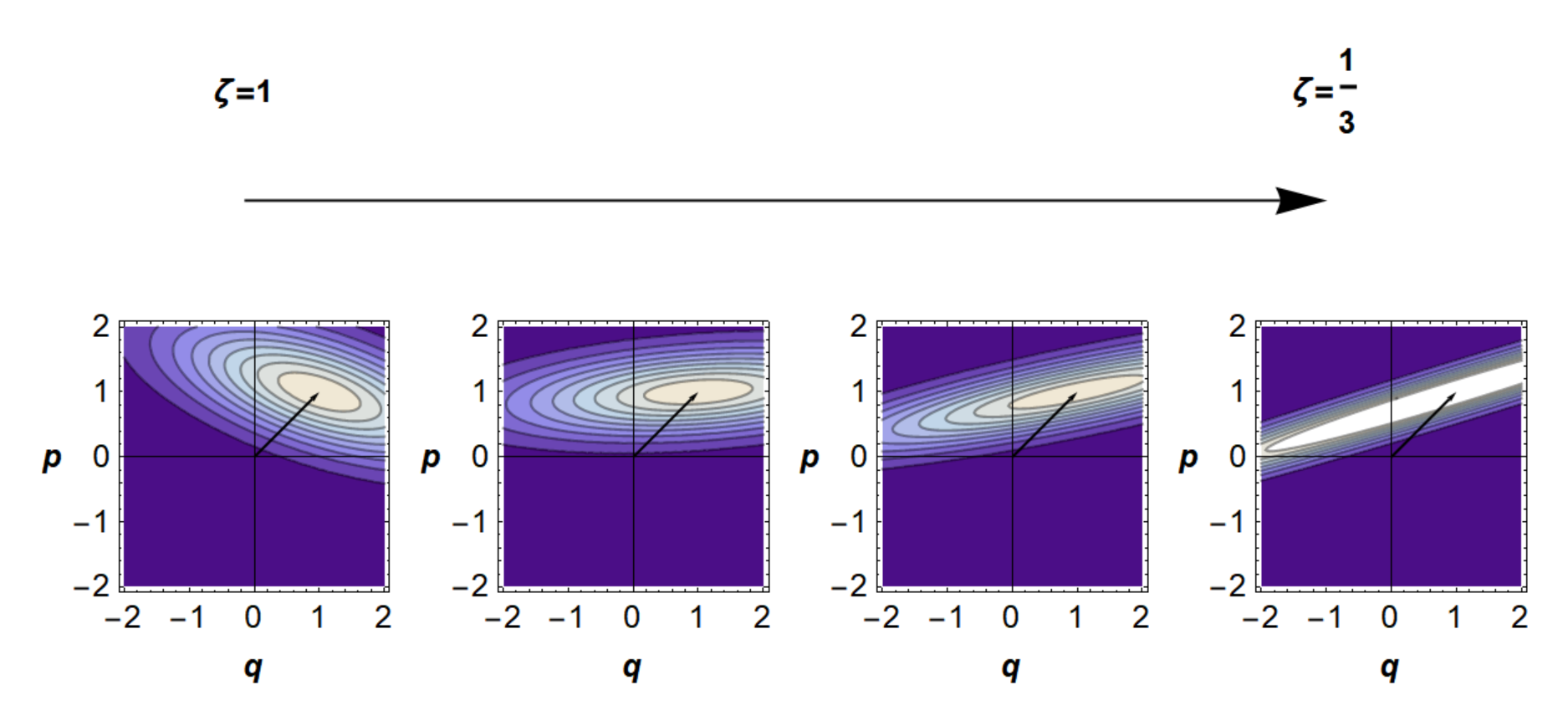}
\begin{doublespace}
\centering{}\caption{\emph{Approximating an infinitely squeezed state.}}
\begin{minipage}[t]{0.9\textwidth}%
The four figures shows how a Gaussian state can be approximately converted
into an infinitely squeezed state under squeezing operations with
increasing amount of squeezing, quantified by the real number $\zeta$.
The initial Gaussian state and the squeezing operation are randomly
chosen.%
\end{minipage}
\end{doublespace}
\end{figure}

\section{Gaussian measurement\label{sec:Gaussian-measurement}}

The Wigner functions can also be applied to describing quantum measurements.
In a discrete variable quantum system, a measurement scheme is described
by a positive-operator-valued measure (POVM), which is a set of positive
operators $\left\{ \hat{P}_{j}\right\} $, satisfying the normalization
condition $\sum_{j}\hat{P}_{j}=\text{Id}$ where the subscript $j$
of each $\hat{P}_{j}$ represents the corresponding measurement outcome.
The probability $p\left(j\right)$ of obtaining such an outcome $j$
by measuring a quantum state $\hat{\rho}$ is determined by
\[
p\left(j\right)=\text{Tr}\left[\hat{\rho}\hat{P}_{j}\right].
\]
In continuous variable systems, the measurement outcomes can be taken
from a continuum\footnote{For example, the outcomes of the heterodyne measurement can be considered
to be real-valued vectors.}. For such measurement schemes, the corresponding POVMs consist of
infintely many positive operators $\hat{P}\left(\eta\right)$, with
measurement outcomes $\eta$ which is in general a real-valued vector,
satisfying $\int\hat{P}\left(\eta\right)d\eta=\text{Id}$\footnote{The formulas for the discrete variable systems can also be expressed
as integrals according to measure theory in mathematics.}. Then the formula
\[
p\left(\eta\right)=\text{Tr}\left[\hat{\rho}\hat{P}\left(\eta\right)\right]
\]
gives a probabilistic distribution of the measurement outcomes. According
to the property of Wigner function in the preceding section, this
formula is equivalent to the following
\[
p\left(\eta\right)=\int_{E}W_{\hat{\rho}}\left(u\right)W_{\hat{P}\left(\eta\right)}\left(u\right)du,
\]
where $W_{\hat{\rho}}$ and $W_{\hat{P}\left(\eta\right)}$ are the
Wigner functions of the corresponding operators.

As we have seen in the preceding section, Gaussian states have easy-to-handle
Wigner functions. Therefore, we are interested in measurement schemes,
POVM of which consists of a family of Gaussian density operators $\hat{\rho}\left(\bar{x}-\eta,V\right)$,
where $\eta$, a $2N$-dimensional real vector, denotes a measurement
outcome. We call such measurement schemes Gaussian measurements\index{Gaussian!measurement}.
In particular, as we mentioned in the preceding section, since Gaussian
states can approximate infinitely squeezed states, Gaussian measurements
also include measurement schemes with POVM consisting of infinitely
squeezed states $\hat{\Pi}_{l_{z}}\left(\eta\right)$ determined by
a fixed Lagrangian plane $l_{z}$. Here the measurement outcome $\eta$
is an $N$-dimensional real vector.
\begin{example}
(Ideal heterodyne measurement) Heterodyne measurement is an example
of Gaussian measurements. The POVM of the ideal heterodyne measurement\index{Ideal!heterodyne measurement}
scheme consists of Gaussian states $\hat{\rho}\left(\eta,\text{Id}\right)$.
\end{example}

\begin{example}
(Ideal homodyne measurement) The ideal homodyne measurement\index{Ideal!homodyne measurement}
is described by the POVM consisting of the infinitely squeezed states
$\hat{\Pi}_{l_{z}}\left(\eta\right)$, with $l_{z}$ being a Lagrangian
plane.
\end{example}

\begin{figure}[h]
\includegraphics[width=0.9\textwidth]{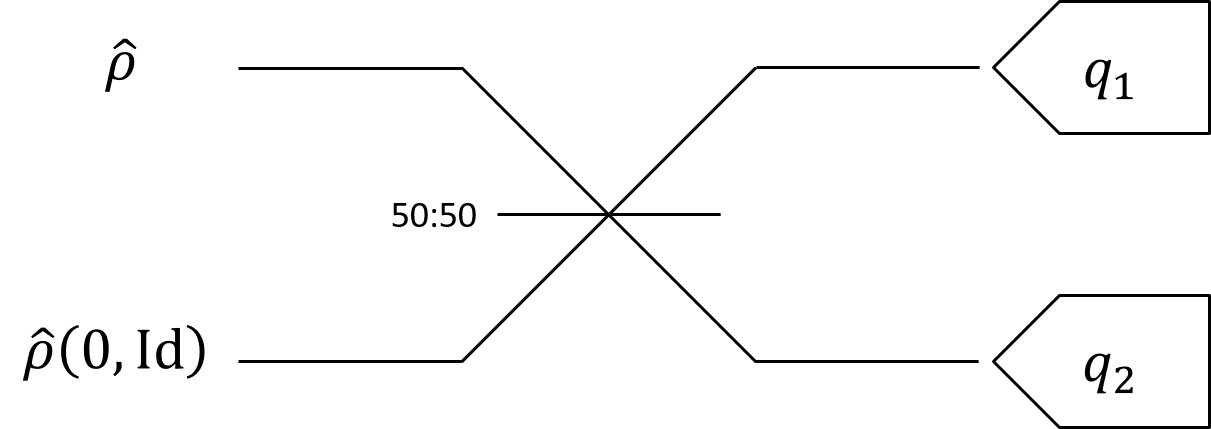}
\begin{doublespace}
\centering{}\label{Ideal heterodyne measurement}\caption{\emph{The ideal heterodyne measurement.}}
\begin{minipage}[t]{0.9\textwidth}%
\begin{doublespace}
The density operator $\hat{\rho}$ represents the quantum state to
be measured. The $\hat{\rho}\left(0,\text{Id}\right)$ is a Vacuum
ancillary state. These states are mixed by a balanced beamsplitter,
of which the output modes are measured by two ideal homodyne measurements
measuring the $\hat{q}$-quadratures. The measurement results, grouped
as a two dimensional real vector or a complex number, are then the
outcome of an ideal heterodyne measurement.
\end{doublespace}
\end{minipage}
\end{doublespace}
\end{figure}

Generally speaking, the theoretical model of a general Gaussian measurement
can be decomposed into the combination of Gaussian states, symplectic
transformations, and ideal homodyne measurements\footnote{For a similar statement, see discussions in Sec. 4.5.2 of \cite{Wiseman2009}
about the ``general dyne'' measurement.}. We start with the ideal heterodyne measurement (Fig.~\ref{Ideal heterodyne measurement})
for an instance. let $\hat{U}_{H(1/2)}$ be a symplectic transformation
with
\[
H(1/2)=\begin{pmatrix}\sqrt{1/2} & 0 & \sqrt{1/2} & 0\\
0 & \sqrt{1/2} & 0 & \sqrt{1/2}\\
-\sqrt{1/2} & 0 & \sqrt{1/2} & 0\\
0 & -\sqrt{1/2} & 0 & \sqrt{1/2}
\end{pmatrix},
\]
describing a balanced beamsplitter. Then the probability of obtaining
the outcome $\eta$ through the ideal heterodyne measurement is\footnote{For convenience, we do not track the normalization constant since
the can always be obtained afterwards.}
\[
p\left(\eta\right)\propto\text{Tr}\left[\hat{\Pi}_{l_{z},0}\left(\eta\right)\hat{U}_{H\left(1/2\right)}\left(\hat{\rho}\otimes\hat{\rho}_{env}\right)\hat{U}_{H\left(\tau\right)}^{\dagger}\right],
\]
where $l_{z}$ is the Lagrangian plane spanned by the $\hat{q}$-quadratures. 

Now we look into a more general situation. Let $\hat{\rho}_{env}$
be a Gaussian state,
\[
H\left(\tau\right)=\begin{pmatrix}\sqrt{1-\tau} & 0 & \sqrt{\tau} & 0\\
0 & \sqrt{1-\tau} & 0 & \sqrt{\tau}\\
-\sqrt{\tau} & 0 & \sqrt{1-\tau} & 0\\
0 & -\sqrt{\tau} & 0 & \sqrt{1-\tau}
\end{pmatrix}
\]
be a beamsplitter-like symplectic transformation with transmittance
satisfying $0\leq\tau\leq1$, and $l_{z}$ a Lagrangian plane. Then
we have
\[
p\left(\eta\right)\propto\text{Tr}\left[\hat{\Pi}_{l,0}\left(\eta\right)\hat{U}_{H\left(\tau\right)}\left(\hat{\rho}\otimes\hat{\rho}_{env}\right)\hat{U}_{H\left(\tau\right)}^{\dagger}\right],
\]
which translate to the language of Winger functions as follows:
\begin{align*}
p\left(\eta\right)\propto & \int_{E\cong E_{1}\oplus E_{2}}\delta\left(u|_{l}-\eta\right)W_{\mathcal{U}_{H}\left(\hat{\rho}\otimes\hat{\rho}_{env}\right)}\left(u\right)du\\
= & \int_{l'}W_{\hat{\rho}}\left(\begin{pmatrix}\sqrt{1-\tau} & 0\\
0 & -\sqrt{\tau}
\end{pmatrix}v+\begin{pmatrix}-\sqrt{\tau} & 0\\
0 & \sqrt{1-\tau}
\end{pmatrix}\eta\right)\\
 & W_{\hat{\rho}_{env}}\left(\begin{pmatrix}\sqrt{\tau} & 0\\
0 & \sqrt{1-\tau}
\end{pmatrix}v+\begin{pmatrix}\sqrt{1-\tau} & 0\\
0 & \sqrt{\tau}
\end{pmatrix}\eta\right)dv\\
\propto & \int_{E_{1}}W_{\hat{\rho}}\left(v\right)W_{\hat{\rho}_{env}}(\begin{pmatrix}\sqrt{\tau/(1-\tau)} & 0\\
0 & -\sqrt{(1-\tau)/\tau}
\end{pmatrix}v\\
 & +\begin{pmatrix}\sqrt{1/(1-\tau)} & 0\\
0 & \sqrt{1/\tau}
\end{pmatrix}\eta)dv.
\end{align*}
Back in the operator representation, it follows that the POVM of this
measurement is the following set
\[
\left\{ \hat{\Pi}_{0,V}\left(\begin{pmatrix}\frac{1}{\sqrt{\tau}} & 0\\
0 & -\frac{1}{\sqrt{1-\tau}}
\end{pmatrix}\eta\right)=\hat{\rho}\left(\begin{pmatrix}\frac{1}{\sqrt{\tau}} & 0\\
0 & -\frac{1}{\sqrt{1-\tau}}
\end{pmatrix}\eta,V\right)\right\} ,
\]
where
\[
V=\begin{pmatrix}\sqrt{\frac{1-\tau}{\tau}} & 0\\
0 & -\sqrt{\frac{\tau}{1-\tau}}
\end{pmatrix}V_{env}\begin{pmatrix}\sqrt{\frac{1-\tau}{\tau}} & 0\\
0 & -\sqrt{\frac{\tau}{1-\tau}}
\end{pmatrix}.
\]
Therefore, we see that by replacing $H(\tau)$ with other symplectic
transformations and properly choosing $V_{env}$, we are able to make
$V$ an arbitrary covariance matrix, and therefore a arbitrary Gaussian
measurement\footnote{This first moments can be easily engineered using displacement operations}.

\section{Gaussian channel}

We have so far focused mainly on unitary operations. Generally speaking,
a quantum process is a CPTP map--quantum channel. A quantum channel
$\mathcal{E}$, is a map between the spaces of bounded operators on
two Hilbert spaces, transforming density operators to density operators.
Certain noisy quantum processes can also transform Gaussian states
into Gaussian states, and are therefore called the Gaussian channel\index{Gaussian!channel}\footnote{We follow the conventions in \cite{Weedbrook2012}.}.

We have seen that the transformation of a Gaussian state is completely
determined by the change of its first moments $\bar{x}$ and covariance
matrix $V$. It turns out that a Gaussian channel can be faithfully
represented by linear transformations describing the corresponding
changes. Specifically, given a Gaussian channel, a Gaussian state
$\hat{\rho}\left(\bar{x},V\right)$ is transformed to another Gaussian
state $\hat{\rho}\left(\bar{x}',V'\right)$ with
\[
\bar{x}'=T\bar{x}+d,
\]
and
\[
V'=T\bar{x}T^{t}+N,
\]
where $d$ is a real vector, and $T$, $N$ are real matrices. All
of $d$, $T$, and $N$ are independent of the input and output states
and thus completely determined by the quantum channel. For this reason,
we denote such a Gaussian channel by $\mathcal{G}_{T,N,d}$. Physically
speaking, the $T$ matrix the distortion of the position, size, and
shape of a Gaussian state in the phase space; and the $N$ matrix,
always a real symmetric matrix, can be considered to be the added
quantum noise, which justifies the noisy nature of the process.
\begin{example}
Symplectic transformations and displacement operations are trivial
examples of Gaussian channels. A symplectic transformation $\hat{U}_{S}$
is the Gaussian channel $\mathcal{G}_{S,0}$ while a displacement
operation $\hat{D}_{v}$ is the Gaussian channel $\mathcal{G}_{\text{Id},0,v}$.
\end{example}

Since the displacement operation are usually omitted in this thesis\footnote{The effect of the omitted displacement operation can be easily added
back through a post-processing operation.}, we will tacitly assume $d=0$ for the Gaussian channels to be discussed,
and therefore denote them by $\mathcal{G}_{T,N}$.

Combining two Gaussian channels is as easy as calculating simple matrix
addition and multiplication. Let $\mathcal{G}_{T_{1},N_{1}}$ and
$\mathcal{G}_{T_{2},N_{2}}$be two Gaussian channels. Then their combination
$\mathcal{G}_{T_{2},N_{2}}\circ\mathcal{G}_{T_{1},N_{1}}$ results
in a Gaussian channel $\mathcal{G}_{T_{3},N_{3}}$ with
\begin{equation}
T_{3}=T_{2}T_{1},
\end{equation}
and
\begin{equation}
N_{3}=T_{2}N_{1}N_{2}^{t}+N_{2}.
\end{equation}
We can also construct a larger-size Gaussian channel by juxtaposing
smaller-size Gaussian channels. Let the two Gaussian channels $\mathcal{G}_{T_{1},N_{1}}$
and $\mathcal{G}_{T_{2},N_{2}}$ be defined in different non-overlapping
bosonic systems. Then we can juxtapose the two systems to form a larger
bosonic system (through the tensor product of their Hilbert spaces),
which leads to a larger symplectic space (the direct sum of the smaller
symplectic spaces). Then on this resulting bosonic system, the juxtaposition
of the given Gaussian channels $\mathcal{G}_{T_{1},N_{1}}\otimes\mathcal{G}_{T_{2},N_{2}}$
is a Gaussian channel $\mathcal{G}_{T_{3},N_{3}}$ with

\begin{align}
T_{3} & =T_{1}\oplus T_{2}\nonumber \\
 & =\begin{pmatrix}T_{1} & 0\\
0 & T_{2}
\end{pmatrix},
\end{align}
and
\begin{align}
N_{3} & =N_{1}\oplus N_{2}\nonumber \\
 & =\begin{pmatrix}N_{1} & 0\\
0 & N_{2}
\end{pmatrix}.
\end{align}

Gaussian channel is an efficient language noisy Gaussian quantum processes.
However, it lacks the critical mathematical features of symplectic
geometry, which should be deeply rooted in its construction. In the
succeeding section, we discuss a systematic method of embedding non-trivial
Gaussian channels in a symplectic transformation.

\section{Symplectic dilation\label{sec:Symplectic-dialtion}}

I would like to conclude this chapter with my finding of the systematic
way to embed a noisy Gaussian process into a symplectic transformation\footnote{Note that some related discussions can be found in \cite{Weedbrook2012},
where it is shown that certain single-mode Gaussian channels can be
embedded in symplectic transformations. In comparison, the method
in section is applicable to arbitrary Gaussian channels.}. The idea is inspired by the fact that any quantum channel can be
dilated to a unitary operation \cite{Stinespring1955}. That is, given
any quantum channel $\mathcal{E}$ defined in the Hilbert space $\mathcal{H}_{a}$,
we can find a density operator $\hat{\rho}_{b}$ on another Hilbert
space $\mathcal{H}_{b}$, and a unitary operator $\hat{U}$ defined
on $\mathcal{H}_{a}\otimes\mathcal{H}_{b}$, such that for any density
operator $\hat{\rho}$ on $\mathcal{H}_{a}$, we have
\begin{equation}
\mathcal{E}(\hat{\rho})=\text{Tr}_{\mathcal{H}_{b}}[\hat{U}(\hat{\rho}\otimes\hat{\rho}_{b})\hat{U}^{\dagger}].\label{eq:Stinespring}
\end{equation}
Note that this the choices of $\mathcal{H}_{b}$, $\hat{U},$and $\hat{\rho}_{b}$
are not unique. This method, known as the Stinespring dilation, indicates
that a quantum channel can be understood as a restriction of a unitary
operation to a subsystem. We will follow the convention and refer
to our embedding method as the symplectic dilation\index{Symplectic!dilation}.
This section is written in a mathematical style in order to keep the
statements short and precise. More physical discussion and applications
will be demonstrated in the later chapters when we deal with practical
physical problems.

Before proceeding, we first introduce how the contrary works. Let
$\hat{U}_{S}$ be a symplectic transformation defined on the joined
Hilbert space $\mathcal{H}=\mathcal{H}_{a}\otimes\mathcal{H}_{b},$
with the joined symplectic space $E=E_{a}\oplus E_{b}$. Let $\hat{\rho}_{b}\left(0,V_{b}\right)$
be a density operator on $\mathcal{H}_{b}$. Then using Eq.~\ref{eq:Stinespring},
we obtain a Gaussian channel $\mathcal{G}_{T,N}$ with
\begin{equation}
T=S_{a,a},
\end{equation}
and
\begin{equation}
N=S_{b,b}V_{b}S_{b,b}^{t}.
\end{equation}
(Here we slightly abuse the notation by representing a symplectic
space using the subscript: $S_{b,b}=S_{E_{b},E_{b}}$.) Therefore,
given a symplectic transformation, one can easily construct a Gaussian
channel using submatrices of the given transformation. In fact, any
Gaussian channel can be obtained in this way.

However, the symplectic dilation is not as straightforward. First,
we need the following mathematical propositions: Let the symplectic
space $E=E_{a}\oplus E_{b}$ be the direct sum of symplectic vector
subspaces $E_{a}$ and $E_{b}$. Let $\Omega$, $\Omega_{a}$, and
$\Omega_{b}$ be the symplectic forms defined on $E$, $E_{a}$, and
$E_{b}$. Let $\mathcal{S}:E_{a}\rightarrow E_{a}$ be a linear transformation,
with $\text{det}\left(\text{Id}-T\right)\neq0$. The we have a matrix
$\mathcal{M}$, given by
\[
\mathcal{M}=\Omega_{a}\left[\frac{1}{2}\text{Id}-\left(\text{Id}-\mathcal{S}\right)^{-1}\right].
\]
Let $\mathcal{M}_{s}=\left(\mathcal{M}+\mathcal{M}^{t}\right)/2$,
$\mathcal{M}_{a}=\left(\mathcal{M}-\mathcal{M}^{t}\right)/2$ be the
symmetric and skew-symmetric part of the $\mathcal{M}$.
\begin{lem}
\label{lem:There-exists-R}There exists a linear transformation $\mathcal{R}:E_{b}\rightarrow E_{a}$,
such that
\[
\mathcal{R}\Omega_{b}\mathcal{R}^{t}=2\Omega_{a}\mathcal{M}_{a}\Omega_{a}.
\]
\end{lem}

\begin{proof}
For any skew-symmetric matrix $\mathcal{M}_{a}$, there exists a matrix
$\mathcal{Q}$, such that $\mathcal{M}_{a}=\mathcal{Q}\Omega\mathcal{Q}^{t}$.
Then we just let $\mathcal{R}=\sqrt{2}\Omega_{a}\mathcal{Q}\Omega$.
\end{proof}
Define linear transformations $\mathcal{L}=-\Omega_{b}\mathcal{R}^{t}\Omega_{a}$,
and $\mathcal{S}'=\text{Id}-\mathcal{L}\left(\Omega_{a}\mathcal{M}+\frac{1}{2}\text{Id}\right)^{-1}\mathcal{R}$.
Then we have
\begin{lem}
\label{lem:first-condition}$\mathcal{S}\Omega_{a}\mathcal{S}^{t}+\left(\text{Id}-\mathcal{S}\right)\mathcal{R}\Omega_{b}\mathcal{R}^{t}\left(\text{Id}-\mathcal{S}\right)^{t}=\Omega_{a}$
\end{lem}

\begin{proof}
Replacing $\mathcal{S}$ with $\mathcal{M}=\Omega_{a}\left[\frac{1}{2}\text{Id}-\left(\text{Id}-\mathcal{S}\right)^{-1}\right]$,
we transform the equation to the equivalent form
\[
\left(\Omega_{a}\mathcal{M}+\frac{1}{2}\text{Id}\right)^{-1}\left(\mathcal{R}\Omega_{b}\mathcal{R}^{t}+\Omega_{a}\right)-\Omega_{a}=\left(\Omega_{a}\mathcal{M}+\frac{1}{2}\text{Id}\right)^{-1}\Omega_{a}\left(\Omega_{a}\mathcal{M}+\frac{1}{2}\text{Id}\right)^{t}.
\]
Since
\begin{align*}
\left(\Omega_{a}\mathcal{M}+\frac{1}{2}\text{Id}\right)^{t} & =-\left(\mathcal{M}_{s}-\mathcal{M}_{a}+\frac{1}{2}\Omega_{a}\right)\Omega_{a},
\end{align*}
and
\begin{align*}
\left(\Omega_{a}\mathcal{M}+\frac{1}{2}\text{Id}\right)^{-1} & =-\left(\mathcal{M}_{s}+\mathcal{M}_{a}-\frac{1}{2}\Omega_{a}\right)^{-1}\Omega_{a},
\end{align*}
the right-hand side of the above equation is equivalent to
\[
\left(\Omega_{a}\mathcal{M}+\frac{1}{2}\text{Id}\right)^{-1}\left(2\Omega_{a}\mathcal{M}_{a}\Omega_{a}+\Omega_{a}\right)-\Omega_{a},
\]
which is equivalent to the left-hand side by the definition of $\mathcal{R}$.
\end{proof}
\begin{lem}
\label{lem:second-condition}$\mathcal{S}\Omega_{a}\left(\text{Id}-\mathcal{S}\right)^{t}\mathcal{L}^{t}+\left(\text{Id}-\mathcal{S}\right)\mathcal{R}\Omega_{b}\left(\mathcal{S}'\right)^{t}=0.$ 
\end{lem}

\begin{proof}
It can be easily verified by throwing in the definition of $\mathcal{L}$
and $\mathcal{S}'$, and expanding each term in the expression.
\end{proof}
\begin{lem}
\label{lem:third-condition}$\mathcal{S}'\Omega_{b}\left(\mathcal{S}'\right)^{t}+\mathcal{L}\left(\text{Id}-\mathcal{S}\right)\Omega_{a}\left(\text{Id}-\mathcal{S}\right)^{t}\mathcal{L}^{t}=\Omega_{b}.$
\end{lem}

\begin{proof}
This equation can be transformed to the equivalent equation
\[
\mathcal{L}\left(\Omega_{a}\mathcal{M}+\frac{1}{2}\text{Id}\right)^{-1}\left(\mathcal{R}\Omega_{b}-\Omega_{a}\mathcal{L}^{t}\right)=\left(\mathcal{L}\Omega_{a}-\Omega_{b}\mathcal{R}^{t}\right)\left[\left(\Omega_{a}\mathcal{M}+\frac{1}{2}\text{Id}\right)^{-1}\right]^{t}\mathcal{L}^{t},
\]
which can be verified by applying Lemma.~\ref{lem:second-condition}
to it.
\end{proof}
\begin{thm}
\label{thm:The-linear-transformation}The linear transformation $S:E\rightarrow E$,
defined as
\[
S=\begin{pmatrix}\mathcal{S} & \left(\text{Id}-\mathcal{S}\right)\mathcal{R}\\
\mathcal{L}\left(\text{Id}-\mathcal{S}\right) & \mathcal{S}'
\end{pmatrix}
\]
is a symplectic transformation.
\end{thm}

\begin{proof}
The linear transformation $S$ is a symplectic transformation if and
only if $S\Omega S^{t}=\Omega$. This equation leads to the verification
of the following set of equations
\begin{align*}
\mathcal{S}\Omega_{a}\mathcal{S}^{t}+\left(\text{Id}-\mathcal{S}\right)\mathcal{R}\Omega_{b}\mathcal{R}^{t}\left(\text{Id}-\mathcal{S}\right)^{t} & =\Omega_{a},\\
\mathcal{S}\Omega_{a}\left(\text{Id}-\mathcal{S}\right)^{t}\mathcal{L}^{t}+\left(\text{Id}-\mathcal{S}\right)\mathcal{R}\Omega_{b}\left(\mathcal{S}'\right)^{t} & =0,\\
\mathcal{S}'\Omega_{b}\left(\mathcal{S}'\right)^{t}+\mathcal{L}\left(\text{Id}-\mathcal{S}\right)\Omega_{a}\left(\text{Id}-\mathcal{S}\right)^{t}\mathcal{L}^{t} & =\Omega_{b},
\end{align*}
which has been shown in Lemmas.~\ref{lem:first-condition},\ref{lem:second-condition}\&\ref{lem:third-condition}.
\end{proof}
With the aforementioned construction of a Gaussian channel out of
a symplectic transformation, the above theorem indicates that the
Gaussian channel $\mathcal{G}_{T,N}$ with
\[
T=\mathcal{S},
\]
and
\[
N=\left(\text{Id}-\mathcal{S}\right)\mathcal{R}V_{b}\mathcal{R}^{t}\left(\text{Id}-\mathcal{S}\right)^{t}
\]
can be constructed using a symplectic dilation $\hat{U}_{S}$ and
a Gaussian state $\hat{\rho}\left(0,V_{b}\right)$, with
\[
S=\begin{pmatrix}\mathcal{S} & \left(\text{Id}-\mathcal{S}\right)\mathcal{R}\\
\mathcal{L}\left(\text{Id}-\mathcal{S}\right) & \mathcal{S}'
\end{pmatrix}
\]
as in the above theorem, and $V_{b}$ a proper covariance matrix.

The symplectic dilation is not unique, since the matrix $\mathcal{R}$
is not uniquely defined (we only showed a special construction of
$\mathcal{R}$ in Lemma.~\ref{lem:There-exists-R}). \footnote{This can be seen by checking the determinant of each factor of the
following

\begin{align*}
\text{Id}-S & =\begin{pmatrix}\text{Id}-\mathcal{S} & -\left(\text{Id}-\mathcal{S}\right)\mathcal{R}\\
-\mathcal{L}\left(\text{Id}-\mathcal{S}\right) & \text{Id}-\mathcal{S}'
\end{pmatrix}\\
 & =\begin{pmatrix}\text{Id}\\
-\mathcal{L}
\end{pmatrix}\left(\text{Id}-\mathcal{S}\right)\begin{pmatrix}\text{Id} & -\mathcal{R}\end{pmatrix},
\end{align*}
} Moreover, for any pair of symplectic transformations $S_{b},S_{b'}$
on $E_{b}$, the symplectic transformation
\[
D=\begin{pmatrix}\text{Id}\\
 & S_{b}
\end{pmatrix}S\begin{pmatrix}\text{Id}\\
 & S_{b}'
\end{pmatrix}=\begin{pmatrix}\mathcal{S} & \left(\text{Id}-\mathcal{S}\right)\mathcal{R}S_{b'}\\
S_{b}\mathcal{L}\left(\text{Id}-\mathcal{S}\right)S_{b}' & S_{b}\mathcal{S}'S_{b}
\end{pmatrix}
\]
is also a symplectic dilation of the same Gaussian channel if we the
Gaussian state operator $\hat{\rho}\left(0,V_{b}\right)$ to $\hat{\rho}\left(0,S_{b'}^{-1}V_{b}S_{b'}^{-1}\right)$.
In fact this non-uniqueness of symplectic dilation implies that any
Gaussian channel $\mathcal{G}_{T,N}$, with $\text{det}\left(\text{Id}-T\right)\neq0$
can be dilated with a proper choice of $\mathcal{R}$ and $S_{b'}$.
Moreover, for $\text{det}\left(\text{Id}-T\right)=0$, we can always
decompose $\mathcal{G}_{T,N}$ as the composition of $\mathcal{G}_{T_{1},N_{1}}$
and $\mathcal{G}_{T_{2},N_{2}}$ with non-vanishing $\text{det}\left(\text{Id}-T_{1}\right)$
and $\text{det}\left(\text{Id}-T_{2}\right)$, so that the composition
of the symplectic dilations of $\mathcal{G}_{T_{1},N_{1}}$ and $\mathcal{G}_{T_{2},n_{2}}$
gives the symplectic dilation of $\mathcal{G}_{T,N}$.

Last, it is easy to check that the symplectic dilation obtained from
the last theorem satisfies $\text{det}\left(\text{Id}-S\right)=0.$
Let $\text{det}\left(\text{Id}-S_{b}S_{b'}\right)\neq0$ in the above
definition of $D$. Then we have
\begin{thm}
There exists a linear transformation $\tilde{\mathcal{M}}:E\rightarrow E$
such that $D=\text{Id}-\left(\Omega\tilde{\mathcal{M}}+\frac{1}{2}\text{Id}\right)^{-1}$
is a symplectic dilation of $\mathcal{S}$.
\end{thm}

\begin{proof}
Let 
\[
\Omega\tilde{\mathcal{M}}+\frac{1}{2}\text{Id}=\begin{pmatrix}\mathcal{A} & \mathcal{B}\\
\mathcal{C} & \mathcal{D}
\end{pmatrix}
\]
with
\begin{align*}
\mathcal{A}= & \mathcal{M}+\mathcal{R}S_{b'}\left(\text{Id}-S_{b}S_{b}'\right)^{-1}S_{b}\mathcal{L},\\
\mathcal{B}= & \mathcal{R}S_{b'}\left(\text{Id}-S_{b}S_{b'}\right)^{-1},\\
\mathcal{C}= & -\left(\text{Id}-S_{b}S_{b'}\right)^{-1}S_{b}\mathcal{L}\\
\mathcal{D} & =\left(\text{Id}-S_{b}S_{b'}\right)^{-1},
\end{align*}
where $\text{Id}-S_{b}S_{b'}$ is invertible. Then we have
\begin{align*}
D= & \begin{pmatrix}\text{Id}-\left(\mathcal{A}-\mathcal{B}\mathcal{D}^{-1}\mathcal{C}\right)^{-1} & \left(\mathcal{A}-\mathcal{B}\mathcal{D}^{-1}\mathcal{C}\right)^{-1}\mathcal{B}\mathcal{D}^{-1}\\
\mathcal{D}^{-1}\mathcal{C}\left(\mathcal{A}-\mathcal{B}\mathcal{D}^{-1}\mathcal{C}\right)^{-1} & \text{Id}-\mathcal{D}^{-1}-\mathcal{D}^{-1}\mathcal{C}\left(\mathcal{A}-\mathcal{B}\mathcal{D}^{-1}\mathcal{C}\right)^{-1}\mathcal{B}\mathcal{D}^{-1}
\end{pmatrix}\\
= & \begin{pmatrix}\text{Id}\\
 & S_{b}
\end{pmatrix}S\begin{pmatrix}\text{Id}\\
 & S_{b}'
\end{pmatrix}
\end{align*}
is a symplectic dilation of $\mathcal{S}$.
\end{proof}
Therefore, $D$ is the symplectic Cayley transform of $\Omega\tilde{\mathcal{M}}$.
Physically speaking, the symplectic dilation $D$ can be viewed as
a noiseless scattering process governed by a dimensionless quadratic
Hamiltonian\index{Dimensionless quadratic Hamiltonian} (see Appx.~\ref{chap:Scattering-process-and})
determined by the following symmetric matrix
\[
\begin{pmatrix}\mathcal{A}-\frac{1}{2}\text{Id} & \mathcal{B}\\
\mathcal{C} & \mathcal{D}-\frac{1}{2}\text{Id}
\end{pmatrix}.
\]

The symplectic dilation allows us to model and analyze the quantum
noise feature of a Gaussian process with limited knowledge of the
Gaussian channel. We will discuss in more detail about the physical
meaning and applications in later chapters.

\chapter{Quantum teleportation and quantum transdcution\label{chap:Quantum-Transduction}}

\section{Introduction to quantum transduction}

In this chapter, we demonstrate our first physical application of
symplectic geometry to an intriguing field of bosonic quantum control,
the quantum transduction\footnote{The discussions in this chapter are based on \cite{Zhang2018}}.
Firstly, we brefeifly what is quantum transduction and why it is of
practical importance in quantum information science.

Nowadays, quantum computation has become a high-profile physical subject
due to its promising potential of drastically improving human being's
ability of solving complicated computationaly problems. Realizing
such prospects requires tremendous efforts to mitigate systematic
computational errors caused by the probabilistic nature of quantum
physics, and to suit quantum computers to running large-scale quantum
programs. Considering the current limitation of controlling quantum
systems in laboratories, the idea of quantum network has been come
up with as a practical way to meet these requirements. A quantum network
is an extendable set of processing units, i.e., small-scale quantum
computers, connected by transmission channels. On a quantum network,
a large-scale quantum program is decomposed into many smaller-scale
sub-programs, each of which is executed on a processing unit. The
outputs of these sub-programs, i.e. quantum states, are then combined
with each other through the transmission channels to constitute the
final output. Therefore, each processing unit or transmission only
needs to be implemented on a small-scale physical platform (e.g.,
several entangled qubits, or several bosonic modes), so that the complexity
of controlling the processing and transmission units can be reduced
greatly.

However, different physical platforms are often suitable for different
tasks in quantum computation. For example, superconducting quantum
circuits (working at microwave frequency) are great at processing
quantum information, while optical fibers are still the best candidates
of transmitting quantum information. The practical realization of
quantum network may be hybrid, with different tasks executed on different
platforms. Therefore, physical devices that can convert quantum states
between different platforms with high fidelity and efficiency becomes
a crucial addition to practical implementation of quantum networks.
In particular, we are particularly interested in such devices that
can faithfully transfer bosonic quantum states, since bosonic modes
have widespread applications in both quantum information processing
and transmission. We refer to these state-conversion devices as quantum
transducers.

Many theoretical and experimental efforts have been made since the
birth of quantum transduction. Many all of the promising proposals
and experimental demonstrations are based on bosonic scattering processes.
For example, as in \cite{Andrews2014}, to transfer a microwave mode
to an optical mode, we first couple the microwave circuit to the optical
cavity directly or indirectly mediated by another mode (e.g., a mechanical
mode). Then we inject the signal into the cavities, and tune the system
parameters to satisfy a so-called matching condition, so that the
transmission of the signal is maximized and the reflection suppressed
completely. Since the physical couplings are linear, i.e. determined
by a quadratic bosonic Hamiltonian, the scattering process is actually
a Gaussian process and therefore can be described by a symplectic
transformation.

However, the matching condition cannot always be satisfied because
of the inevitable presence of side bands. Specifically, a (over-)simplified
quadratic Hamiltonian for the process mentioned above is
\begin{alignat*}{1}
\hat{H} & =\omega_{o}\hat{a}_{o}^{\dagger}\hat{a}_{o}+\omega_{m}\hat{a}_{m}^{\dagger}\hat{a}_{m}+\omega_{e}\hat{a}_{e}^{\dagger}\hat{a}_{e}\\
 & +g_{om}\hat{a}_{o}^{\dagger}\hat{a}_{m}+g_{om}\hat{a}_{m}^{\dagger}\hat{a}_{o}+g_{em}\hat{a}_{e}^{\dagger}\hat{a}_{m}+g_{em}\hat{a}_{m}^{\dagger}\hat{a}_{e}\\
 & +g_{om}\hat{a}_{o}^{\dagger}\hat{a}_{m}^{\dagger}+g_{om}\hat{a}_{o}\hat{a}_{m}+g_{em}\hat{a}_{e}^{\dagger}\hat{a}_{m}^{\dagger}+g_{em}\hat{a}_{e}\hat{a}_{m},
\end{alignat*}
with $\omega_{o}$, $\omega_{m}$ and $\omega_{e}$ being the frequencies
of the optical, mechanical and microwave modes, $g_{om}$ and $g_{em}$
being the corresponding coupling strengths. The excitation-number-non-preserving
terms in the the third line of the above equation are known as the
counter-rotating terms and are usually omitted in rotating wave approximation
in quantum optics given that $\omega_{o},\omega_{m},\omega_{e}\gg g_{om},g_{oe}$
(i.e. physically speaking, these terms will not make any substantial
contribution to the quantum process). This approximation cannot be
carried out in this scenario, since $\omega_{m}$, the frequency of
the mechanical mode, is comparable with the coupling strengths. Therefore,
$\hat{H}$ will not preserve the total excitation number, $\hat{a}_{o}^{\dagger}\hat{a}_{o}+\hat{a}_{m}^{\dagger}\hat{a}_{m}+\hat{a}_{e}^{\dagger}\hat{a}_{e}$,
due tot the failure of the rotating wave approximation. Since excitation-number-preserving
is a prerequisite of the matching condition, we see that the previously
expected perfect quantum transdcution cannot be realized using this
quadratic Hamiltonian. More intuitively, the irremovability of the
counter-rotating terms leads to the presence of side bands, additional
bosonic modes, in the frequency spectrum. These side bands will interact
with the quantum signal, causing leakage of quantum information during
the scattering process.

From the perspective of Gaussian process, the above amounts to the
following abstraction: Consider a bosonic system consisting of bosonic
modes of different physical characteristics (frequencies, carriers,
etc.). The symplectic space will be divided in the following two ways:
(1) Let $E$ be the symplectic space of this system, $E_{in}$ be
the canonical symplectic subspace associated with the modes storing
the input quantum signal, and $E_{anc}$ the canonical symplectic
subspace associated with the unwanted environmental modes (e.g., the
side bands); (2) Let $E_{out}$ be the canonical symplectic subspace
associated with the modes storing the output quantum signal, and $E_{idl}$
be the canonical symplectic subspace associated with the rest of the
system. Now let $S$ be the symplectic transformation describing the
Gaussian process. Then the transduction is perfect if the $S_{out,anc}=S_{idl,in}=0$;
and imperfect if the either the submatrix $S_{out,anc}$ or $S_{idl,in}$
is non-vanishing.

\begin{figure}[h]
\includegraphics[width=0.9\textwidth]{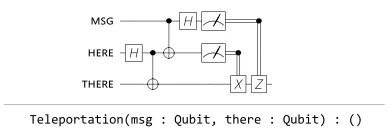}
\begin{doublespace}
\centering{}\caption{\emph{Direct quantum transduction.}}
\begin{minipage}[t]{0.9\textwidth}%
\begin{doublespace}
The input quantum state is mixed with the with the ancillary state
by a symplectic transformation $S$. The idler modes are dropped and
the direct output modes are used as the the transduced quantum signal.
\end{doublespace}
\end{minipage}
\end{doublespace}
\end{figure}

Based on this abstraction (which will be referred to as the direct
quantum transduction\index{Direct quantum transduction} for later
convenience), in the following sections, we will demonstrate how a
generalization of continuous-variable quantum teleportation protocol
turns out to be a promising candidate to address the above mentioned
problems concerning perfect quantum transduction and its robustness
against practical system imperfections. We will also see symplectic
geometry plays a crucial in inspiring such an application-oriented
scheme.

\section{Generalization of continuous-variable quantum teleportation\label{sec:Generalized-continuous-variable-}}

\begin{figure}
\begin{doublespace}
\centering{}\includegraphics[width=0.9\textwidth]{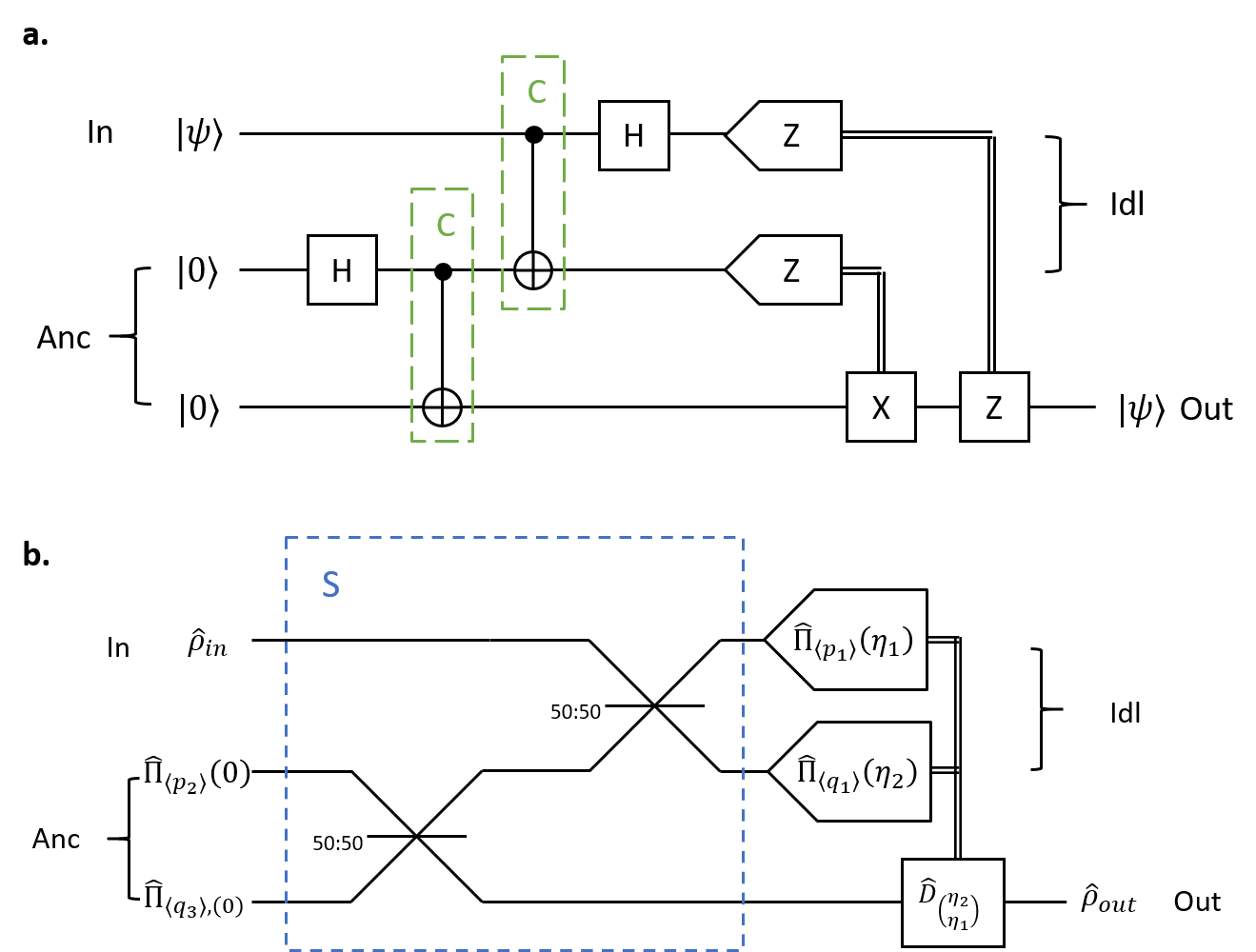}\caption{\emph{Quantum teleportation circuits.}\label{fig:CV_Teleportation} }
\begin{minipage}[t]{0.9\textwidth}%
(a) Illustration of the qubit quantum teleportation circuit. The boxes
labeled by H are Hadamard gates, labled by X and Z are Pauli-X and
Pauli-Z operators. The two-qubit gates, encircled by the green dashed
boxes and labeled by C, are the CNOT gates. The solid double lines
represent classical communication channels. The classical communication
channels connects the syndrome measurement, i.e. the Pauli-Z measurements,
to the corresponding Pauli operators. (b) Illustration of continuous
variable quantum teleportation circuit. It shows the resemblance between
the two teleportation schemes: The CNOT gates are replaced by the
50:50 (balanced) beam-splitters, the eigenstates $|0\rangle$'s by
the infinitely squeezed states, the syndrome measurement by the homodyne
measurements, and the Pauli operators by the displacement operator.
Note that the Hadmard gates are absorbed into the basis of squeezed
ancillary states and the homodyne measurements.%
\end{minipage}
\end{doublespace}
\end{figure}

The task of quantum transduction, transferring quantum states between
different physical platforms, reminds us of a well-known physical
concept, quantum teleportation. We first briefly introduce quantum
teleportation in bosonic system. Continuous-variable quantum teleportation
\cite{Braunstein1998,Furusawa1998} is an analogue of the discrete-variable
quantum teleportation in bosonic systems \cite{Bennett1993}. The
system considered in quantum teleportation consists of three qubits,
which can be tagged as an input qubit and two ancilla qubits or as
an output qubit and two idler qubits. The input qubit, storing the
quantum state to transfer, and the output qubit, storing the transferred
quantum state, are different qubits, possibly located at rather distant
places. To implement the quantum teleportation protocol, we first
maximally entangle the two ancilla qubits, which are prepared as eigenstates
of the Pauli-Z operator, using a CNOT gate; then entangle the input
qubit with one of the ancillary qubits (the other one is the output
qubit) using anothre CNOT gate; then measure the entangled input-ancillia
pair using Pauli measurements; and at last perform adaptive Pauli
operations on the output qubit, based on the syndrome measurement
outcomes, to obtain a perfectly transferred output quantum state.
People found the counterparts of the required components of discrete-variable
quantum teleportation in continuous variable system: infinitely squeezed
states for the Pauli-operator eigenstates, balanced beam-splitters
for CNOT gates, homodyne measurements for Pauli measurements, and
displacement operators for Pauli-operators, as shown in Fig.~\ref{fig:CV_Teleportation}.
By combing these counterparts in the same order as in the discrete-variable
quantum teleportation, we obtain the continuous-variable quantum teleportation.{]}

We notice that beam-splitters are symplectic transformations. Then
the composition of the two balanced beam-splitters in continuous-variable
quantum teleportation can be described by a single symplectic transformation
$S$, which can be viewed as an imperfect quantum transducer since
obviously both $S_{out,anc}$ and $S_{idl,in}$ are non-vanishing,
as mentioned in the last section. In other words, we see that continuous
variable quantum teleportation can be considered as a way to convert
a specific imperfect quantum transducer, a combination of two beam-splitters,
into a perfect transducer, using only single-mode Gaussian controls
including infinite single-mode squeezing, homodyne measurements, and
displacement operations. Then it is natural to ask: Can we devise
a similar protocol that can be applied to general imperfect quantum
transducers requiring the same single-mode Gaussian mode controls?

It turns out that such a generalization exists and can be understood
intuitively. Given a symplectic transformation $S$ defined on a multi-mode
bosonic system associated with the symplectic space $E$. As mentioned
in the previous section, we decompse $E$ as $E_{in}\oplus E_{anc}$
and $E_{idl}\oplus E_{out}$, where $E_{in}$ supports the input quantum
signal, $E_{anc}$ supports the ancillary modes, $E_{idl}$ supports
the modes to measure later on, and $E_{out}$ supports the output
quantum signal. Then we prepare an ancillary infinitely squeezed state
$\hat{\Pi}_{l_{z}}\left(0\right)$ with $l_{z}$ being a Lagrangian
plane in $E_{anc}$ (e.g., $l_{z}$ can be spanned by all $\hat{q}$
quadratures in $E_{anc}$). Then, the input and the infinitely squeezed
ancillary modes are mixed by the given symplectic transformation $S$.
Then the we perform a multi-mode homodye measurement, described by
the POVM $\{\hat{\Pi}_{l_{h}}\left(\eta\right)\}$ on the $E_{idl}$,
with $l_{h}$ being a Lagrangian plane of $E_{idl}$ spanned by the
measured quadratures (e.g., all $\hat{q}$-quadratures in $E_{idl}$)
and $\eta$ a vector of the measurement outcomes of the corresponding
quadratures. At last, we transform the vector $\eta$ using a real
matrix $F$ and displace the output signal using a displacement operation
$\hat{D}_{F\eta}$. We found if $F$ is chosen properly, the output
state will be equivalent to the input state up to a symplectic transformation
$\tilde{S}$, i.e., 
\[
\hat{\rho}_{out}=\hat{U}_{\tilde{S}}\hat{\rho}_{in}\hat{U}_{\tilde{S}}^{\dagger}.
\]
Then how to choose $F$? The idea is based on noise cancellation.
By squeezing the ancillary modes, we concentrate the quantum noise
to the anti-squeezed quadratures, represented by a Lagrangian plane
$l_{z'}$ of $E_{anc}$. When mixed with the input quantum signal,
the quantum noise is distributed to both the idler modes and the output
modes. The proportion of the noise going into the idler modes can
be quantified by the submatrix $S_{idl,z'}$ while the proportion
going into the output by the submatrix $S_{out,z'}$\footnote{We can imagine the original quantum noise in the ancillary modes to
be a real vector, say $v$. Then the noise going into the the idler
modes can be imagined as the vector $S_{idl,z}v$ and the noise going
into the output as $S_{out,z'}v$, etc..}. Then the homomedyne measurement captures the proportion, quantified
by $S_{h,z'}$ of noise coming into the idler modes. Therefore, the
proper matrix $F$ should be able to transform the measured noise
to match the noise in the output, i.e., we require
\[
FS_{h,z'}=S_{out,z'}.
\]
Or equivalently, we require $F=S_{out,z'}\left(S_{h,z'}\right)^{-1}$.

\begin{figure}[h]
\includegraphics[width=0.9\textwidth]{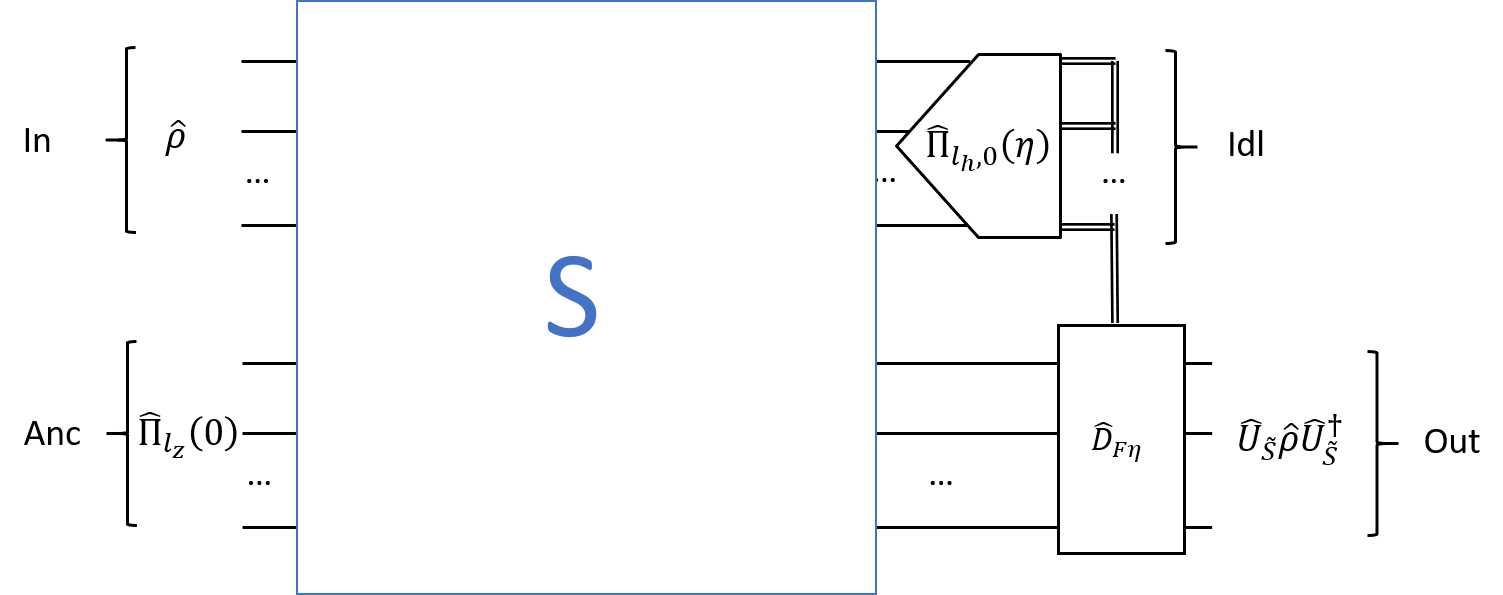}
\begin{doublespace}
\centering{}\caption{\emph{Generalized teleportation protocol.}}
\begin{minipage}[t]{0.9\textwidth}%
The multi-mode input state $\hat{\rho}$ is mixed with the infinitely
squeezed multi-mode ancillary state $\hat{\Pi}_{l_{z}}\left(0\right)$(determined
by a Lagrangian plane $l_{z}$) by the multi-mode symplectic transformation
$S$. Then the idler modes are measured by ideal homodyne measurements,
described by $\hat{\Pi}_{l_{h}}\left(\eta\right)$. The measurement
outcomes, represented by a real vector $\eta$, are used to determine
the displacement operation $\hat{D}_{F\eta}$ with $F$ the matrix
given in the main text. At last, the density operator of the output
state is given by $\hat{U}_{\tilde{S}}\hat{\rho}\hat{U}_{\tilde{S}}^{\dagger}$,
which is unitarily equivalent to the input state $\hat{\rho}$.%
\end{minipage}
\end{doublespace}
\end{figure}

This intuitive picture can be rigorously proved. Our generalization
is equivalent to the following equation
\begin{equation}
\hat{\rho}_{out}=\hat{U}_{\tilde{S}}\hat{\rho}_{in}\hat{U}_{\tilde{S}}^{\dagger}=\int_{l_{h}}\hat{D}_{F\eta}\left(\text{Tr}_{\mathcal{H}_{idl}}[\hat{\Pi}_{l_{h}}\left(\eta\right)\hat{U}_{S}\left(\hat{\rho}_{in}\otimes\hat{\Pi}_{l_{z}}\left(0\right)\right)\hat{U}_{S}^{\dagger}]\right)\hat{D}_{F\eta}^{\dagger}d\eta.\label{eq:CV teleportation-1}
\end{equation}
Note that we have abused the notation by representing symplectic
subspaces using their subscripts (e.g., $S_{idl,z'}=S_{E_{idl},l_{z'}}$).
For convenience, we list the subspaces that will be useful later:
canonical symplectic subspaces $E_{in}$, $E_{anc}$, $E_{idl}$ and
$E_{out}$ satisfying $E=E_{in}\oplus E_{anc}$ and $E=E_{idl}\oplus E_{out}$,
Lagrangian planes $l_{z}$, $l_{z'}$, $l_{h}$, and $l_{h'}$ satisfying
$E_{anc}=l_{z}\oplus l_{z'}$ and $E_{idl}=l_{h}\oplus l_{h'}$. Translating
to Wigner functions by letting
\begin{equation}
F=\tilde{S}\left[\left(S^{-1}\right)_{in,h}-\left(S^{-1}\right)_{in,h'}\left(\left(S^{-1}\right)_{z,h'}\right)^{-1}\left(S^{-1}\right)_{z,h'}\right],
\end{equation}
and
\[
\tilde{S}=\left(S^{-1}\right)_{in,out}-\left(S^{-1}\right)_{in,h'}\left(\left(S^{-1}\right)_{z,h'}\right)^{-1}\left(S^{-1}\right)_{z,out},
\]
the above equation amounts to
\[
W_{\hat{\rho}_{out}}\left(u\right)=W_{\hat{U}_{\tilde{S}}\hat{\rho}_{in}\hat{U}_{\tilde{S}}^{\dagger}}\left(u\right)=W\left(\tilde{S}^{-1}u\right)
\]
if we $\tilde{S}$ is a symplectic transformation. Therefore, we conclude
this section with the following theorem, where we prove $F$ thus
defined is equal to what we obtained from the noise cancellation picture,
and $\tilde{S}$ is indeed a symplectic transformation.
\begin{thm}
\label{thm:generalized-teleportation}Let $S:E\rightarrow E$ be a
symplectic transformation with $S_{h,z'}$ being invertible. Then
the following statements hold:

\emph{(i)} $\tilde{S}:=S_{out,in}-S_{out,z'}\left(S_{h,z'}\right)^{-1}S_{h,in}:E_{in}\rightarrow E_{out}$
is a symplectic transformation;

\emph{(ii)} $\check{S}:=\left(S^{-1}\right)_{in,out}-\left(S^{-1}\right)_{in,h'}\left(\left(S^{-1}\right)_{z,h'}\right)^{-1}\left(S^{-1}\right)_{z,out}:E_{out}\rightarrow E_{in}$
is a symplectic transformation; 

\emph{(iii)} $\tilde{S}=\left(\check{S}\right)^{-1}$;

\emph{(iv)} (forward transformation)
\begin{equation}
F=-S_{out,z'}\left(S_{h,z'}\right)^{-1}=\tilde{S}\left[\left(S^{-1}\right)_{in,h}-\left(S^{-1}\right)_{in,h'}\left(\left(S^{-1}\right)_{z,h'}\right)^{-1}\left(S^{-1}\right)_{z,h'}\right];\label{eq:feedforward}
\end{equation}

\emph{(v) }(backward transmission)
\begin{equation}
B=-\left(S^{-1}\right)_{in,h'}\left(\left(S^{-1}\right)_{z,h'}\right)^{-1}=\tilde{S}^{-1}\left[S_{out,z}-S_{out,z'}\left(S_{h,z'}\right)^{-1}S_{h,z}\right].\label{eq:feedbackward}
\end{equation}
\end{thm}

\begin{proof}
(i): For all $w,w'\in E_{in}$, let $u,u'\in E_{in}\oplus E_{anc}$
be vectors with
\begin{align*}
u_{in} & =w,\\
u_{z} & =0,\\
u_{z'} & =-\left(S_{h,z'}\right)^{-1}S_{h,in}w,\\
u'_{in} & =w',\\
u'_{z} & =0,\\
u'_{z'} & =-\left(S_{h,z'}\right)^{-1}S_{h,in}w'.
\end{align*}
Then we let $v=Su$ and $v'=Su'$. It is easy to show that
\begin{align*}
v_{out} & =\tilde{S}w,\\
v_{h} & =0,\\
v_{h'} & =S''w,\\
v'_{out} & =\tilde{S}w',\\
v'_{h} & =0,\\
v'_{h'} & =S''w',
\end{align*}
with $S''=S_{h',in}-S_{h',z'}\left(S_{h,z'}\right)^{-1}S_{h,in}$.
It then follows that
\[
\sigma_{E_{in}\oplus E_{anc}}\left(u,u'\right)=\sigma_{E_{idl}\oplus E_{out}}\left(v,v'\right),
\]
\begin{align*}
\sigma_{E_{idl}\oplus E_{out}}\left(v,v'\right) & =\sigma_{E_{out}}\left(\tilde{S}w,\tilde{S}w'\right)+\sigma_{E_{idl}}\left(v|_{l_{h'}},v'|_{l_{h'}}\right)\\
 & =\sigma_{E_{out}}\left(\tilde{S}w,\tilde{S}w'\right),
\end{align*}
and
\begin{align*}
\sigma_{E_{in}\oplus E_{anc}}\left(u,u'\right) & =\sigma_{E_{in}}\left(w,w'\right)+\sigma_{E_{anc}}\left(u|_{l_{z'}},u'|_{l_{z'}}\right)\\
 & =\sigma_{E_{in}}\left(w,w'\right).
\end{align*}
Therefore, we have
\[
\sigma_{E_{out}}\left(\tilde{S}w,\tilde{S}w'\right)=\sigma_{E_{in}}\left(w,w'\right),
\]
which, by definition, means that $\tilde{S}$ is a symplectic transformation.

(ii): Recalling the symplectic form $\Omega$, the inverse of a symplectic
transformation $S$ is
\[
S^{-1}=-\Omega S^{t}\Omega.
\]
Then we have
\begin{align*}
\left(S^{-1}\right)_{z,h'} & =-\Omega_{z,z'}\left(S^{t}\right)_{z',h}\Omega_{h,h'}\\
 & =-\Omega_{z,z'}\left(S_{h,z'}\right)^{t}\Omega_{h,h'},
\end{align*}
which shows the invertibility of $(S^{-1})_{z,h'}$ given the invertibility
of $S_{h,z'}$. Using similar arguments as in (i), it then can be
proven easily that the linear transformation $\check{S}$ is symplectic.
The details of the proof are omitted here.

(iii): As what we defined in (i), $v=Su$, so we also have $u=S^{-1}v$.
This leads to the following equations:
\[
w=u_{in}=\left[\left(S^{-1}\right)_{in,out}\tilde{S}+\left(S^{-1}\right)_{in,h'}S''\right]w,
\]
and
\[
0=u_{z}=\left[\left(S^{-1}\right)_{z,out}\tilde{S}+\left(S^{-1}\right)_{z,h'}S''\right]w
\]
for all $w\in E_{in}$. Then, we have
\[
\left(S^{-1}\right)_{in,out}\tilde{S}+\left(S^{-1}\right)_{in,h'}S''=\text{Id},
\]
and
\[
\left(S^{-1}\right)_{z,out}\tilde{S}+\left(S^{-1}\right)_{z,h'}S''=0.
\]
Since $\left(S^{-1}\right)_{z,h'}$ is invertible, combining the above
two equations gives the following:
\[
\check{S}\tilde{S}=\text{Id}.
\]
Therefore, $\check{S}$ is an inverse of $\tilde{S}$.

(iv) \&(v): As shown in (iii), we have
\[
\left(S^{-1}\right)_{z,out}\tilde{S}+\left(S^{-1}\right)_{z,h'}S''=0.
\]
It can be rewritten as
\[
-\left(\left(S^{-1}\right)^{t}\right)_{out,z}\left(\left(\left(S^{-1}\right)^{t}\right)_{h',z}\right)^{-1}=\left(\tilde{S}^{t}\right)^{-1}\left(S''\right)^{t}.
\]
We can view $\tilde{\Box}$ and $\Box''$ as mappings between sets
of linear transformations. It is easy to check that both $\tilde{\Box}$
and $\Box''$ commute with the transposition operation. Thus we have
\[
-\left(\left(S^{t}\right)^{-1}\right)_{out,z}\left(\left(\left(S^{t}\right)^{-1}\right)_{h',z}\right)^{-1}=\left(\tilde{S^{t}}\right)^{-1}\left(S^{t}\right)''.
\]
Then (4) is proved by replacing $S^{t}$ with $S^{-1}$, and exchanging
the subscripts $z$ with $z'$, and $h$ with $h'$; (5) is proved
by replacing $S^{t}$ with $S$, and exchanging the subscripts $in$
with $out$, $z$ with $h'$ and $z'$ with $h$. It is also easy
to check these replacements will not require invertibility of the
sub-linear-transformation of $S$ other than that of $S_{h,z'}$.
\end{proof}

\section{\label{sec:Gaussian-channel-representation}Gaussian channel representation
of adaptive control}

In Sec.~\ref{sec:Generalized-continuous-variable-}, we demonstrated
the generalized teleportation protocol assuming ideal and noiseless
quantum controls. In this and the following sections, we investigate
the robustness of this scheme against practical experimental imperfections.
First, we need to prepare necessary theoretical models of some of
the key components in the language of Gaussian channel. In this section,
we discuss how to represent the crucial adaptive displacement operation
as a Gaussian channel.

We follow the convention in the last section, except that we assume
the matrix $F$, governing the displacement operation. is not of the
specific form as discussed previously, but rather an arbitrary linear
transformation. Specifically, we will show how to represent the quantum
channel defined by
\begin{equation}
\mathcal{A}_{F}(\hat{\rho}):=\int\hat{D}_{F\eta}\text{Tr}_{\mathcal{H}_{idl}}\left[\hat{\Pi}_{l_{h}}\left(\eta\right)\hat{\rho}\right]\hat{D}_{F\eta}^{\dagger}
\end{equation}
 as a Gaussian channel (i.e. we will find the two matrices determining
this channel). Note that here the density operator $\hat{\rho}$ is
defined on the whole symplectic space.

Firstly, we represent the above equation by Winger functions, which
is given by the following formula
\[
W_{\mathcal{A}_{F}(\hat{\rho})}\left(u\right)\propto\int_{l_{h}}\int_{E_{idl}}W_{\hat{\rho}}\left(\left(u-F\eta\right)\oplus v\right)\delta\left(v-\eta\right)dvd\eta.
\]
Then we let
\begin{equation}
X=-\Omega_{h',h}F^{t}\Omega_{out},
\end{equation}
and
\begin{equation}
Y=\frac{1}{2}\Omega_{h',h}F^{t}\Omega_{out}F,
\end{equation}
in the linear transformation $A_{F}$ defined by
\begin{equation}
A_{F}=\begin{pmatrix}\text{Id} & F & 0\\
0 & \text{Id} & 0\\
-X & Y & \text{Id}
\end{pmatrix}.
\end{equation}
It turns out that $A_{F}$ is a symplectic transformation with its
inverse being
\begin{equation}
A_{F}^{-1}=\begin{pmatrix}\text{Id} & -F & 0\\
0 & \text{Id} & 0\\
E & G & \text{Id}
\end{pmatrix}.
\end{equation}
Then straightforward calculation shows that
\begin{align}
W_{\mathcal{A}_{F}(\hat{\rho})}\left(u\right)= & \int_{E_{idl}}W_{\hat{\rho}}\left(A_{F}^{-1}\left(u\oplus v\right)\right)dv\nonumber \\
= & \int_{E_{idl}}W_{\hat{U}_{A_{F}}\hat{\rho}\hat{U}_{A_{F}}^{\dagger}}\left(u\oplus v\right)dv.
\end{align}
In other words, this shows that the quantum channel $\mathcal{A}_{F}$
can be equivalently defined as
\[
\mathcal{A}_{F}\left(\hat{\rho}\right)=\text{Tr}_{\mathcal{H}_{idl}}\left[\hat{U}_{A_{F}}\hat{\rho}\hat{U}_{A_{F}}^{\dagger}\right],
\]
which is at the same time a Gaussian channel $\mathcal{G}_{\left(A_{F}\right)_{out,tot},0}$.

This will be done as follows: Let $\hat{\rho}$ be arbitrary density
operator, $l_{h}$ a Lagrangian plane, and $l_{h'}$ the symplectic
conjugate of $l_{h}$, such that $l_{h}\oplus l_{h'}=E_{idl}$ which
is the symplectic subspace with associated to the idle subsystem.
Let $F:l_{h}\rightarrow E_{out}$ be an arbitrary linear transformation
which will be used to construct the adaptive displacement operation.
We will show how to represent the quantum channel
\begin{equation}
\mathcal{A}_{F}(\hat{\rho}):=\int\hat{D}_{-F\eta}\text{Tr}_{\mathcal{H}_{idl}}\left[\hat{\Pi}_{l_{h}}\left(\eta\right)\hat{\rho}\right]\hat{D}_{-F\eta}^{\dagger}
\end{equation}
as a Gaussian channel. First, we translate it to the Wigner function
representation (we will frequently omit the constant prefactor in
such computations):
\begin{equation}
W_{\mathcal{A}_{F}(\hat{\rho})}\left(u\right)\propto\int\int_{E_{idl}}W_{\hat{\rho}}\left(\left(u-F\eta\right)\oplus v\right)\delta\left(v|_{l_{h}}-\eta\right)dvd\eta.
\end{equation}
Then by letting
\begin{equation}
X=-\Omega_{h',h}F^{t}\Omega_{out},
\end{equation}
and
\begin{equation}
Y=\frac{1}{2}\Omega_{h',h}F^{t}\Omega_{out}F,
\end{equation}
we have a linear transformation $A:E_{tot}=E_{out}\oplus E_{idl}\rightarrow E_{tot}$,
with the explicit form
\begin{equation}
A_{F}=\begin{pmatrix}\text{Id} & F & 0\\
0 & \text{Id} & 0\\
-X & Y & \text{Id}
\end{pmatrix}.
\end{equation}
It is easy to check that $A$ is a symplectic transformation, the
inverse of which is the following:
\begin{equation}
A_{F}^{-1}=\begin{pmatrix}\text{Id} & -F & 0\\
0 & \text{Id} & 0\\
E & G & \text{Id}
\end{pmatrix}.
\end{equation}
After several steps of straightforward calculations, we have
\begin{align}
W_{\mathcal{A}_{F}(\hat{\rho})}\left(u\right)= & \int_{E_{idl}}W_{\hat{\rho}}\left(A_{F}^{-1}\left(u\oplus v\right)\right)dv\nonumber \\
= & \int_{E_{idl}}W_{\hat{U}_{A_{F}}\hat{\rho}\hat{U}_{A_{F}}^{\dagger}}\left(u\oplus v\right)dv.
\end{align}
Therefore the map $\mathcal{A}$ is a Gaussian channel defined by
\begin{equation}
\mathcal{A}_{F}\left(\hat{\rho}\right)=\text{Tr}_{\mathcal{H}_{idl}}\left[\hat{U}_{A_{F}}\hat{\rho}\hat{U}_{A_{F}}^{\dagger}\right],
\end{equation}
or can equivalently be denoted as $\mathcal{G}_{\left(A_{F}\right)_{out,tot},0}$
with the submatrix of $A_{F}$ being
\begin{equation}
\left(A_{F}\right)_{out,tot}=\begin{pmatrix}\text{Id}, & -F, & 0\end{pmatrix}.
\end{equation}
Here $tot$ represents the whole symplectic space $E$.

\section{Adaptive quantum transduction}

Following the discussion in the last section, we continue representing
more components of the generalized teleportation by Gaussian channels
and at the same time incorporate models of quantum noise into the
new representation. To mark the ``practicality'' of this model and
emphasize the final goal, we now officially rebrand the generalized
teleportation as the adaptive quantum transduction\index{Adaptive quantum transduction}.

We have seen The successful implementation of the adaptive quantum
trasndcution protocol relies on the availability of infinitely-squeezed
ancillary states and the availability of perfect homodyne measurements
of the given transducer\footnote{For simplicity, we ignore other non-unitary imperfections such as
the intrinsic losses or gains in the bosonic system, since they are
not as an urgent issue as the these two major imperfections. However,
the method introduced in this section can be easily adapted to incorporating
these imperfections into our adaptive quantum transduction scheme. }. However, in practical experimental environments, these conditions
cannot be totally satisfied: Infinitely-squeezed states are purely
theoretical assumptions which can only be approximated by finitely-squeezed
Gaussian states; homodyne measurements are usually inefficient; intrinsic
losses are inevitable. Here, we will discuss in detail how to model
these experimental imperfections and incorporate them into our adaptive
quantum transduction scheme.

The conventions in Secs.~\ref{sec:Generalized-continuous-variable-}\&\ref{sec:Gaussian-channel-representation}
will be followed in this section, with the addition of the environmental
modes, associated with the symplectic space $E_{env}$. These environmental
modes will be used to model the inefficeicy of the homodyne measurement.
Therefore, now the symplectic space $E$ of the whole system can be
decomposed as follows
\begin{align*}
E & =E_{in}\oplus E_{anc}\oplus E_{env}\\
 & =E_{out}\oplus E_{idl}\oplus E_{env}.
\end{align*}
First, the imperfect ancillary state will be modeled by a possibly
thermal and finitely squeezed Gaussian state, denoted by $\hat{\rho}\left(0,V_{anc}\right)$,
on $E_{anc}$. As we have seen in Sec.~\ref{sec:Gaussian-measurement},
the imperfect homodyne measurement, still a Gaussian measurement,
will be modeled by the combination of a thermal state in $E_{env}$,
a symplectic transformation on $E_{idl}\oplus E_{env}$ and perfect
homodyne measurements on $E_{idl}$. Here we would like to explain
in more detail how we will model the imperfect homodyne measurement.
For a single-quadrature homodyne measurement, we will model it as
a combination of a fictitious beamsplitter and an ideal homodyne detector,
so that the signal will first be mixed with a fictitious environmental
mode by the beamsplitter before being measured. Then, by changing
the transmittance of the fictitious beamsplitter and the thermal excitation
number of the environmental mode, this mode can neatly simulate the
imperfections in a practical homodyne measurement. For the multi-quadrature
situation, we will simply group many copies of this single-mode model.
That is, we will use a fictitious multi-mode beamsplitter, denoted
by $\tilde{H}$, and a multi-mode environmental Gaussian state $\hat{\rho}\left(0,V_{env}\right)$
to model the imperfections\footnote{Specifically, we let
\begin{align*}
\tilde{H} & =\oplus_{i=1}^{M}H(\tau)
\end{align*}
with $M$ the number of modes supported by $\mathcal{H}_{idl}$, $\tau$
the transmittance of a single two-mode beamsplitter as in Sec.~\ref{sec:Gaussian-measurement}.}.

\begin{figure}[h]
\includegraphics[width=0.9\textwidth]{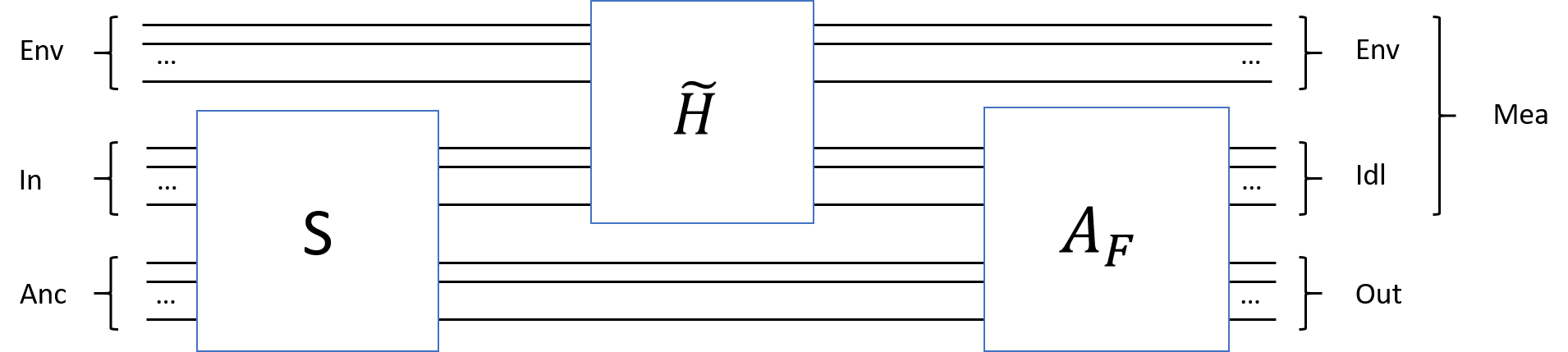}
\begin{doublespace}
\centering{}\caption{\label{fig:Adaptive-quantum-transduction}\emph{Adaptive quantum transduction.}}
\begin{minipage}[t]{0.9\textwidth}%
The abstraction of adaptive quantum transduction with imperfections.
The box labeled by $S$ represents the given imperfect transducer
with the corresponding symplectic transformation. The box $\tilde{H}$
represents the fictitious multi-mode beam-splitter modeling the imperfections
of of the homodyne measurements. The symplectic transformation $A_{F}$
represents the adaptive control as explained in the preceding section.%
\end{minipage}
\end{doublespace}
\end{figure}

As explained in the last section, we model the adaptive displacement
operation determined by a linear transformation $F$ using the symplectic
transformation $A_{F}$. Then the adaptive quantum transduction scheme
amounts to the following Gaussian channel (as shown in Fig.~\ref{fig:Adaptive-quantum-transduction}):
\[
\mathcal{G}_{T,N}\left(\hat{\rho}\right)=\text{Tr}_{\mathcal{H}_{idl}\otimes\mathcal{H}_{ign}}\left[\mathcal{G}_{A_{F}\oplus\text{Id},0}\circ\mathcal{G}_{\text{Id}\oplus\tilde{H},0}\circ\mathcal{G}_{S\oplus\text{Id},0}\left(\hat{\rho}\otimes\hat{\rho}_{anc}\otimes\hat{\rho}_{env}\right)\right]
\]
with
\begin{align}
T= & S_{out,in}+F\tilde{H}_{h,idl}S_{idl,in},
\end{align}
and
\[
N=\tilde{B}V_{anv}\tilde{B}^{t}+\left(F\tilde{H}_{h,env}\right)V_{env}\left(F\tilde{H}_{h,env}\right)^{t},
\]
where
\[
\tilde{B}=S_{out,anc}+F\tilde{H}_{F,idl}S_{idl,anc}.
\]

Furthermore, we let $n_{z}$ denote the thermal excitation in
the ancillary state, $\xi$ be the degree of squeezing. Then the covariance
matrix, arranged in the alternative order as mentioned in Sec.~\ref{symplectic_transformation},
of the ancillary state is given by

\[
V_{anc}=\begin{pmatrix}e^{-2\xi}\left(2n_{z}+1\right)\text{Id} & 0\\
0 & e^{2\xi}\left(2n_{z}+1\right)\text{Id}
\end{pmatrix}.
\]
In addition, we assume the transmittance of the beamsplitter $\tilde{H}$
to be $\tau$, and the thermal excitation number of the environmental
state to be $n_{h}$. The the covariance matrix of the environmental
state is given by
\[
V_{env}=\left(2n_{h}+1\right)\text{Id}.
\]
Shortly after, we will see that these two imperfections can be
quantified by two real numbers: (1) the ancillary imperfection coefficient\index{Ancillary imperfection coefficient}
$\nu$, defined as
\[
\nu=e^{-2\xi}(2n_{z}+1),
\]
(2) and the measurement imperfection coefficient\index{Measurement imperfection coefficient}
$\mu$, defined as
\[
\mu=\frac{\tau}{1-\tau}(2n_{h}+1).
\]
 Now we put these data in the adaptive quantum transduction channel
$\mathcal{G}_{T,N}$. We obtain that
\[
T=S_{out,in}+\sqrt{1-\tau}FS_{h,in},
\]
and
\[
N=\tilde{B}V_{anc}\tilde{B}^{t}+\tau FF^{t}
\]
with
\[
\tilde{B}=S_{out,anc}+\sqrt{1-\tau}FS_{h,anc}.
\]
According to Theorem.~\ref{thm:generalized-teleportation}, if now
we choose $F$ to be $F_{\star}=-S_{out,z'}\left(S_{h,z'}\right)^{-1}/\sqrt{1-\tau}$,
then it follows that
\[
T=\tilde{S},
\]
is a symplectic transformation, and
\[
N=\nu\tilde{S}B_{\star}B_{\star}^{t}\tilde{S}^{t}+\mu F_{\star}F_{\star}^{t},
\]
with 
\[
B_{\star}=B=-\left(S^{-1}\right)_{in,h'}\left(\left(S^{-1}\right)_{z,h'}\right)^{-1}.
\]
Then, composing this Gaussian channel $\mathcal{G}_{\tilde{S},N}$
with a post-processing unitary Gaussian channel $\mathcal{G}_{\tilde{S}^{-1},0}$,
we obtain the Gaussian channel
\[
\mathcal{G}_{\text{Id},N'}=\mathcal{G}_{\tilde{S}^{-1},0}\circ\mathcal{G}_{\tilde{S},N}
\]
with
\[
N'=\nu B_{\star}B_{\star}^{t}\left(\tilde{S}\right)+\mu\left[\tilde{S}^{-1}S_{out,z'}\left(S_{h,z'}\right)^{-1}\right]\left[\tilde{S}^{-1}S_{out,z'}\left(S_{h,z'}\right)^{-1}\right]^{t},
\]
which will be referred to as the Gaussian channel of the adaptive
quantum transduction. Now we see that this channel is completely determined
by the transducer $S$, and the two imperfection coefficients $\mu$
and $\nu$. 

\section{Average fidelity for single-mode Gaussian channels}

In the preceding section, we defined the Gaussian channel for adaptive
quantum transduction, which is the model of the noise caused by the
experimental imperfections. In this section, we will use this knowledge
to verify the robustness of the adaptive quantum transduction against
these experimental imperfections. We will demonstrate, with respect
to certain quantum benchmark, the adaptive quantum transduction is
a promising quantum transduction scheme even in presence of practical
experimental limitations.

The quantity we use to benchmark the adaptive quantum transduction
scheme is the average fidelity. The performance of a quantum channel
for quantum transduction depends largely on how close the channel
is to a unitary operation. One natural way to measure this ``distance''
is to send an ensemble of quantum states into the given quantum channel
$\mathcal{E}$, and then calculate the average overlap between the
ensemble of the output quantum states and the original input ensemble.
Specifically, let $\left\{ \hat{\rho}_{i}\right\} $ be an ensemble
quantum states associated with a probability distribution $\left\{ p_{i}\right\} $,
so that $\left\{ \mathcal{E}(\hat{\rho}_{i})\right\} $ is the output
ensemble. The ``overlapping'' of the two ensembles, i.e. the average
fidelity\index{Average fidelity}, is given by the following formula:
\[
\bar{F}\left(\mathcal{E};p\right)=\sum_{i}p_{i}F\left(\hat{\rho}_{i},\mathcal{E}\left(\hat{\rho}_{i}\right)\right),
\]
where 
\[
F\left(\hat{\rho},\hat{\rho}'\right)=\text{Tr}\left[\sqrt{\hat{\rho}}\hat{\rho}'\sqrt{\hat{\rho}}\right]
\]
 is the quantum fidelity\emph{ }between two arbitrary quantum states
$\hat{\rho}$ and $\hat{\rho}'$. Now supposing that the input states
are sampled from a uniformly-distributed ensemble of single-mode coherent
states (i.e. every coherent state in the phase space has the same
probability to be sampled out), and that the quantum channel $\mathcal{E}$
is a single-mode Gaussian channel $\mathcal{G}_{T,N}$, the average
fidelity can be calculated by the following simple formula:
\[
\bar{F}\left(\mathcal{G}_{T,N}\right)=\begin{cases}
\frac{2}{\sqrt{\text{det}\left(2\text{Id}+N\right)}}, & \text{if }T=\text{Id},\\
0, & \text{if }T\neq\text{Id},
\end{cases}
\]

\begin{example}
This formula makes sense in the special situation of an identity Gaussian
channel $\mathcal{G}_{\text{Id,0}}$, with $\bar{F}(\mathcal{G}_{\text{Id},0})=1$
as it should be. 
\end{example}

In addition, the average fidelity not only allows us to quantify the
difference between a quantum channel and an identity channel, but
also serves as a measure of the ``quantumness''. To see it, we consider
the following process: Instead of being sent through a quantum channel
$\mathcal{E}$, the ensemble of input states are measured in the beginning.
Then the measurement outcomes are transferred through a classical
communication channel to the output. Based only on the classical information
transferred to the output, we prepare the output state to be as close
to the input state as possible.We refer to such a process as classical
transduction. For instance, considering the uniformly-distributed
coherent-state ensemble, the average fidelity of any classical transduction
scheme, regardless of what measurement scheme or state preparation
method is applied, can only approach but not exceed the $1/2$ threshold
\cite{Braunstein2001}. Therefore, given a quantum channel $\mathcal{E}$,
the ``quantumness'' of $\mathcal{E}$ as a transduction protocol
will only be justified if $\bar{F}(\mathcal{E})>1/2$. In particular,
for a single-mode Gaussian channel $\mathcal{G}_{\text{Id},N}$, this
amounts to
\[
\text{det}\left(\text{Id}+\frac{N}{2}\right)<4.
\]

Later on, we will use the average fidelity to compare our adaptive
transduction protocol with the direct transduction protocol and the
classical threshold, given a symplectic transformation $S$ involving
two bosonic modes. Specifically, we will need to calculate the average
fidelity for the Gaussian channels associated with the adaptive quantum
transduction protocol and the direct quantum transduction protocol.
The former has been displayed in the previous section. For the latter,
we will assume the ancillary state is the vacuum state so that the
added noise to the output is minimized. This leads to the Gaussian
channel $\mathcal{G}_{T,N}$ with
\[
T=S_{out,in},
\]
and
\[
N=S_{out,anc}(S_{out,anc})^{t}.
\]
Furthermore, when $T$ is invertible, another Gaussian channel $\mathcal{G}_{T^{-1},N'}$
will be appended, leading to the Gaussian channel
\begin{align*}
\mathcal{G}_{T'',N''}= & \mathcal{G}_{T^{-1},N'}\circ\mathcal{G}_{T,N}\\
= & \mathcal{G}_{\text{Id},T^{-1}N\left(T^{-1}\right)^{t}+N'},
\end{align*}
which will be referred to as the Gaussian channel of the direct transduction
protocol later on. Note that the matrix $N'$ will be optimized to
maximize the average fidelity $\bar{F}(\mathcal{G}_{\text{Id},N''})$.

\begin{figure}[h]
\includegraphics[width=0.9\textwidth]{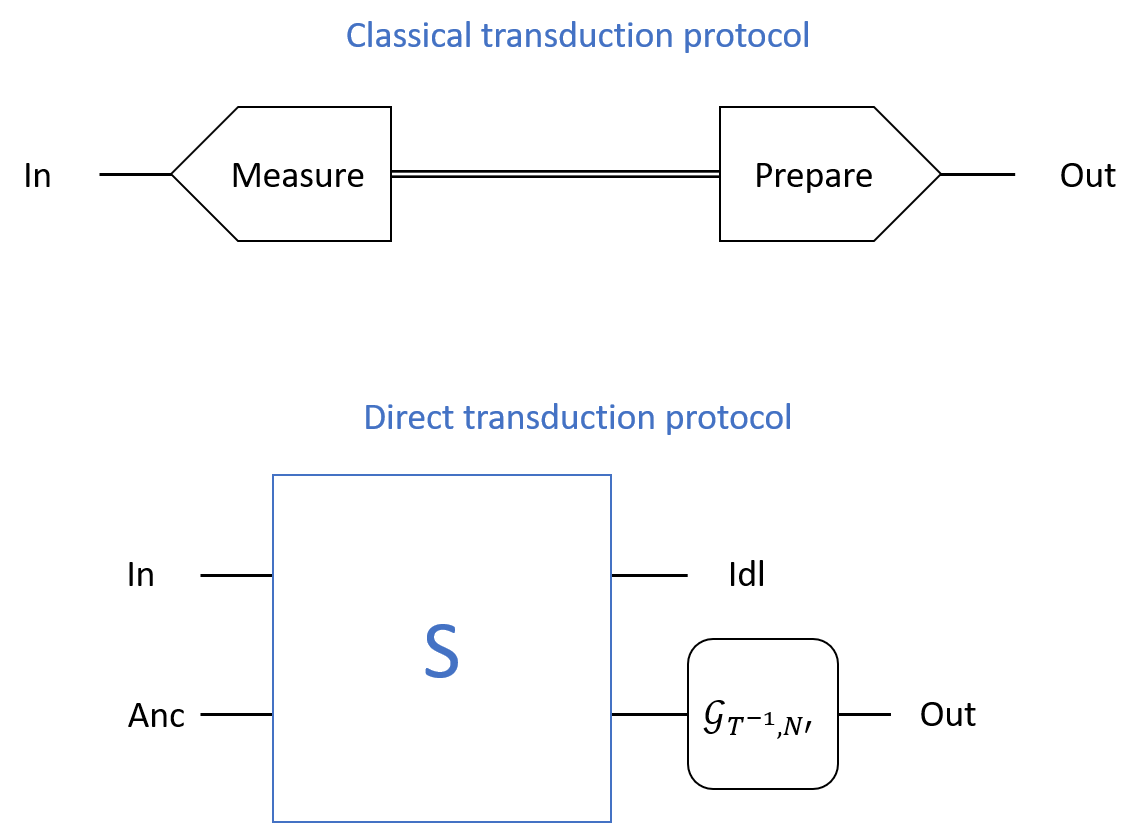}
\begin{doublespace}
\centering{}\caption{\emph{Classical and direct quantum transduction protocols.}}
\begin{minipage}[t]{0.9\textwidth}%
The classical transduction protocol: The input quantum state is measured
by a given measurement scheme. Then the measurement outcome is transmitted
to the output through a classical communication channel, and is used
to prepare the output quantum state.The Direct transduction protocol:
The input quantum state and the ancillary noisy environment state
are mixed by the symplectic transformation $S$. Then the idler modes
are dropped and the output modes are modified by a Gaussian channel
$\mathcal{G}_{T^{-1},N'}$. This modifying Gaussian channel is applied
to optimize the average fidelity at the end. %
\end{minipage}
\end{doublespace}
\end{figure}

Finally, we can apply the theoretical tools established in this and
the previous sections to analyze the transdcution performance of Gaussian
processes. We will demonstrate the method in the following two examples:
\begin{example}
(Passive two-mode coupling\emph{.})\emph{ }First, we consider a two-mode
bosonic system, dynamics of which are determined by the following
total-excitation-number-preserving Hamiltonian
\[
\hat{H}=g\left(\hat{a}_{1}^{\dagger}\hat{a}_{2}+\hat{a}_{2}^{\dagger}\hat{a}_{1}\right).
\]
Each of the two modes is coupled to an external propagating mode.
i.e. the mode $\hat{a}_{1(2)}$ is coupled to the propagating mode
$\hat{A}_{1(2)}$ with coupling rate $\kappa_{1(2)}$. We identify
the mode $\hat{A}_{1}$ as the input and $\hat{A}_{2}$ as the output.
Then this process amounts to a Gaussian process associated with the
following symplectic transformation (see Appx.~\ref{chap:Scattering-process-and}
for more detail)
\[
S=\begin{pmatrix}0 & -t & r & 0\\
t & 0 & 0 & r\\
r & 0 & 0 & -t\\
0 & r & t & 0
\end{pmatrix}
\]
with
\[
r=\frac{C-1}{C+1},\quad t=\frac{2\sqrt{C}}{C+1}
\]
satisfying $r^{2}+t^{2}=1$, where
\[
C=\frac{g^{2}}{\kappa_{1}\kappa_{2}}.
\]
The details of the calculation will be elaborated in later chapters.
It follows that the direct quantum transduction Gaussian channel of
this process is $\mathcal{G}_{\text{Id},N_{D}}$ with
\[
N_{D}=2\frac{1-t^{2}}{t^{2}}\text{Id},
\]
average fidelity of which is
\[
\bar{F}\left(\mathcal{G}_{\text{Id},N_{D}}\right)=t^{2}.
\]
In the meantime, with the $\hat{q}$-quadrature of the ancillary state
being squeezed and the idler mode being measured, we obtain the adaptive
quantum transduction Gaussian channel (after performing the post-processing
symplectic transformation $\tilde{S}$), $\mathcal{G}_{\text{Id},N_{A}}$
with
\[
N_{A}=\begin{pmatrix}\mu\left(1-t^{2}\right)\\
 & \nu\frac{1-t^{2}}{t^{2}}
\end{pmatrix},
\]
average fidelity of which is
\[
\bar{F}\left(\mathcal{G}_{\text{Id},N_{D}}\right)=\frac{1}{\sqrt{\left(1+\frac{1-t^{2}}{2}\mu\right)\left(1+\frac{1-t^{2}}{2t^{2}}\nu\right)}},
\]
where $\mu$, $\nu$ are the measurement and ancillary imperfection
coefficients, as is introduced earlier in this chapter.
\end{example}

\begin{figure}[h]
\includegraphics[width=0.9\textwidth]{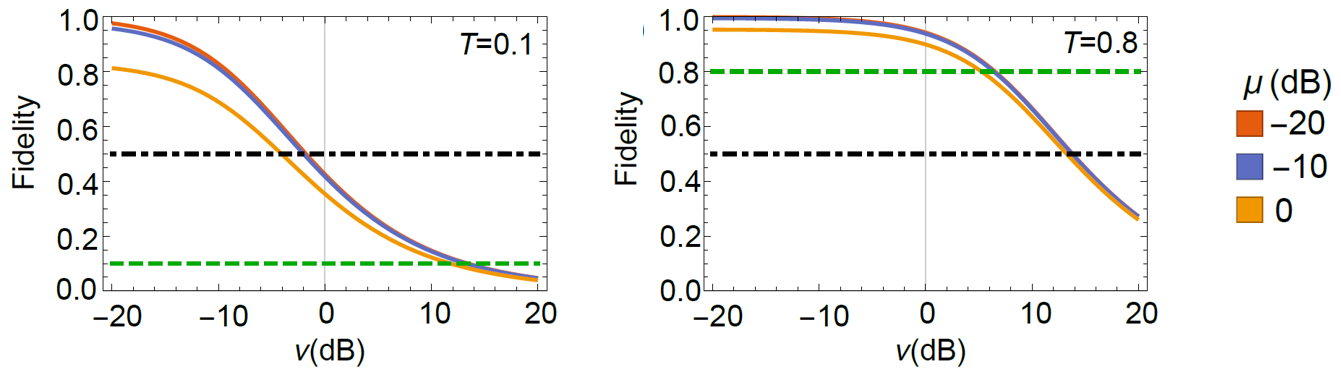}
\begin{doublespace}
\centering{}\caption{\label{fig:Average-fidelity-in}\emph{Average fidelity in the passive
system.}}
\begin{minipage}[t]{0.9\textwidth}%
The figures show the average fidelity of adaptive quantum transduction
as a function of ancillary imperfection coefficient $\nu$, given
the measurement imperfection coefficient $\mu=-20$, $-10$, and $0$
dB. The simulated imperfect transducers are passive two-mode scattering
processes with the transmittance being $T=t^{2}=0.8$ and $T=t^{2}=0.1$
respectively, as explained in the example. The dark-dotted dashed
lines correspond to the threshold fidelity of $1/2$. The green-dashed
lines correspond to the fidelity achieved by the optimized direct
quantum transduction protocol. Therefore, we see that adaptive quantum
transduction can greatly improve the average fidelity with reasonable
imperfection coefficients, and can even convert a non-quantum transducer
(with average fidelity lower that $1/2$) into a quantum transducer.%
\end{minipage}
\end{doublespace}
\end{figure}

\begin{example}
(Active two-mode coupling.)\emph{ }The second system we consider is
described by the following Hamiltonian:
\[
\hat{H}=g\left(\hat{a}_{1}^{\dagger}\hat{a}_{2}^{\dagger}+\hat{a}_{1}\hat{a}_{2}\right),
\]
which couples the two modes like a two-mode squeezer. With other components
unchanged, the Gaussian process is associated with the symplectic
matrix (see Appx.~\ref{chap:Scattering-process-and} for more detail),
\[
S=\begin{pmatrix}0 & t' & r' & 0\\
t' & 0 & 0 & r'\\
r' & 0 & 0 & t'\\
0 & r' & t' & 0
\end{pmatrix},
\]
with
\[
r'=\frac{C+1}{C-1},\quad t'=\frac{2\sqrt{C}}{1-C},
\]
satisfying $r'^{2}-t'^{2}=1$, and
\[
C=\frac{g^{2}}{\kappa_{1}\kappa_{2}}.
\]
. It follows that the direct quantum transduction Gaussian channel
is $\mathcal{G}_{\text{Id},N_{D}}$ with
\[
N_{D}=2\frac{1+t'^{2}}{t'^{2}}\text{Id}.
\]
The average fidelity of $\mathcal{G}_{\text{Id},N_{D}}$ is
\[
\bar{F}\left(\mathcal{G}_{\text{Id},N_{D}}\right)=\frac{t'^{2}}{1+2t'}<\frac{1}{2}.
\]
The adaptive quantum transduction Gaussian channel is $\mathcal{G}_{\text{Id},N_{A}}$
with
\[
N_{A}=\begin{pmatrix}\mu\left(1+t'^{2}\right)\\
 & \nu\frac{1+t'^{2}}{t^{2}}
\end{pmatrix}.
\]
Its average fidelity is
\[
\bar{F}\left(\mathcal{G}_{\text{Id},N_{A}}\right)=\frac{1}{\sqrt{\left(1+\frac{1+t'^{2}}{2}\mu\right)\left(1+\frac{1+t'^{2}}{2t^{2}}\nu\right)}}
\]
\end{example}

\begin{figure}[h]
\includegraphics[width=0.9\textwidth]{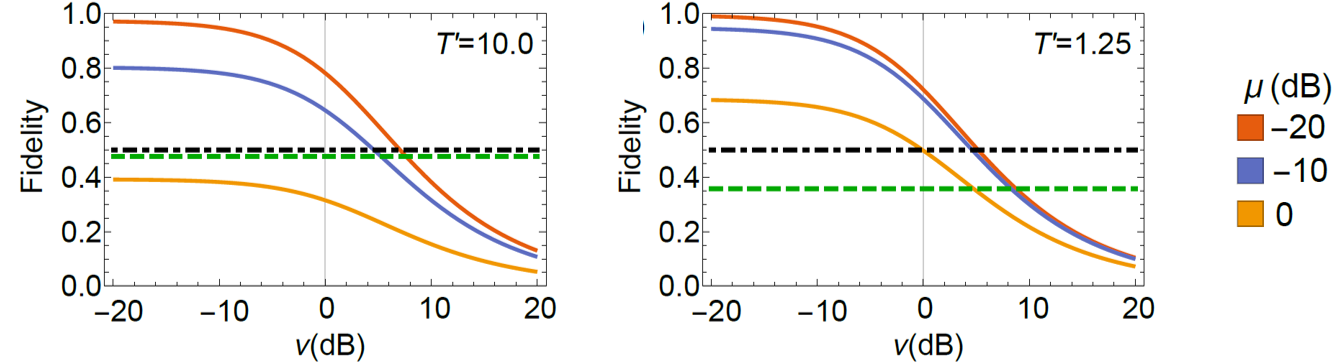}
\begin{doublespace}
\centering{}\caption{\label{fig:Average-fidelity-in-1}\emph{Average fidelity in the active
system.}}
\begin{minipage}[t]{0.9\textwidth}%
The figures show the average fidelity of adaptive quantum transduction
as a function of ancillary imperfection coefficient $\nu$, given
the measurement imperfection coefficient $\mu=-20$, $-10$, and $0$
dB. The simulated imperfect transducers are active two-mode scattering
processes with the transmittance being $T=t^{2}=10.0$ and $T=t^{2}=1.25$
respectively, as explained in the example. The dark-dotted dashed
lines correspond to the threshold fidelity of $1/2$. The green-dashed
lines correspond to the fidelity achieved by the optimized direct
quantum transduction protocol. In both figures, the direct quantum
transduction protocols are beaten by classical transduction protocols,
since the average fidelities are below the $1/2$ threshold, while
the adaptive quantum transduction protocols beat both of them, given
reasonable $\mu$ and $\nu$.%
\end{minipage}
\end{doublespace}
\end{figure}

Clearly, as shown in Figs.~\ref{fig:Average-fidelity-in}\&\ref{fig:Average-fidelity-in-1},
when $\mu$ and $\nu$ are reasonably small, we can always obtain
a ``quantum'' transducer from the adaptive quantum transduction
protocol in both of the above examples, while the direct transdcution
production protocol fails to be quantum when $t^{2}\le\frac{1}{2}$
in the passive situation or for any $t'$ in the active situation.

Furthermore, the symplectic transformation $\tilde{S}$ in each of
the two examples can be viewed as a single-mode squeezing operation.
Specifically, in the first example, we have
\[
\tilde{S}=-Z\left(t\right)\Omega=-\begin{pmatrix}t\\
 & t^{-1}
\end{pmatrix}\begin{pmatrix}0 & 1\\
-1 & 0
\end{pmatrix}=\begin{pmatrix}0 & -t\\
t^{-1} & 0
\end{pmatrix}
\]
while in the second example, we have
\[
\tilde{S}=-Z\left(t'\right)\Omega=\begin{pmatrix}0 & -t'\\
t'^{-1} & 0
\end{pmatrix}.
\]
Thus, the adaptive quantum transdcution protocol also provides a way
to squeeze bosonic modes using squeezing resources in the other modes.
In general, the generalized teleportation protocol allows us generating
controllable few-body Gaussian operation out of a many-body Gaussian
operation using only local Gaussian processes and classical communications.

\section{Conclusion and outlook}

In this chapter, we demonstrated a generalized interpretation of continuous
variable quantum teleportation and its application in quantum transduction.
This interpretation also applied to discrete variable quantum teleportation,
which we will see in the last chapter of this thesis. Here we summarize
the advantages of this theoretical framework: (1) It applies to, in
general, arbitrary symplectic transformations; (2) It imposes no constraints
on the number of bosonic modes, meaning that we can apply it to a
large bosonic system comprising a huge amount of modes; (3) It requires
on local Gaussian processes and classical communications. Therefore,
its flexibility renders this theoretical framework a promising candidate
for practical solution to quantum transduction and possibly more bosonic
engineering problems.

However, as we have demonstrated, the success of this framework relies
heavily on the high quality of squeezing and homodyne measurement,
which are often elusive in laboratories. In the succeeding chapter,
we discuss another possibility for practical quantum transduction
without assuming these usually impractical resources of bosonic control.

\chapter{Interference-based Gaussian control\label{chap:Interference-based-Gaussian-cont}}

\section{Introduction}

Interference, an essential feature of quantum mechanics, is the fountain
of countless applications. As we know, interference can be easily
calculated, if we represent quantum states as vectors and quantum
operations as matrices. Therefore, we are wondering if this feature
can be useful in the phase space, where the quadrature operators are
treated as vectors and the symplectic transformations as matrices.
It turns out that interference not only exists in the phase space
but is also a powerful tool for solving practical bosonic control
problems, including quantum transduction\footnote{This chapter is based on the results in \cite{Zhang2020}.}.

\begin{figure}[h]
\includegraphics[width=0.9\textwidth]{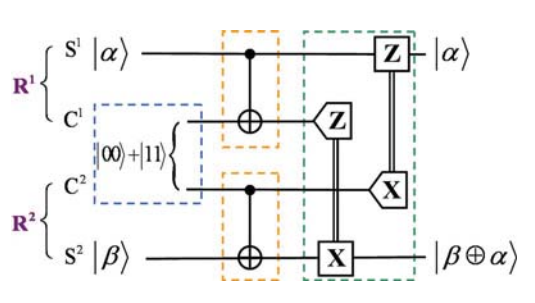}
\begin{doublespace}
\begin{centering}
\caption{\label{fig:Interference-based-swapping-of}\emph{Interference-based
swapping of two bosonic modes.}}
\begin{minipage}[t]{0.9\textwidth}%
The blue boxes labeled by $S$ represent identical copies of a given
two-mode symplectic transformation $S$. The other blue boxes represent
the local (single-mode) symplectic transformations which are carefully
designed to engineer the interference of the quadratures. Once properly
constructed, this sequence will faithfully convert the input quantum
state to the output port, up to a unitary transformation.%
\end{minipage}
\par\end{centering}
\end{doublespace}
\end{figure}

Alluded to our conclusion of the preceding chapter, adaptive quantum
transduction suffers the unavailability of infinite squeezing and
perfect homodyne measurement. Therefore, finding a transduction scheme
independent of these impractical resources is much needed. An attempt
to solve this problem is demonstrated in \cite{Lau2019}, based on
the interference of symplectic transformations. Simply speaking, suppose
we have many identical copies of a given imperfect two-mode quantum
transducer, which is described by a symplectic transformation, and
we are able to control both of the involved modes using arbitrary
single-mode unitary Gaussian operations (not including infinite squeezing);
Then we can construct a sequence of symplectic transformations interspersed
by carefully-chosen single-mode symplectic transformations, such that
the whole sequence effectively swaps the two modes up to single-mode
symplectic transformations\footnote{This means, the input quantum state $\hat{\rho}$ of one mode will
be transformed to the output state $\hat{U}_{S}\hat{\rho}\hat{U}_{S}^{\dagger}$
of the other mode, with $\hat{U}_{S}$ a single-mode symplectic transformation.}, as hown in Fig.~\ref{fig:Interference-based-swapping-of}. In other
words, we are able to convert an imperfect quantum transducer to a
perfect quantum transducer, without requiring infinite squeezing or
any measurement.

The above scheme demonstrates the power of the interference of symplectic
transformations and its promising application in quantum transduction.
However, we notice that scheme works only for the two-mode situation.
Thus, we cannot apply this scheme to addressing the inevitable side
band effects, which is a major challenge for usual transduction proposals
and our motivation of adaptive quantum transduction as explained in
the beginning section of the preceding chapter.

\begin{figure}[h]
\begin{centering}
\includegraphics[width=0.9\textwidth]{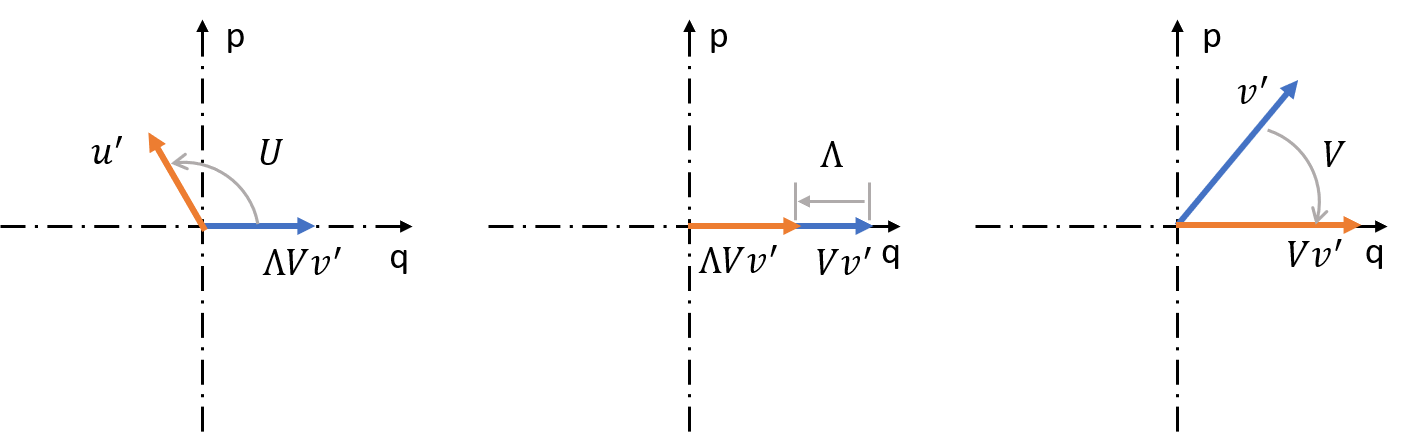}
\par\end{centering}
\begin{doublespace}
\centering{}\caption{\label{fig:Existence-of-.}\emph{Transitivity of single-mode symplectic
transformations.}}
\begin{minipage}[t]{0.9\textwidth}%
These figures demonstrate a random quadrature can be transformed to
another random quadrature by symplectic transformations in the single-mode
symplectic space. We start with a random quadrature, i.e. the blue
vector in the right figure, which can be symplectically rotated to
the $\hat{q}$-axis (the orange vector). Then this orange vector can
be dilated to a proper length using a symplectic squeezing operation,
as shown in the middle figure. At last, in the left figure, the dilated
vector is transformation to the target quadrature via a simple rotation.
Denoting the rotations and the squeezing operation by $U,V$ and $\Lambda$,
we conclude that the same result can be achieved by a single-mode
symplectic operation $U\Lambda V$. Since a multi-mode quadrature
can be decomposed as multi-mode single-mode chuncks, each of which
can be freely controlled by single-mode symplectic transformations,
this single-mode conclusion can be easily extended to multi-mode situations.%
\end{minipage}
\end{doublespace}
\end{figure}

In this chapter, we will show that, by fully exploiting the mathematical
properties of symplectic transformations, we can swap a pair of bosonic
modes by construcing a sequence of symplectic transformations, with
similar structure as of the above scheme but fundamentally different
engineering strategies, for arbitrary-mode quantum transduction scenarios.
Specifically, we still assume we have many copies of an imperfect
quantum transducer, now described by a multi-mode symplectic transformation
$S$, and arbitrary single-mode unitary Gaussian control of each mode
involved in the process, which will be referred to as the local Gaussian
operations. The whole idea will be based on a surprisingly simple
but useful property of the local Gaussian operations: We can transform
any given quadrature operator to any other quadrature operator\footnote{We have to exclude some exceptional cases to make this statement rigorous:
The statement may not hold if a considered quadrature does not act
on every canonical quadrature operator. Specifically, the quadrature
$\hat{q}_{1}+\hat{p}_{1}$ cannot be transformed to $\hat{q}_{2}+\hat{p}_{2}$
by local Gaussian operations, since $\hat{q}_{1}+\hat{p}_{1}$ does
not influence the second bosonic mode.} using local Gaussian operations. This can be easily seen form the
single-mode situation as shown in Fig.~\ref{fig:Existence-of-.}.
For some technical reasons, this property provides for us great flexibility
in manipulating multi-mode symplectic transformations: Now we can
construct a three-component sequence $S'=SL'S$, which we call a sandwich
operation, using the given symplectic transformation and a local operation
$L'$ to decouple $\hat{q}_{1}$ with the other canonical quadrature
operators; similarly, we can construct a sandwich operation $S^{*}=SL^{*}S$
to decopule $\hat{q}_{N}$ (suppose there are $N$-modes) with the
other canonical quadrature operators. Then another layer of sandwich
operation, $S^{td}=S^{*}L^{td}S'$, will be an effective symplectic
transformation perfectly transducing the first mode to the last mode.
Moreover, we can obtain another sandwich operation $S^{dc}=S'L^{dc}S'$
to decouple the first mode from the rest of the system. It turns out,
the sequence $\left(S^{td}L^{a}S^{dc}\right)L^{b}\left(S^{td}L^{c}S^{dc}\right)$
swaps the first and the last modes of system. Since which mode is
the first mode and which mode is the last only depends on how we label
the bosonic modes, we therefore obtain universal swap and universal
decoupling ($S^{dc}$) for multi-mode bosonic systems.

To give a more intuitive picture of the above mechanism, we would
like to demonstrate its application to the simplest two-mode situation
in the following example.
\begin{example}
(Two-mode application) Consider a symplectic transformation
\begin{equation}
S=\begin{pmatrix}S_{11} & S_{12} & S_{13} & S_{14}\\
S_{21} & S_{22} & S_{23} & S_{24}\\
S_{31} & S_{32} & S_{33} & S_{34}\\
S_{41} & S_{42} & S_{43} & S_{44}
\end{pmatrix}.
\end{equation}
Now assume a local operation $L'$ transforming the quadrature\footnote{As we know in euclidean geometry, the columns (and rows) of an orthogonal
matrix form a orthonormal basis of the euclidean space. Similarly
in symplectic geometry, the columns (and rows) of a symplectic transformation
form a symplectic basis of the symplectic space, which is a direct
result of the definition of symplectic transformation. Therefore,
we can safely treat columns (and rows) of a symplectic transformation
as quadratures.} $\left(S_{11},S_{21},S_{31},S_{41}\right)^{t}$ (the first column
of $S$) to the quadrature $\Omega\left(S_{11},S_{12},S_{13},S_{14}\right)^{t}$
(the first row of $S$)\footnote{Here $\Omega$ is the symplectic form.}.
Then the definition of $S$, i.e. the CCRs, guarantees that the sandwich
operation $S'=SL'S$ to be of the following form: 
\[
S'=\begin{pmatrix}0 & 1 & 0 & 0\\
-1 & S_{22}' & S_{23}' & S_{24}'\\
0 & S_{32}' & S_{33}' & S_{34}'\\
0 & S_{42}' & S_{43}' & S_{44}'
\end{pmatrix}.
\]
Similarly, let $L^{*}$ be a local operation transforming the quadrature
$\Omega\left(S_{11},S_{21},S_{31},S_{41}\right)^{t}$ (the first column
of $S$) to $\left(S_{31},S_{32},S_{33},S_{34}\right)^{t}$ (the third
row of $S$). The sandwich operation $S^{*}=SL^{*}S$ is of the form
\[
S^{*}=\begin{pmatrix}0 & S_{12}^{*} & S_{13}^{*} & S_{14}^{*}\\
0 & S_{22}^{*} & S_{23}^{*} & S_{24}^{*}\\
0 & 1 & 0 & 0\\
-1 & S_{42}^{*} & S_{43}^{*} & S_{44}^{*}
\end{pmatrix}.
\]
Then using a local operation $L^{td}$ transforming properly transforming
the second column of $S'$ to be a linear combination, multiplied
by the symplectic form, of $S^{*}$, of rows of $S^{*}$, i.e.
\[
\begin{pmatrix}1\\
S_{22}'\\
S_{32}'\\
S_{42}'
\end{pmatrix}\rightarrow-2S_{42}^{*}\Omega\begin{pmatrix}0\\
1\\
0\\
0
\end{pmatrix}+\Omega\begin{pmatrix}-1\\
S_{42}^{*}\\
S_{43}^{*}\\
S_{44}^{*}
\end{pmatrix},
\]
we obtain the sandwich opeartion
\[
S^{td}=S^{*}L^{td}S'=\begin{pmatrix}0 & 0 & S_{13}^{td} & S_{14}^{td}\\
0 & 0 & S_{23}^{td} & S_{24}^{td}\\
0 & 1 & 0 & 0\\
-1 & S_{42}^{td} & 0 & 0
\end{pmatrix}.
\]
Since each off-diagonal $2\times2$ block of $S^{td}$ is a symplectic
transformation, the symplectic transformation $S^{td}$ swaps the
two modes up to local symplectic transformations.
\end{example}

We will provide more technical details and rigorous proofs of this
idea in the succeeding section. However, although this scheme is widely
applicable to physically practical symplectic transformations, there
are still some exceptional situations where more careful discussions
are necessary. To stress these exceptions, we say the above scheme
is applicable to generic symplectic transformations and call the non-generic
symplectic transformations the edge cases. We will discuss the edge
cases, which are possibly of more mathematical interest than practical
importance, in detail in the third section of this chapter, to complete
our discussion.

\section{Permuting bosonic modes using generic symplectic transformations}

Symplectic transformations can be viewed as coordinate changes on
the phase space. Therefore, for a symplectic transformation $S:E\rightarrow E$,
we have symplectic bases $\{y_{j}\}$ and $\{z_{j}\}$ of $E$, such
that
\begin{align*}
S & =\sum_{j}y_{j}x_{j}^{t},\\
 & =\sum_{j}x_{j}z_{j}^{t},
\end{align*}
with $\{x_{j}\}$ be the canonical symplectic basis of $E$. We allude
to our previous discussions on symplectic subspaces that
\[
E=\oplus_{j}E_{j}
\]
with each $E_{j}$ a canonical symplectic subspace, spanned by $\{x_{2j-1},x_{2j}\}$.
Then we can denote the restriction and projection of an element $u$
in $E$ to a symplectic subspace $E_{j}$ by
\[
u_{E_{j}}:=\pi_{j}u.
\]
We suppose the symplectic space $E$ is $2N$-dimensional. Then we
have the following:
\begin{lem}
\label{lem:replicate-quadratures}For all $u,v\in E$ which satisfy
\[
u_{E_{j}}^{t}u_{E_{j}}+v_{E_{j}}^{t}v_{E_{j}}=0,\ \text{if and only if }u_{E_{j}}=v_{E_{j}}=0,\text{for all }j\in\{1,\dots,N\},
\]
there exists a symplectic transformation
\[
L=\oplus_{j=1}^{N}L_{j},
\]
with each $L_{j}:E_{j}\rightarrow E_{j}$ being a symplectic transformation,
such that
\[
Lu=cv,
\]
for any non-vanishing real number $c$.
\end{lem}

\begin{proof}
For each pair of non-vanishing $u',v'$ in each symplectic subspace
$E_{j}$, there are orthogonal matrices $U$, $V$ and a diagonal
matrix
\[
\Lambda=\begin{pmatrix}\lambda & 0\\
0 & \lambda^{-1}
\end{pmatrix},
\]
such that
\[
u'=U\Lambda Vv'.
\]
This proves the existence of $L_{j}$ for all $j\in\{1,\dots,N\}$,
and hence the existence of $L$.
\end{proof}
Symplectic transformations on $E$ that are direct sum of symplectic
transformations on the canonical symplectic subspaces, such as $L$
in the above lemma, will be referred to as the local symplectic transformations.

Now let $\{x_{j}\}$ be the canonical symplectic basis, $\{y_{j}\}$
and $\{z_{j}\}$ be two other symplectic bases of $E$, satisfying
\[
\sigma(y_{j},y_{k})=\Omega_{j,k},\ \forall j,k\in\{1,2,\dots,2N\},
\]
and
\[
\sigma(z_{j},z_{k})=\Omega_{j,k},\ \forall j,k\in\{1,2,\dots,2N\}.
\]
Then we have two symplectic transformations
\[
S=\sum_{j=1}^{2N}y_{j}x_{j}^{t},
\]
and
\[
S'=\sum_{j=1}^{2N}x_{j}z_{j}^{t}.
\]
Furthermore, these symplectic transformations are generic if they
are randomly sampled from the ensemble of all the symplectic transformations.
In particular, for generic $S$ and $S'$, each pair of vectors $y_{j}$
and $z_{j}$ are expected to satisfy the condition in the lemma above,
for the probability of sampling a non-generic symplectic transformation
from the ensemble vanishes\footnote{We can simply assume a generic symplectic transformation is a matrix
with no vanishing element.}. Then the following holds:
\begin{lem}
Let $S$ and $S'$ be generic symplectic transformation. For any non-vanishing
real number $c$, there is a sympletic transformation $L$, such that
\[
S''=S'LS
\]
is a symplectic transformation satisfying
\[
(S'')_{2a-1,2b}=c^{-1},
\]
\[
(S'')_{2a,2b-1}=-c,
\]
and
\[
(S'')_{j,2b-1}=(S'')_{2a-1,k}=0,
\]
for any $j\neq2b$ and $k\neq2a$, for any $a,b\in\{1,2,\dots,N\}$.
\end{lem}

\begin{proof}
According to Lemma.~\ref{lem:replicate-quadratures}, for any non-vanishing
real number $c$, there always exists a local symplectic transformation
$L$, such that
\[
Ly_{2b-1}=c\Omega z_{2a-1},
\]
or equivalently
\[
z_{2a-1}=-c^{-1}\Omega Ly_{2b-1}.
\]
It follows that
\begin{align*}
S'' & =S'LS\\
 & =\sum_{j,k=1}^{2N}(z_{j}^{t}Ly_{k})x_{j}x_{k}^{t}\\
 & =\sum_{j=1}^{2N}(z_{j}^{t}Ly_{2b-1})x_{j}x_{2b-1}^{t}+\sum_{j=2}^{2N}(z_{2a-1}^{t}Ly_{j})x_{2a-1}x_{j}^{t}+\sum_{j,k=2}^{2N}(z_{j}^{t}Ly_{k})x_{j}x_{k}^{t}\\
 & =-cx_{2a}x_{2b-1}^{t}+c^{-1}x_{2a-1}x_{2b}^{t}++\sum_{j,k=2}^{2N}(z_{j}^{t}Ly_{k})x_{j}x_{k}^{t},
\end{align*}
which leads to the following results
\[
(S'')_{2a-1,2b}=x_{2a-1}^{t}S''x_{2b}=c^{-1},
\]
\[
(S'')_{2a,2b-1}=x_{2a-1}^{t}S''x_{2b}=-c,
\]
\[
(S'')_{2a-1,j}=x_{2a-1}^{t}S''x_{j}=0,\ \forall j\neq2,
\]
and
\[
(S'')_{j,2b-1}=x_{j}^{t}S''x_{2b-1}=0,\ \forall j\neq2.
\]
This finishes the proof.
\end{proof}
Let $S$, $S'$ be symplectic transformations as introduced above,
with
\[
y_{2b-1}=-cx_{2b}
\]
and
\[
z_{2a-1}=c^{-1}x_{2a},
\]
for some non-vansihing real number $c$, and $a,b\in\{1,2,\dots,N\}$.
Then according to the definition of symplectic transformations, we
also have
\[
S_{j,2b-1}=S_{2a-1,j}=(S')_{j,2b-1}=(S')_{2a-1,j}=0,
\]
for any $j\neq2$. Furthermore, they are generic in the sense that
Lemma.~\ref{lem:replicate-quadratures} can be applied to $y_{2b}$
and $z_{2a}$. We denote the conjugate symplectic subspace of $E_{a}$
by $E_{a'}$ and $E_{b}$ by $E_{b'}$, then we have:
\begin{lem}
Consider generic symplectic transformations $S$ and $S'$ as introduced
right above. Then there exists a symplectic transformation $L$ such
that
\[
S''_{E_{a},E_{b}}=S_{1},
\]
\[
S''_{E_{a'},E_{b'}}=S_{r},
\]
and
\[
S''_{E_{a},E_{b'}}=S''_{E_{a'},E_{b}}=0
\]
with
\[
S''=S'LS,
\]
where $S_{1}:E_{b}\rightarrow E_{a}$ and $S_{r}:\rightarrow E_{b'}\rightarrow E_{a'}$
are symplectic transformations.
\end{lem}

\begin{proof}
According to Lemma.~\ref{lem:replicate-quadratures}, there exists
a local symplectic transformation
\[
L=L_{1}\oplus L_{2}\oplus\cdots\oplus L_{N},
\]
 such that
\[
Ly_{2b}=c^{-2}\Omega z_{2a}-(z_{2a-1}^{t}z_{2a}-y_{2b-1}^{t}y_{2b})\Omega z_{2a-1}.
\]
This can be satisfied if we let $L_{a}=\Omega_{a}$ which leads to
\[
Ly_{2b-1}=-c^{2}\Omega z_{2a-1}.
\]
The above amounts to the following
\[
z_{2a-1}=c^{-2}\Omega Ly_{2b-1},
\]
and
\[
z_{2a}=(z_{2a-1}^{t}z_{2a}-y_{2b-1}^{t}y_{2b})\Omega Ly_{2b-1}-c^{2}\Omega Ly_{2b}.
\]
It follows that
\begin{align*}
x_{j}^{t}S''x_{k} & =x_{j}^{t}S'LSx_{k}\\
 & =z_{j}^{t}Ly_{k}\\
 & =0,
\end{align*}
for $j\neq2a-1,2a$ and $k=2b-1,2b$ or $j=2a-1,2a$ $k\neq2b-1,2b$,
which finishes the proof. Besides, we have the following explicit
form:
\begin{align*}
S_{1} & =\begin{pmatrix}S_{2a-1,2a-1} & S_{2a-1,2b}\\
S_{2a,2b-1} & S_{2a,2b}
\end{pmatrix}\\
 & =\begin{pmatrix}x_{2a-1}^{t}\\
x_{2a}^{t}
\end{pmatrix}S\begin{pmatrix}x_{2b-1} & x_{2b}\end{pmatrix}\\
 & =\begin{pmatrix}z_{2a-1}^{t}Ly_{2b-1} & z_{2a-1}^{t}Ly_{2b}\\
z_{2a}^{t}Ly_{2b-1} & z_{2a}^{t}Ly_{2b}
\end{pmatrix}\\
 & =\begin{pmatrix}0 & c^{-2}\\
-c^{2} & z_{2a-1}^{t}z_{2a}-y_{2b-1}^{t}y_{2b}
\end{pmatrix}.
\end{align*}
\end{proof}
The lemmas we just proved directly lead to the following:
\begin{thm}
\label{thm:Interference-based-decoupling}Let $\left\{ v_{i}\right\} $
be a symplectic basis on $E$. For generic symplectic transformations
$S_{a},S_{b},S_{c},S_{d}:E\rightarrow E$, there exist three local
symplectic transformations $L^{\left(1\right)}$, $L^{(2)}$and $L^{(3)}$,
such that
\[
S_{a}L^{\left(3\right)}S_{b}L^{\left(2\right)}S_{c}L^{\left(1\right)}S_{d}=S_{1}\oplus S_{r},
\]
with symplectic transformations $S_{1}:E_{1}\rightarrow E_{1}$, $S_{r}:E_{2}\oplus E_{3}\oplus\cdots\oplus E_{N}\rightarrow E_{2}\oplus E_{3}\oplus\cdots\oplus E_{N}$.
\end{thm}

And:
\begin{thm}
\label{thm: interference-based-transduction}Let $\left\{ v_{i}\right\} $
be a symplectic basis on $E$. For generic symplectic transformations
$S_{a},S_{b},S_{c},S_{d}:E\rightarrow E$, there exist local symplectic
trasnformations $L^{(1)}$, $L^{(2)}$, and $L^{(3)}$ such that
\[
S_{a}L^{\left(3\right)}S_{b}L^{\left(2\right)}S_{c}L^{\left(1\right)}S_{d}=\begin{pmatrix}0 & S_{r}\\
S_{1} & 0
\end{pmatrix},
\]
where $S_{1}:E_{1}\rightarrow E_{N}$, $S_{r}:E_{1'}\rightarrow E_{N'}$
are symplectic transformations.
\end{thm}

Furthermore, by combing the above two theorems we see that:
\begin{thm}
\label{thm: interference-based-swapping}Let $\left\{ v_{i}\right\} $
be a symplectic basis on $E$. For generic symplectic transformations
$S^{(1)},S^{(2)},\dots,S^{(16)}:E\rightarrow E$, there local symplectic
transformations exist $L^{(1)},L^{(2)},\dots,L^{(15)}$, such that
\[
S^{(16)}L^{(15)}S^{(15)}\cdots S^{(3)}L^{(2)}S^{(2)}L^{(1)}S^{(1)}=\begin{pmatrix}0 & 0 & S_{1}\\
0 & S_{r} & 0\\
S_{2} & 0 & 0
\end{pmatrix},
\]
where $S_{1}:E_{N}\rightarrow E_{1}$, $S_{2}:E_{1}\rightarrow E_{N}$,
$S_{r}:E_{1'}\rightarrow E_{N'}$ are symplectic transformations.
\end{thm}

\begin{proof}
It suffices to consider the following product of symplectic transformations
\begin{align*}
 & \begin{pmatrix}S_{g} & 0\\
0 & S_{h}
\end{pmatrix}\begin{pmatrix}L_{c} & 0\\
0 & \text{Id}
\end{pmatrix}\begin{pmatrix}S_{e} & 0\\
0 & S_{f}
\end{pmatrix}\begin{pmatrix}L_{b} & 0\\
0 & \text{Id}
\end{pmatrix}\begin{pmatrix}S_{c} & 0\\
0 & S_{d}
\end{pmatrix}\begin{pmatrix}L_{a} & 0\\
0 & \text{Id}
\end{pmatrix}\begin{pmatrix}0 & S_{a}\\
S_{b} & 0
\end{pmatrix}\\
= & \begin{pmatrix}0 & S_{g}L_{c}S_{e}L_{b}S_{c}L_{a}S_{a}\\
S_{h}S_{f}S_{d}S_{b} & 0
\end{pmatrix}
\end{align*}
with
\[
S_{a}:E_{1'}\rightarrow E_{N'},
\]
\[
S_{b}:E_{N}\rightarrow E_{1},
\]
\[
S_{c},S_{e},S_{g}:E_{N'}\rightarrow E_{N'},
\]
\[
S_{h},S_{f},S_{d}:E_{N}\rightarrow E_{N},
\]
being symplectic transformations and
\[
L_{a},L_{b},L_{c}:E_{N'}\rightarrow E_{N'}
\]
being local symplectic transformations. Therefore, $S_{h}S_{f}S_{d}S_{b}$
should be equivalent to $S_{2}$. According to the previous theorem,
we then choose the proper forms of $L_{a}$, $L_{b}$, and $L_{c}$,
such that the product of symplectic transformations $S_{g}L_{c}S_{e}L_{b}S_{c}L_{a}S_{a}$
amounts to the following transformation
\[
\begin{pmatrix}0 & S_{1}\\
S_{r} & 0
\end{pmatrix},
\]
which finishes the proof.
\end{proof}
This result implies that we can swap any two bosonic modes (up to
local symplectic transformations) using generic symplectic transformations.
Therefore, it enables us to arbitrarily permute bosonic modes using
a sequence of generic symplectic transformations and local symplectic
transformations, since any permutation is the product of swappings.

\section{Edge cases: the non-generic symplectic transformations}

The proofs in the last section are based on the assumption of generic
symplectic transformations, which is only vaguely defined by now.
Although a direct and explicit definition of generic symplectic transformations
may be elusive, it is possible to specify what are non-generic symplectic
transformations, which are referred to as the edge cases. A quick
example of a non-generic symplectic transformations is a symplectic
transformation permuting multiple bosonic modes (up to local symplectic
transformations), e.g.
\[
S=\begin{pmatrix}0 & 0 & 0 & 0 & 1 & 0\\
0 & 0 & 0 & 0 & 0 & 1\\
1 & 0 & 0 & 0 & 0 & 0\\
0 & 1 & 0 & 0 & 0 & 0\\
0 & 0 & 1 & 0 & 0 & 0\\
0 & 0 & 0 & 1 & 0 & 0
\end{pmatrix}.
\]
Apparently, the product of such bosonic permutations cannot be of
arbitrary form, which conflicts with our promise at the end of the
last section.

Before proceeding, we first define a map $f$, mapping an arbitrary
symplectic transformation $S$ to a matrix with each of its elements
equals 1. Specifically, denoting
\[
\mathbb{S}_{j,k}=\begin{pmatrix}S_{2j-1,2k-1} & S_{2j-1,2k}\\
S_{2j,2k-1} & S_{2j,2k}
\end{pmatrix},
\]
we have
\[
f(S)_{j,k}=\begin{cases}
1, & \text{if }\mathbb{S}_{j,k}\neq0,\\
0, & \text{if }\mathbb{S}_{j,k}=0,
\end{cases}
\]
for any $j,k\in\{1,2,\dots,N\}$. Then we note that an arbitrary symplectic
transformation $S$ will satisfy one of the following three conditions:
\begin{enumerate}
\item $f(S)=P+Q$, with $P$ being a cyclic permutation matrix and $Q$
a full-ranked matrix, each element of which is equal to $1$;
\item There are decomposition of the symplectic space
\begin{align*}
E & \cong E_{S}^{(1)}\oplus E_{S}^{(2)}\oplus\cdots\oplus E_{S}^{(M)}\\
 & \cong E_{S}^{'(1)}\oplus E_{S}^{'(2)}\oplus\cdots\oplus E_{S}^{'(M)}
\end{align*}
with each $E^{(j)}$ being a direct sum of canonical symplectic subspaces
for all $j\in\{1,2,\dots,M\}$, and a permutation $\mathcal{P}$ of
the set $\{1,2,\dots,M\}$, such that $S_{E'^{(j)},E^{(k)}}$ is not
a zero matrix if and only if $j=\mathcal{P}(k)$ for any $j,k\in\{1,2,\dots,M\}$.
And each non-zero $S_{E'^{(j)},E^{(k)}}$ satisfies the first condition.
\end{enumerate}
This implies that we can associate any given symplectic transformation
$S$ with a pair of collectioncs of symplectic spaces decomposition
$(\left\{ E_{S}^{\left(j\right)}\right\} ,\left\{ E_{S}^{'\left(k\right)}\right\} )$,
and a permutation of the set $\{1,2,\dots,M\}$ denoted by $\mathcal{P}_{S}$.
In particular for a symplectic transformation satisfying the first
condition, we have $M=1$ and $\mathcal{P}_{S}$ is the trivial identity
operation.

Since we can perform local symplectic transformations before and after
any symplectic transformations, from now on, we will assume $S$ is
fully randomized, i.e. each $\mathbb{S}_{j,k}$ will be considered
to be a randomly-generated $2\times2$ matrix. Then for any such symplectic
transformation $S$, we have
\[
f(S^{c})=f(f(S)^{c}),
\]
where $c$ is an integer.

It turns out that there will always be a large enough integer $c$,
such that the $c$-th power $S^{c}$ of an symplectic transformation
$S$ satisfies the following stabilizing condition: Let $(\left\{ E_{S^{c}}^{\left(j\right)}\right\} ,\left\{ E_{'S^{c}}^{\left(k\right)}\right\} )$
be the symplectic space decompositions and $\mathcal{P}_{S^{c}}$
the permutation associated to $S^{c}$. Then we have $E_{S^{c}}^{\left(j\right)}=E_{'S^{c}}^{\left(j\right)}$
for all $j\in\{1,2,\dots,M\}$ and $\mathcal{P}_{S^{c}}$ is the trivial
identify operation. Moreover,  there is an integer $d\ge c$, such
that the associated permutations of $S^{c},S^{c+1}$,...,$S^{d}$
forms a permutation group\footnote{This implies that any $S^{e}$ with $e>d$ will also be associated
to an element in this group.}. Also note that $S^{c},S^{c+1},\dots,S^{d}$ will have the same associated
symplectic space decomposition. When $N=2$, it is easy to check the
stabilizing condition can be satisfied with $c\le2$. When $N>2$,
a rough upper bound of such an integer $c$ is $N\left(N+1\right)$,
which is a result of the observation that the total number of non-zero
$\mathbb{S}_{j,k}$ decreases as $c$ increases. 
\begin{example}
Consider the symplectic transformation
\[
S=\begin{pmatrix}0 & 0 & 1 & 0 & 0 & 0 & 1 & 0\\
0 & 0 & 0 & 1 & 0 & 0 & 0 & 1\\
1 & 0 & 0 & 0 & 1 & 0 & 0 & 0\\
0 & 1 & 0 & 0 & 0 & 1 & 0 & 0\\
0 & 0 & 1 & 0 & 0 & 0 & 1 & 0\\
0 & 0 & 0 & 1 & 0 & 0 & 0 & 1\\
1 & 0 & 0 & 0 & 1 & 0 & 0 & 0\\
0 & 1 & 0 & 0 & 0 & 1 & 0 & 0
\end{pmatrix}.
\]
Then
\[
f(S)=Q+P=\begin{pmatrix}0 & 1 & 0 & 1\\
1 & 0 & 1 & 0\\
0 & 1 & 0 & 1\\
1 & 0 & 1 & 0
\end{pmatrix}
\]
with
\[
Q=\begin{pmatrix}0 & 0 & 0 & 1\\
1 & 0 & 0 & 0\\
0 & 0 & 0 & 0\\
0 & 0 & 1 & 0
\end{pmatrix}\text{ and}\ P=\begin{pmatrix}0 & 1 & 0 & 0\\
0 & 0 & 1 & 0\\
0 & 0 & 0 & 1\\
1 & 0 & 0 & 0
\end{pmatrix},
\]
then we have
\[
f\left(S^{2}\right)=\left(\begin{array}{cccc}
1 & 0 & 1 & 0\\
0 & 1 & 0 & 1\\
1 & 0 & 1 & 0\\
0 & 1 & 0 & 1
\end{array}\right)=\begin{pmatrix}1 & 1\\
1 & 1
\end{pmatrix}\oplus\begin{pmatrix}1 & 1\\
1 & 1
\end{pmatrix},
\]
which satisfies the stabilizing condition above.
\end{example}

\begin{example}
Let
\[
f(S)=Q+P=\begin{pmatrix}0 & 1 & 1 & 1\\
1 & 0 & 1 & 0\\
0 & 1 & 0 & 1\\
1 & 0 & 1 & 0
\end{pmatrix}
\]
with
\[
Q=\begin{pmatrix}0 & 0 & 1 & 1\\
1 & 0 & 0 & 0\\
0 & 1 & 0 & 0\\
0 & 0 & 1 & 0
\end{pmatrix},\text{ and}\ P=\begin{pmatrix}0 & 1 & 0 & 0\\
0 & 0 & 1 & 0\\
0 & 0 & 0 & 1\\
1 & 0 & 0 & 0
\end{pmatrix},
\]
then we have
\[
f\left(S^{2}\right)=\left(\begin{array}{cccc}
1 & 1 & 1 & 1\\
0 & 1 & 1 & 1\\
1 & 0 & 1 & 0\\
0 & 1 & 1 & 1
\end{array}\right),
\]
\[
f\left(S^{3}\right)=\left(\begin{array}{cccc}
1 & 1 & 1 & 1\\
1 & 1 & 1 & 1\\
0 & 1 & 1 & 1\\
1 & 1 & 1 & 1
\end{array}\right),
\]
and
\[
f\left(S^{4}\right)=\left(\begin{array}{cccc}
1 & 1 & 1 & 1\\
1 & 1 & 1 & 1\\
1 & 1 & 1 & 1\\
1 & 1 & 1 & 1
\end{array}\right),
\]
which satisfies the stabilizing condition above. 
\end{example}

\begin{example}
The last example: Let
\[
f(S)=\begin{pmatrix}0 & 0 & 0 & 1 & 1\\
0 & 0 & 1 & 0 & 1\\
0 & 0 & 1 & 1 & 0\\
1 & 0 & 0 & 0 & 0\\
0 & 1 & 0 & 0 & 0
\end{pmatrix}.
\]
Then we have
\[
f(S^{7})=\begin{pmatrix}1 & 1 & 1 & 1 & 1\\
1 & 1 & 1 & 1 & 1\\
1 & 1 & 1 & 1 & 1\\
1 & 1 & 1 & 1 & 1\\
1 & 1 & 1 & 1 & 1
\end{pmatrix}
\]
which satisfying the stabilizing condition.
\end{example}

Therefore, supposing that we only have access to a given symplectic
transformation and arbitrary local symplectic transformations, the
edge cases can now be easily identified, as well as the resulting
bosonic mode permutation achieved via our results in the previous
section. In particular, when the first stabilizing condition is satisfied,
we are able to permute arbitrary pair of bosonic modes and thus construct
arbitrary bosonic mode permutations. Specifcally, let $S^{c}$ satisfy
the stabilizing condition above, associated to a symplectic space
decomposition
\[
E\cong E_{S^{c}}^{(1)}\oplus E_{S^{c}}^{(2)}\oplus\cdots\oplus E{}_{S^{c}}^{(M)},
\]
and let integer $d\ge c$, such that the associated permutations of
$S^{c},S^{c+1}$,...,$S^{d}$ form a permutation group. Then we have
the following conclusions:
\begin{enumerate}
\item With a proper randomized prior-processing based on the discussions
in this section, any mode in the bosonic system can be decoupled using
the algorithm introduced in the last section. 
\item Moreover, the algorithms in the last section can be applied to swap
the $j$th and $k$th modes in a $N$-mode system using copies of
a given symplectic transformation $S$, if and only if either both
canonical symplectic spaces $E_{j}$ and $E_{k}$are contained in
the same direct summand $E_{S^{c}}^{\left(l\right)}$ for some $l\in\{1,2,\dots,M\}$
or there exist integers $l,m\in\{1,2\dots,N\}$ and $c\le n\le d$,
such that $E_{j}\subseteq E_{S^{c}}^{\left(l\right)}$, $E_{k}\subseteq E_{S^{c}}^{\left(m\right)}$,
and $\mathcal{P}_{S^{n}}$ swaps the $j$ and $k$.
\end{enumerate}

\section{Conclusion and outlook}

In this chapter, we discussed how to construct universal swap and
decoupling, and hence permutations of bosonic modes using interference
of symplectic transformations. We demonstrated our scheme, derived
from the fundamental CCRs of quantum mechanics, are widely applicable
to bosonic systems and unitary Gaussian processes involving arbitrary
many bosonic modes. When applied to quantum transduction, we obtain
a scheme that convert imperfect quantum transducers to perfect quantum
transducers only using reasonable single-mode Gaussian unitary operations. 

This chapter concludes our attempts of applying symplectic transformation
to solving bosonic quantum control, especially quantum transduction.
Next, we discuss the role of symplectic transformations in another
intriguing field--bosonic sensing. We will see, with the knowledge
of symplectic geometry, the quantum noise model of a bosonic sensing
scheme can be easily established and utilized to provide an ultimate
bound of the sensing precision of the corresponding scheme.

\chapter{Applications to bosonic sensing\label{chap:Applications-to-bosonic}}

\section{Introduction}

We discussed the application of symplectic geometry in quantum information
processing in the previous chapters. In this chapter\footnote{This chapter is mainly based on our publication \cite{Zhang2019}.},
our focus is shifted to another intriguing field--exotic bosonic
classical sensing\footnote{For readers who are familiar with quantum sensing, here, we would
like to stress that although tools of quantum metrology will be heavily
used in this chapter, the schemes we will investigate remain classical
insofar as the entanglement of probes are not used as a boost of the
sensing precision.} with drastic precision enhancement. We will utilize the theory of
symplectic geometry to provide a thorough and systematic analysis
for the exotic bosonic sensing schemes.

What motivated us to invest our time in this field is the recent rapid
development of exceptional point sensing. Exceptional point sensing\index{Exceptional point sensing}
has recently attracted a lot of attention because of its potential
to drastically improve signal sensing precision, which is of broad
interest in physics with many important applications. Based on this,
people proposed~\cite{Wiersig2014,Wiersig2016,Liu2016} and experimentally
demonstrated~\cite{Chen2017,Hodaei2017} the novel sensing scheme
enhanced by exceptional points\index{Exceptional point}, where the
whole frequency spectrum is measured and the frequency splitting is
used to estimate the signal which is treated as a small perturbation
to the exceptional point system. However, these theoretical proposals
and experimental demonstrations lack the consideration of measurement
uncertainty caused by quantum noise, which may be possibly amplified
in presence of the exceptional points and hence deteriorate the sensing
enhancement predicted by the classical analysis. Therefore, it is
questionable that this precision enhancement is genuine in a serious
quantum model of the whole process.

To answer this question, we must first find a suitable quantum noise
model for exceptional point sensing. Fortunately, as classical bosonic
sensing, the physical process of exceptional point sensing can be
described a Gaussian process. Specifically, the physical process of
an exceptional point sensing sensing can be described as follows:
People first add the quantity to estimate into a linearly-coupled
bosonic system at exceptional point; and then inject classical probes
into this coupled system; at last, the scattered probes will be measured
and the outcomes will be used to provide an estimate of the value
of the quantity. For example, the change of temperature may cause
the change of the resonant frequency of optical modes. Therefore,
the small temperature change can be viewed as a perturbation to the
Hamiltonian of a linearly-coupled optical system. Now we put the unperturbed
the system at the exceptional point. And send probing light to interact
with these optical modes. Then, the amplitude and phase of the scattered
probing light will depend on the small temperature change. Therefore,
to estimate this temperature change, we only need to measure the scattered
probing light and extract the wanted information from the outcomes.
This linear scattering process picture of exceptional point sensing
reveals its Gaussian process nature. Moreover, alluded to Sec.~\ref{sec:Symplectic-dialtion},
a noisy Gaussian process can be embedded in a unitary Gaussian process,
which can thus be modeled by a symplectic transformation. Physically
speaking, this is equivalent to a conclusion in quantum noise theory
\cite{gardiner_quantum_2004}: bosonic losses and gains, as long as
they are generated by linear coupling between the system and the reservoir,
can be modeled as a linear interaction between fictitious probing
modes and the system modes. This observation lays the ground of our
theoretical framework in this chapter.

Now we know exceptional sensing is closely related to Gaussian processes.
It remains unknown how to theoretically justify the sensing precision
enhancement. Since now we model exceptional sensing as a quantum process,
the justification must be done using a theory compatible with quantum
mechanics. This theory is quantum metrology, originally developed
for justifying sensing precision of all quantum sensing schemes. In
particular, we will utilize the notion of quantum Fisher information,
a quantity that is used to quantify the ultimate bound of the sensing
power of a sensing scheme. The calculation of quantum Fisher information
only depends on the physical process, and is by its definition optimized
over all possible physical measurement schemes. Although the calculation
of quantum Fisher information is complicated and non-analytical in
general, for Gaussian processes and Gaussian input states, we have
convenient closed-form formulas which yield analytical results. We
will see in the succeeding section more details about quantum Fisher
information and its calculation and demonstrate in later sections
how this tool can justify the sensing enhancement induced by exceptional
points.

After the discussions of exceptional point sensing, we extend our
discussions to the general ideal of exotic bosonic sensing. We are
interested in the following question: Is exceptional point sensing
the only classical bosonic sensing scheme that can provide unexpected
drastic precision enhancement, beyond any conventional classical sensing
schemes? This question is partly answered in \cite{Lau2018}, where
a non-exceptional-point classical bosonic sensing scheme is demonstrated
to have a genuine precision enhancement even in presence of quantum
noise. In this chapter, by using our knowledge of symplectic dilation
of noisy Gaussian processes, we would like to provide a unified theoretical
framework for any such exotic bosonic sensing schemes based on Gaussian
process. Our investigation delivers a short message: by dilating the
noisy Gaussian process to a symplectic transformation, one can see
the exotic precision enhancement, if it exists, can be understood
as a phenomenon of squeezing. We hope this novel perspective can help
deepen our understanding of classical bosonic sensing schemes and
inspire more promising applications in the future.

\section{Gaussian Fisher information}

The process of sensing a classical physical quantity $\theta$ can
be described as follows: We let physical objects, the probes, interact
with a system described by a Hamiltonian containing this quantity.
Then we measure the scattered probes, described by the quantum state
$\hat{\rho}_{\theta}$, to obtain a measurement outcome. Usually,
we will perform multiple rounds, say $N$ rounds, of measurement,
such that there will be than one outcomes, denoted by $\eta_{j}$,
for $j\in\{1,2,\dots,N\}$. Then based on our knowledge of the physical
process, we will use a function $f(N,\eta_{1},\eta_{2},\dots,\eta_{N})$
to obtain an estimate $\hat{\theta}_{N}$ of the true value $\theta$.
This function will be wisely chosen, so that as $N$ approaches infinity,
the difference between the estimate and the true value tends to vanishing.

When the dependence of the measured quantum state on $\theta$ has
as good properties as what we focus on in the rest of this chapter,
we know how to construct the optimal estimate in general. Note that
the state $\hat{\rho}_{\theta}$ and the measurement scheme together
yields a probability distribution of the measurement outcome, $p_{\theta}(\eta)$.
Then the optimal estimate can be given by
\begin{equation}
\hat{\theta}_{N}=\text{argmax}_{\phi}\prod_{i=1}^{N}p_{\phi}\left(\eta_{i}\right),
\end{equation}
which is known as the maximum likelihood estimation method \cite{Amari2007}. 

This estimate is optimal in the following sense: With a large $N$,
the value of $\hat{\theta}_{N}$ can be considered to be sampled from
a normal distribution centered at the true value $\theta$. Then letting
\begin{equation}
\delta\theta_{N}=\sqrt{\mathbb{E}(\hat{\theta}_{N}-\theta)}
\end{equation}
be the deviation of the estimate from the true value, we have
\begin{equation}
\delta\theta_{N}=\frac{1}{N}I(\theta)^{-1}+o(\frac{1}{N^{2}}),
\end{equation}
which is the smallest deviation one can obtain no matter how we estimate
the value of $\theta$. The function $I(\theta)$ is known as the
Fisher information\index{Fisher information} and is given by the
following equation \cite{Amari2007}
\begin{equation}
I\left(\theta\right)=\int\left(\frac{\partial p_{\theta}\left(\eta\right)}{\partial\theta}\right)^{2}p_{\theta}\left(\eta\right)d\eta
\end{equation}
which is completely determined by the quantum state $\hat{\rho}_{\theta}$
and the measurement scheme. Furthermore, if we are allowed tor choose
the optimal measurement scheme, we can construct such a function of
$\theta$:
\begin{equation}
\mathcal{I}(\theta)=\text{sup}\{I(\theta)\,|\,\text{all measurement schemes}\}.
\end{equation}
This value is known as the quantum Fisher information\index{Quantum Fisher information}
\cite{Helstrom1969,Holevo1982,Braunstein1994,Braunstein1996}.

Therefore, the power of a physical process for sensing a quantity
$\theta$ is characterized by the Cram\'{e}r-Rao bound\index{Cramr-Rao bound@Cram\'{e}r-Rao bound}
\cite{Amari2007}
\begin{equation}
\delta\theta_{N}\ge\frac{1}{\sqrt{NI\left(\theta\right)}},
\end{equation}
if the measurement scheme is fixed, and the quantum Cram\'{e}r-Rao
bound
\begin{equation}
\delta\theta_{N}\ge\frac{1}{\sqrt{N\mathcal{I}\left(\theta\right)}}
\end{equation}
if we have the freedom to perform the optimal measurement scheme.

Although the calculation of both the Fisher information and the quantum
Fisher information can be difficult in general, it turns out that
given a Gaussian state $\hat{\rho}_{\theta}$, there exist explicit
formulae that allow us to compute these quantities. First of all,
what can be easily proved is that the probability distribution yielded
by an arbitrary Guassian state $\hat{\rho}(\bar{x}_{\theta},V_{\theta})$
and an arbitrary Gaussian measurement scheme is a Gaussian function
with first moments $\mu_{\theta}$ and covariance matrix $\Sigma_{\theta}$.
In particular, when the ideal heterodyne measurement is performed
to measure each mode in the probes, we have
\[
\mu_{\theta}=\bar{x}_{\theta},
\]
and
\[
\Sigma_{\theta}=V_{\theta}+\text{Id}.
\]
Then the Fisher information is given by
\begin{equation}
I(\theta)=I_{\Sigma}(\theta)+I_{\mu}(\theta)
\end{equation}
with
\begin{equation}
I_{\Sigma}\left(\theta\right)=\frac{1}{2}\text{Tr}\left[\Sigma_{\theta}^{-1}\frac{d\Sigma_{\theta}}{d\theta}\Sigma_{\theta}^{-1}\frac{d\Sigma_{\theta}}{d\theta}\right]
\end{equation}
and
\begin{equation}
I_{\mu}\left(\theta\right)=\left(\frac{d\mu_{\theta}}{d\theta}\right)^{t}\Sigma_{\theta}^{-1}\frac{d\mu_{\theta}}{d\theta}.
\end{equation}
In addition, the quantum Fisher information of a Gaussian state is
determined by a similar formula \cite{Monras2013,Jiang2014}
\begin{equation}
\mathcal{I}\left(\theta\right)=\mathcal{I}_{V}\left(\theta\right)+\mathcal{I}_{\bar{x}}\left(\theta\right)
\end{equation}
where
\begin{equation}
\mathcal{I}_{V}\left(\theta\right)=\frac{1}{2}\text{Tr}\left[\Phi_{\theta}\frac{dV_{\theta}}{d\theta}\right]
\end{equation}
and
\begin{equation}
\mathcal{I}_{\bar{x}}\left(\theta\right)=\left(\frac{d\bar{x}_{\theta}}{d\theta}\right)^{t}V_{\theta}^{-1}\left(\frac{d\bar{x}_{\theta}}{d\theta}\right).
\end{equation}
The matrix $\Phi_{\theta}$ above is implicitly determined by the
following equation:
\begin{equation}
\frac{dV_{\theta}}{d\theta}=V_{\theta}\Phi_{\theta}V_{\theta}-\Omega\Phi_{\theta}\Omega^{t},
\end{equation}
with $\Omega$ the symplectic form. In general, there is no explicit
analytical formula for the matrix $\Phi_{\theta}$. However, in situations
where $\text{det}\left(V_{\theta}\right)\gg1$, $\Phi_{\theta}$ can
be approximated by \cite{Jiang2014}
\begin{equation}
\Phi_{\theta}\approx V_{\theta}^{-1}\left(\frac{dV_{\theta}}{d\theta}\right)V_{\theta}^{-1}.
\end{equation}
This leads to
\begin{equation}
\mathcal{I}_{V}\left(\theta\right)\approx\frac{1}{2}\text{Tr}\left[V_{\theta}^{-1}\left(\frac{dV_{\theta}}{d\theta}\right)V_{\theta}^{-1}\frac{dV_{\theta}}{d\theta}\right].
\end{equation}
As a matter of fact, under such assumptions, the value of $\mathcal{I}(\theta)$
can also be directly approximated by the value of $I(\theta)$ with
all the probe modes measured by the ideal heterodyne measurement scheme.

\section{Quantum theory of exceptional point sensing\label{sec:Exceptional-point-systems}}

In this section, we establish the quantum noise model for exceptional
sensing and apply the results of the preceding section to bounding
its ultimate sensing precision. The existence of exceptional points,
as a feature of the non-Hermitian Hamiltonians, requires losses and
gains from the surround environment. To establish the suitable quantum
model for exceptional point sensing, we must theoretically simulate
the losses and gains using quantum noise theory. The idea is that
we model each loss or gain channel using a linear interaction between
a fictitious probing channel and the corresponding system mode. Then
we assume vacuum state inputs for these fictitious probing channel
as the simulation of the quantum noise associated to the corresponding
losses or gains. Then the whole process can be viewed as a symplectic
transformation involving the original system modes and these fictitious
modes. By doing so, the output state, which is a Gaussian state given
that all the inputs are Gaussian states, can be easily calculated.
The output states, which carries the information of the parameter
to estimate, will be used to calculate the quantum Fisher information.
Let us demonstrate this idea in the following example.

\begin{figure}[h]
\includegraphics[width=0.9\textwidth]{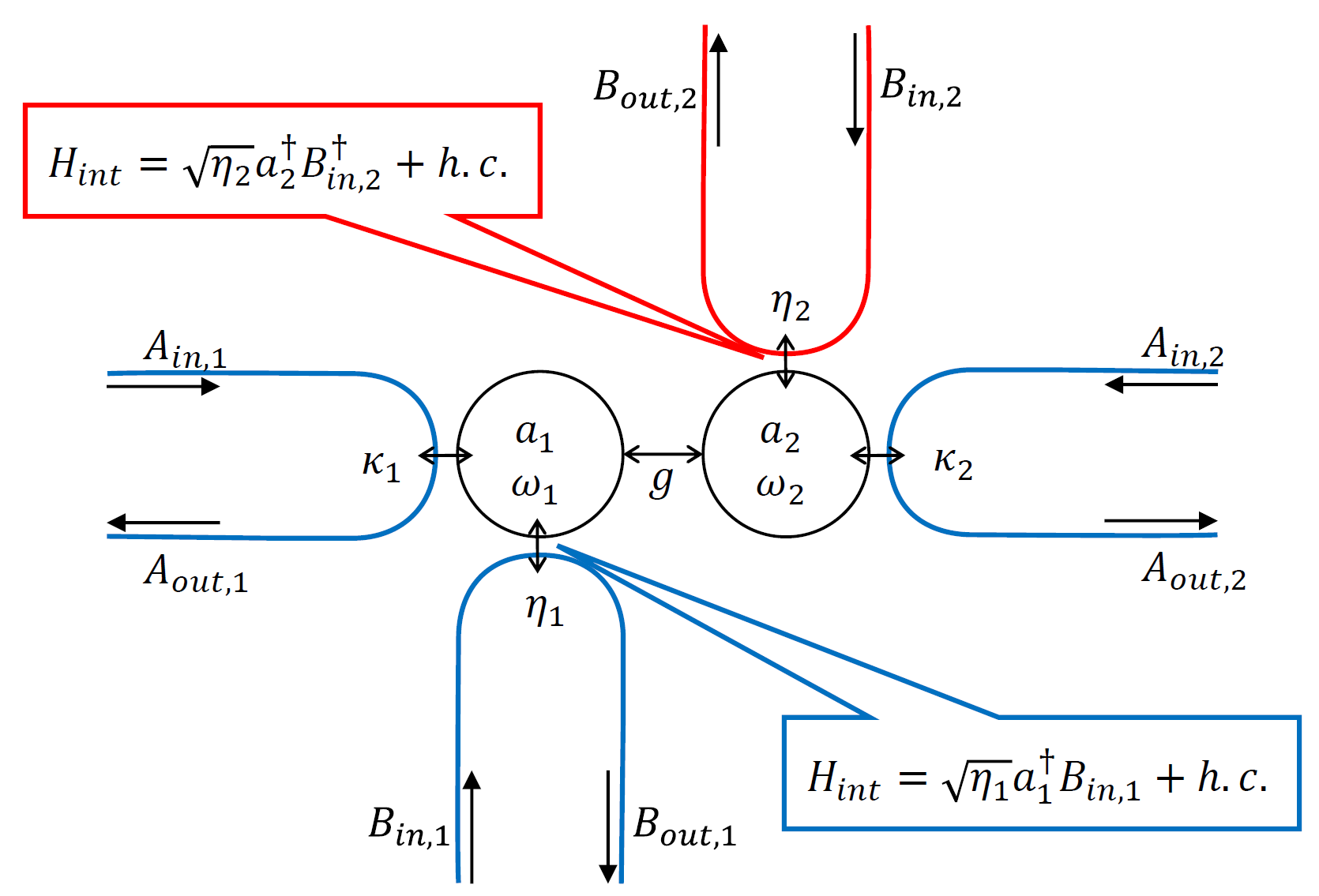}
\begin{doublespace}
\centering{}\caption{\label{fig:Modeling-a-two-bosonic-mode}\emph{Modeling a two-bosonic-mode
system.}}
\begin{minipage}[t]{0.9\textwidth}%
This figure shows the schematic of a two-bosonic-mode system with
a full quantum description of the losses and gains. The labeled circles
represent the two system modes $\hat{a}_{1}$ and $\hat{a}_{2}$ with
frequency $\omega_{1}$ and $\omega_{2}$. They are coupled to the
two probing channels, $\hat{A}_{1}$ and $\hat{A}_{2}$ with coupling
rates $\kappa_{1}$ and $\kappa_{2}$ respectively. In addition to
the probing channels, the system mode $\hat{a}_{1}$ is intrinsically
dissipated by the fictitious channel \textbf{$\hat{B}_{1},$} and
$\hat{a}_{2}$ intrinsically amplified by the fictitious channel $\hat{B}_{2}$.
Here the dissipation is modeled by the fictitious interaction $\hat{H}_{int}=\sqrt{\eta_{1}}\hat{a}_{1}^{\dagger}\hat{B}_{in,1}+h.c.$
and the amplification is modeled by the fictitious interaction $\hat{H}_{int}=\sqrt{\eta_{2}}\hat{a}_{2}^{\dagger}\hat{B}_{in,2}^{\dagger}+h.c.$,
with $\eta_{1}$, $\eta_{2}$ being the corresponding intrinsic loss
and gain. The arrows in the channels which point towards the system
modes represent the input modes, while those with opposite direction
represent the output modes, as defined in the quantum noise theory
\cite{gardiner_quantum_2004}.%
\end{minipage}
\end{doublespace}
\end{figure}

\begin{example}
\label{exa:Consider-two-bosonic}Consider two bosonic modes $\hat{a}_{1}$
and $\hat{a}_{2}$ linear coupled with each other, through the Hamiltonian\footnote{The readers may remember our discussion of side bands in the previous
chapters. In this example, as we may assume the bosonic modes to be
optical (so the rotating wave approximation works well), the side
bands can be safely omitted.}
\[
\hat{H}=\omega_{1}\hat{a}_{1}^{\dagger}\hat{a}_{1}+\omega_{2}\hat{a}_{2}^{\dagger}\hat{a}_{2}+g\hat{a}_{1}^{\dagger}\hat{a}_{2}+g\hat{a}_{2}^{\dagger}\hat{a}_{1}.
\]
Each system mode will be coupled with a probing channel--a propagating
bosonic mode: $\hat{a}_{1}$ coupled with $\hat{A}_{1}$, and $\hat{a}_{2}$
coupled with $\hat{A}_{2}$. The coupling rates are given by $\kappa_{1}$
and $\kappa_{2}$, respectively. In addition, an additional loss $\eta_{1}$
is introduced to $\hat{a}_{1}$ while an additional gain $\eta_{2}$
is introduced to $\hat{a}_{2}$, as shown in Fig.~\ref{fig:Modeling-a-two-bosonic-mode}.
The loss is modeled by a linear coupling between the fictitious $\hat{B}_{1}$
channel with $\hat{a}_{1}$ through a passive interaction $\sqrt{\eta_{1}}\hat{a}_{1}^{\dagger}\hat{B}_{in,1}+h.c.$,
and the gain is modeled by a linear coupling between the fictitious
$\hat{B}_{2}$ channel through an active interaction $\sqrt{\eta_{2}}\hat{a}_{2}^{\dagger}\hat{B}_{in,2}^{\dagger}+h.c.$.
The non-Hermitian Hamiltonian associated to the system is the matrix
\[
\begin{pmatrix}\omega_{1}-i\frac{\eta_{1}+\kappa_{1}}{2} & g\\
g & \omega_{2}+i\frac{\eta_{2}-\kappa_{2}}{2}
\end{pmatrix},
\]
which is at an exceptional point (i.e. with coalesced eigen-vectors)
when we let
\[
\omega_{1}=\omega_{2},\,\kappa_{1}=\kappa_{2}=\kappa\,\text{and }g=\frac{\eta_{1}+\kappa_{1}}{2}=\frac{\eta_{2}-\kappa_{2}}{2}.
\]
Therefore, for simplicity, from now on, we let $\omega_{1}=\omega_{2}=0$\footnote{This is valid if we are working in the interaction picture.}
and $\kappa_{1}=\kappa_{2}=\kappa$. Now assume we want to sense a
dimensionless parameter $\theta$, which is added to the system as
a perturbation of the resonance frequencies, i.e. resulting in a perturbed
Hamiltonian
\[
\begin{pmatrix}\theta\kappa-ig & g\\
g & \theta\kappa+ig
\end{pmatrix}.
\]
Then the dynamics of this system is determined by the following quantum
Langevin equation\footnote{Note that the mode $\hat{B}_{2,in}$ appears as $\hat{B}_{2,in}^{\dagger}$
in the following equation, due to its active interaction with the
system mode.}
\[
\frac{d}{dt}\begin{pmatrix}\hat{a}_{1}\\
\hat{a}_{2}
\end{pmatrix}=-i\begin{pmatrix}\theta\kappa-ig & g\\
g & \theta\kappa+ig
\end{pmatrix}\begin{pmatrix}\hat{a}_{1}\\
\hat{a}_{2}
\end{pmatrix}+\sqrt{\kappa}\begin{pmatrix}\hat{A}_{1,in}\\
\hat{A}_{2,in}
\end{pmatrix}+\begin{pmatrix}\sqrt{\eta_{1}}\hat{B}_{1,in}\\
-\sqrt{\eta_{2}}\hat{B}_{2,in}^{\dagger}
\end{pmatrix}.
\]
Then the symplectic transformation associated with this scattering
process can be easily calculated using the formulas introduced in
Appx.~\ref{chap:Scattering-process-and}. Then we denote the input
probe, a Gaussian state on the joined symplectic space of $\hat{A}_{1}$
and $\hat{A}_{2}$, by $\hat{\rho}\left(\bar{x}_{in},V_{in}\right)$,
which is independent of the parameter $\theta$; and the input noise,
a Gaussian state on the joined symplectic space of $\hat{B}_{1}$
and $\hat{B}_{2}$, by $\hat{\rho}\left(0,V_{env}\right)$. Then the
output of the probing channels $\hat{A}_{1}$ and $\hat{A}_{2}$,
after tracing out hte noise channels $\hat{B}_{1}$ and $\hat{B}_{2}$,
is a Gaussian state $\hat{\rho}\left(\bar{x}_{out}\left(\theta\right),V_{out}\left(\theta\right)\right)$
with
\[
\bar{x}_{out}\left(\theta\right)=\text{Id}-G_{\theta},
\]
and
\[
V_{out}\left(\theta\right)=\left(\text{Id}-G_{\theta}\right)V_{in}\left(\text{Id}-G_{\theta}\right)^{t}+G_{\theta}RV_{in}'R^{t}G_{\theta}.
\]
Here we have
\[
G_{\theta}=\frac{\kappa}{2g}\begin{pmatrix}-1 & 0 & -\theta & 1\\
0 & 1 & 1 & -\theta\\
\theta & -1 & -1 & 0\\
-1 & \theta & 0 & 1
\end{pmatrix}^{-1},
\]
and
\[
R=\begin{pmatrix}\sqrt{\frac{\eta_{1}}{\kappa}} & 0 & 0 & 0\\
0 & -\sqrt{\frac{\eta_{2}}{\kappa}} & 0 & 0\\
0 & 0 & \sqrt{\frac{\eta_{1}}{\kappa}} & 0\\
0 & 0 & 0 & \sqrt{\frac{\eta_{2}}{\kappa}}
\end{pmatrix},
\]
where the matrices are arranged according to the alternative order
of the symplectic basis. Applying formulas of the quantum Fisher information
to this output Gaussian state, we obtain that
\[
\mathcal{I}_{\bar{x}}\left(\theta\right)=\left(\frac{d}{d\theta}\bar{x}_{out}\left(\theta\right)\right)^{t}V_{out}\left(\theta\right)^{-1}\left(\frac{d}{d\theta}\bar{x}_{out}\left(\theta\right)\right)=\text{constant}\cdot\theta^{-4}
\]
 which implies that the smaller the parameter $\theta$, the larger
the quantum Fisher information and hence the better the sensing precision\footnote{This is a big edge over the traditional sensing schemes, as one can
check that $\mathcal{I}_{\bar{x}}\left(\theta\right)$ for a traditional
sensor does not have such a singular dependence of the parameter $\theta$.
Moreover, the $\theta^{-4}$ scaling of $\mathcal{I}_{\bar{x}}\left(\theta\right)$
guarantees that the relative error $\delta\theta/\theta$ can be infinitely
close to zero as the parameter $\theta$ approaching zero, which cannot
happen even with the a $\theta^{-2}$ scaling.}. Moreover, we can check that $\bar{x}_{out}\left(\theta\right)\sim\theta^{-2}$
and $V_{out}\left(\theta\right)\sim\theta^{-4}$, which means the
noise, indicated by the covariance matrix $V_{out}\left(\theta\right)$,
is actually of the same magnitude as the signal, indicated by the
first moments $\bar{x}_{out}\left(\theta\right)$. The reason that
$\mathcal{I}_{\bar{x}}\left(\theta\right)$ has such an unexpectedly
singular dependence on $\theta$ is the special mathematical structure
of the quantity: It has a differential-signal-to-noise ratio structure
but in a inner product form. In other words, unlike the usual signal-to-noise
ration which only involves scalars, both signal and noise here are
represented by non-scalars, and combined by matrix multiplications.
Matrix multiplication is the fountain of interference. This leads
to the unexpected underlying amplification of signals and cancellation
of noise, which together result in the drastic boost of sensing precision.
\end{example}

In the rest of this section, we further extend the protocol of the
above example to the general situation, in a more rigorous and formal
sense. The ideas are essentially the same. Only that in the general
situation where more complicated exceptional points can be constructed
in a multi-mode bosonic system, the precision boost can be further
increased (i.e. $\mathcal{I}_{\bar{x}}\left(\theta\right)\sim\theta^{-2k}$),
depending on the number (i.e. $k$) of the coalesced eigenvectors
of a non-Hermitian system at an exceptional point. The readers can
skip this part and move on to the next section since the following
technical details will not add much to the general picture.

Now we consider non-unitary Gaussian processes associated with exceptional
point systems. For any $M\in\mathbb{C}^{m\times m}$ and any integer
$m$, there always exists a $m$-dimensional invertible matrix $P$,
such that
\[
M=P\Lambda P^{-1},
\]
where
\[
\Lambda=\oplus_{k=1}^{r}J_{n_{k}}\left(\lambda_{k}\right),
\]
with $r\le m$, $\sum_{k=1}^{r}n_{k}=m$, $\lambda_{k}\in\mathbb{C}$,
and
\[
J_{n_{k}}\left(\lambda_{k}\right)=\begin{pmatrix}\lambda_{k} & 1\\
 & \lambda_{k} & \ddots\\
 &  & \ddots & 1\\
 &  &  & \lambda_{k}
\end{pmatrix}\in\mathbb{C}^{n_{k}\times n_{k}}.
\]
This is well-known as the Jordan decomposition\emph{, }and $\Lambda$
is called the Jordan normal form of $A$ and is uniquely defined insofar
as any two Jordan normal forms of $A$ have the same set of $J_{n_{k}}\left(\lambda_{k}\right)$,
known as the Jordan blocks. We say a Jordan block $J_{n}\left(\lambda\right)$
is trivial if $n=1$, and $\Lambda$ is trivial if all its Jordan
blocks are trivial.

Physical properties of the linearly-coupled bosonic oscillators are
determined by matrices. The calculation of the evolution of the associated
quadrature operators relies on a real dynamical matrix $M$ as in
the equation
\[
\frac{d}{dt}\begin{pmatrix}\langle\hat{q}_{1}\rangle\\
\langle\hat{p}_{1}\rangle\\
\vdots\\
\langle\hat{q}_{N}\rangle\\
\langle\hat{p}_{N}\rangle
\end{pmatrix}=M\begin{pmatrix}\langle\hat{q}_{1}\rangle\\
\langle\hat{p}_{1}\rangle\\
\vdots\\
\langle\hat{q}_{N}\rangle\\
\langle\hat{p}_{N}\rangle
\end{pmatrix}.
\]
Moreover, when probed by extrinsic bosonic modes, the scattering process
is determined by a real scattering matrix $\mathcal{S}$, as in the
following relation
\[
\begin{pmatrix}\langle\hat{Q}_{1}^{(out)}\rangle\\
\langle\hat{P}_{1}^{(out)}\rangle\\
\vdots\\
\langle\hat{Q}_{N}^{(out)}\rangle\\
\langle\hat{P}_{N}^{(out)}\rangle
\end{pmatrix}=\mathcal{S}\begin{pmatrix}\langle\hat{Q}_{1}^{(in)}\rangle\\
\langle\hat{P}_{1}^{(in)}\rangle\\
\vdots\\
\langle\hat{Q}_{N}^{(in)}\rangle\\
\langle\hat{P}_{N}^{(in)}\rangle
\end{pmatrix}.
\]
Therefore, there exist bosonic systems where either $M$ or $\mathcal{S}$
have non-trivial Jordan normal forms and we say such systems are at
exceptional points\emph{ }(EPs)\emph{. }Unlike the non-EP systems,
where the dynamics can be well-understood by tracking the phase change
of the linear combinations of the quadratures since the $M$ or $\mathcal{S}$
matrices are diagonlizable, more complicated dynamical features may
emerge in EP systems. One of them is the exotic behavior under perturbations.
Let $A\left(s\right)$ be a smooth-matrix-valued function dependent
on a complex parameter $s$. Let $A(0)$ be at the exceptional point
(or equivalently, let $s=0$ be the exceptional point). Then the eigenvalues
$\epsilon(s)$ of $A(s)$ may scale as $\epsilon(s)\sim s^{\alpha}$
with $0<\alpha<1$ when $s$ is near zero, which cannot be observed
in any non-EP system.

Bosonic exceptional-point systems are usually constructed by adding
extra losses and gains. However, these added losses or gains will
inevitably introduce extra quantum noise into the physical process,
which is often not taken into consideration in classical EP studies
and is crucial to what we will discuss here in this chapter. These
losses and gains, in most cases, can be viewed as the result of linear
coupling between the system and certain bosonic modes in the environment. 

Let $\mathcal{S}:E_{in}\rightarrow E_{out}$ be a scattering matrix
at some exceptional point, satisfying $\text{det}\left(\text{Id}-\mathcal{S}\right)\neq0$.
In Section.~\ref{sec:Symplectic-dialtion}, we showed for any such
$\mathcal{S}$, no matter symplectic or not, there always exists a
symplectic transformation $S:E_{in}\oplus E_{env}\rightarrow E_{out}\oplus E_{env}$,
of which $\mathcal{S}$ is a submatrix. Denoting $\text{Id}-\mathcal{S}$
by $G$, and let $\mathcal{R}:E_{env}\rightarrow E_{out}$ be the
linear transformation as defined in Theorem.~\ref{thm:The-linear-transformation}.
Then for a input state Gaussian state $\hat{\rho}\left(\bar{x}_{in},V_{in}\right)\otimes\hat{\rho}\left(0,V_{env}\right)$,
we have the output state being $\hat{\rho}\left(\bar{x}_{out},V_{out}\right)$
with
\begin{equation}
\bar{x}_{out}=\left(\text{Id}-G\right)\bar{x}_{in},
\end{equation}
and
\begin{equation}
V_{out}=\left(\text{Id}-G\right)V_{in}\left(\text{Id}-G\right)^{t}+G\mathcal{R}V_{env}\mathcal{R}^{t}G^{t}.
\end{equation}
In other words, we thus obtain a Gaussian channel $\mathcal{G}_{\text{Id}-G,G\mathcal{R}V_{env}\mathcal{R}^{t}G^{t}}$.

Now we consider a multi-mode bosonic system, each mode of which is
coupled with an external detectable channel. Then suppose that we
send a Gaussian signal $\hat{\rho}\left(\bar{x}_{in},V_{in}\right)$
into the system through these channels, and detect the scattered signal
using the ideal heterodyne measurement. As we have mentioned, the
scattering process can be described by the Gaussian channel $\mathcal{G}_{\text{Id}-G,G\mathcal{R}V_{env}\mathcal{R}^{t}G^{t}}$,
if we assume the source of the undetectable losses and gains are intelligible
and hence can be simulated by the matrix $\mathcal{R}$ and a Gaussian
state $\hat{\rho}\left(0,V_{env}\right)$. Now we further assume the
system under investigation depends on a real parameter $\theta$ in
the form of $G=G_{\theta}=\left(\theta\Pi-M\right)^{-1}$ with $\Pi$
a full-rank matrix. Physically, this describes a system whose Hamiltonian
is perturbed by a small interaction proportional to $\theta$. Furthermore,
we assume $\Pi^{-1}M$ has a non-trivial Jordan normal form $\oplus_{l}J_{m_{l}}\left(0\right)$.
Particularly the unperturbed system is at an exceptional point when
$\Pi=\text{Id}$. Let $k=\max_{l}\left\{ m_{l}\right\} $, then 
\[
G_{\theta}=\theta^{-k}C_{0}+o\left(\theta^{-k}\right),
\]
with $C_{0}$ constant matrix.

With these assumptions, we have the following equations
\begin{align*}
G_{\theta}^{-1}V_{out}\left(G_{\theta}^{-1}\right)^{t}= & G_{\theta}^{-1}\left[\left(\text{Id}-G_{\theta}\right)V_{in}\left(\text{Id}-G_{\theta}^{t}\right)+G_{\theta}\mathcal{R}V_{env}\mathcal{R}^{t}G_{\theta}^{t}\right]\left(G_{\theta}^{-1}\right)^{t}\\
= & \left(\theta\Pi-M-\text{Id}\right)V_{in}\left(\theta\Pi-M-\text{Id}\right)^{t}+\mathcal{R}V_{env}\mathcal{R}^{t}\\
= & C_{1}+o\left(\theta\right),
\end{align*}
and
\begin{align*}
G_{\theta}^{-1}\left(V_{out}+\text{Id}\right)\left(G_{\theta}^{-1}\right)^{t}= & C_{1}+MM^{t}+o\left(\theta\right)
\end{align*}
with the positive-definite matrix
\[
C_{1}:=\left(\text{Id}+M\right)V_{in}\left(\text{Id}+M\right)^{t}+\mathcal{R}V_{env}\mathcal{R}^{t}.
\]
In addition, we also have
\begin{align*}
\frac{d\bar{x}_{out}}{d\theta}= & \frac{d\left(\text{Id}-G_{\theta}\right)}{d\theta}\bar{x}_{in}\\
= & G_{\theta}\Pi G_{\theta}\bar{x}_{in}.
\end{align*}

Then it is easy to calculate
\begin{equation}
\mathcal{I}_{\bar{x}}\left(\theta\right)=\theta^{-2k}\bar{x}_{in}^{t}C_{0}^{t}C_{1}C_{0}\bar{x}_{in}+o\left(\theta^{-2k}\right),
\end{equation}
\begin{equation}
I_{\mu}\left(\theta\right)=\theta^{-2k}\bar{x}_{in}^{t}C_{0}^{t}\left(C_{1}+MM^{t}\right)C_{0}\bar{x}_{in}+o\left(\theta^{-2k}\right).
\end{equation}
Thus, we can already conclude that when the perturbation $\theta$
is close to zero,
\[
I\left(\theta\right),\mathcal{I}\left(\theta\right)\apprge\theta^{-2k}\cdot\text{constant}.
\]
Moreover, we can remove the ``$>$'' sign above since the similar
calculations shows
\[
I_{\Sigma}\left(\theta\right)=\theta^{-2k}\cdot\text{constant},
\]
and, when the output Gaussian state is very noisy(i.e., $\text{det}\left(V_{out}\right)\gg1$),
\[
\mathcal{I}_{V}\left(\theta\right)\approx I_{\Sigma}\left(\theta\right),
\]
as explained in the last section. This means we may be able to infer
the value of the parameter $\theta$ up to an arbitrary precision,
whenever the system is still stable. The enhancement can be further
boosted by sending stronger input signal (i.e., with larger $\bar{x}_{in}$),
and/or by constructing higher-order exceptional points (i.e., increasing
the magnitude of $k$). However, a strong input signal may saturate
the bosonic system, causing the failure of the linearly-coupled-mode
theory which serves as the bedrock of our analysis. This problem may
be addressed by sending a vacuum signal with $\bar{x}_{in}=0$ and
detecting the spontaneous emission of the bosonic system afterwards,
because the quantities $I_{0}\left(\theta\right)$ and $\mathcal{I}_{0}\left(\theta\right)$
does not depend on $\bar{x}_{in}$ at all. 

In the scenario above, the (quantum) Fisher information is enhanced
when the parameter to be inferred is approaching zero. Often times,
we are allowed to tune one or more than one system parameters to improve
the sensor performance. Suppose the parameter $\theta$ to be inferred
has the value of zero. Let $\alpha$ be a controllable system parameter.
We assume
\[
G_{\theta=0}=\left(\alpha\Pi+M\right)^{-1}\sim\alpha^{-k}
\]
with $\Pi$ a full rank matrix. Then we have
\begin{align*}
G_{0}^{-1}V_{out}\left(G_{0}^{-1}\right)^{t}= & C_{1}+o\left(\alpha\right),
\end{align*}
\begin{align*}
G_{0}^{-1}\left(V_{out}+\text{Id}\right)\left(G_{0}^{-1}\right)^{t}= & C_{1}+MM^{t}+o\left(\alpha\right),
\end{align*}
\begin{align*}
\left(\frac{d\bar{x}_{out}}{d\theta}\right)_{\theta=0}= & G_{0}\left(\frac{dG_{\theta}^{-1}}{d\theta}\right)_{\theta=0}G_{0}\bar{x}_{in},
\end{align*}
and thus
\begin{align*}
I_{\mu}\left(\theta=0\right),\mathcal{I}_{\bar{x}}\left(\theta=0\right),I_{\Sigma}\left(\theta=0\right)\sim & \alpha^{-2k},
\end{align*}
as well as,
\[
\mathcal{I}_{V}\left(\theta=0\right)\sim\alpha^{-2k},
\]
when the output state is highly thermalized.

Moreover, sometimes it is easier tune the system parameter $\alpha$
to an arbitrarily large magnitude. For this situation, we can design
a system with
\[
G_{\theta=0}=\left(\alpha M+\Pi\right)^{-1}\sim\alpha^{k-1},
\]
such that
\begin{align*}
G_{0}^{-1}V_{out}\left(G_{0}^{-1}\right)^{t}= & \alpha^{2}\cdot\text{constant}+o\left(\alpha^{2}\right),
\end{align*}
\begin{align*}
G_{0}^{-1}\left(V_{out}+\text{Id}\right)\left(G_{0}^{-1}\right)^{t}= & \alpha^{2}\cdot\text{constant}+o\left(\alpha^{2}\right),
\end{align*}
\begin{align*}
\left(\frac{d\bar{x}_{out}}{d\theta}\right)_{\theta=0}= & G_{0}\left(\frac{dG_{\theta}^{-1}}{d\theta}\right)_{\theta=0}G_{0}\bar{x}_{in},
\end{align*}
Thus
\begin{align}
I_{\mu}\left(\theta=0\right),\mathcal{I}_{\bar{x}}\left(\theta=0\right),I_{\Sigma}\left(\theta=0\right)\sim & \alpha^{2k-4},
\end{align}
as well as, when the output state is very noisy,
\begin{equation}
\mathcal{I}_{V}\left(\theta=0\right)\sim\alpha^{2k-4}.
\end{equation}

We can freely combine the techniques listed above to construct even
better exceptional point sensors.

\section{Unitary bound of non-unitary Gaussian sensing}

Motivated by the surprising potential of exceptional point sensing,
we are wondering whether there are other bosonic sensing schemes described
by carefully engineered Gaussian processes exhibit the same ability
to drastically enhance the sensing precision. We believe the answer
to this question lies at its fundamental physical origin: What is
the dominating physical mechanism behind this phenomenon? In the preceding
section, we used the symplectic dilation to embed a specific noisy
Gaussian process into a symplectic transformation. Here, we replace
the specific process, exceptional point sensing, with an arbitrary
Gaussian process corresponding to an arbitrary bosonic sensing scheme,
and dilate it symplectically. Then we assume not only the modes involved
in the orignial Gaussian process can be accessed, but also the modes
added later to dilate this Gaussian process (say, the $\hat{B}$ modes
in Example.~\ref{sec:Exceptional-point-systems}). In other words,
we are allowed to measure all ports of the dilated symplectic transformation.
For simplicity and without loss of generality, let us assume the considered
symplectic transformation $S$ satisfies $\text{det}\left(\text{Id}-S\right)\neq0$
and therefore can be generated by a symmetric matrix $M$ via the
symplectic Cayley transform:
\[
S=\text{\ensuremath{\left(\Omega M-\frac{1}{2}\text{Id}\right)^{-1}}}\left(\Omega M+\frac{1}{2}\text{Id}\right).
\]
That is, $S$ is generated from a scattering process governed by
a quadratic Hamiltonian described by the symmetric matrix $M$ (see
Appx.~\ref{chap:Scattering-process-and}).

Let $S_{\theta}$ be the symplectic dilation of a bosonic sensing
scheme with $\theta$ the parameter to be sensed. Then let the $\theta$-dependent
quantum state, i.e. the output state of the process, be a Gaussian
state
\[
\hat{\rho}(S_{\theta}\bar{x},S_{\theta}VS_{\theta}^{t})=\hat{U}_{S_{\theta}}\hat{\rho}(\bar{x},V)\hat{U}_{S_{\theta}}^{\dagger}.
\]
It follows that
\begin{equation}
\mathcal{I}_{\bar{x}}\left(\theta\right)=\bar{x}^{t}\left[S_{\theta}^{-1}\left(\frac{dS_{\theta}}{d\theta}\right)\right]^{t}V\left[S_{\theta}^{-1}\left(\frac{dS_{\theta}}{d\theta}\right)\right]\bar{x}.
\end{equation}
Therefore, the scaling of $\mathcal{I}_{1}\left(\theta\right)$ with
respect to the parameter $\theta$ should be determined by the matrix
\[
W_{\theta}=S_{\theta}^{-1}\left(\frac{dS_{\theta}}{d\theta}\right).
\]
Supposing $S_{\theta}=\text{Id}-\left(\Omega M_{\theta}+\frac{1}{2}\text{Id}\right)^{-1}$
with $M$ as symmetric matrix\footnote{As explained in Sec.~\ref{sec:Symplectic-dialtion}, such a dilation
$S_{\theta}$ always exists}, then we have
\[
W_{\theta}=\left(\Omega M_{\theta}-\frac{1}{2}\text{Id}\right)^{-1}\Omega\frac{dM_{\theta}}{d\theta}\left(\Omega M_{\theta}+\frac{1}{2}\text{Id}\right)^{-1}.
\]

We note that for certain choices of $M_{\theta}$, either of the factors
\[
(\Omega M_{\theta}-\frac{1}{2}\text{Id})\text{or }(\Omega M_{\theta}+\frac{1}{2}\text{Id})
\]
can be singular and therefore $W_{\theta}$ can blow up. Specifically,
$W_{\theta}$ blows up when 
\[
M_{\theta}=S_{\theta}(\oplus_{j}M_{\theta}^{(j)})S_{\theta}^{t}
\]
with $S_{\theta}$ being an arbitrary symplectic transformation and
at least one of the direct summands, say $M_{\theta}^{(m)}$, corresponding
to the a Hamiltonain of the following squeezing-like form \footnote{We can obtain this result by writing down the matrix representation
for the quadratic Hamiltonians classified in Appendix 6 of \cite{Arnold2013}. }
\[
i\sum_{kl}g_{kl}(\theta)\left(\hat{a}_{k}\hat{a}_{l}-\hat{a}_{k}^{\dagger}\hat{a}_{l}^{\dagger}\right).
\]

This observation states the following: Let $M_{\theta}$ also depends
on a bunch of other real parameters $\{\xi_{j}\}$with $\xi_{0}:=\theta$,
and the limit 
\[
\text{lim}_{\xi_{k}}M_{\theta}
\]
 be a matrix satisfying the above conditions so that either of
\[
(\Omega\text{lim}_{\xi_{k}}M_{\theta}-\frac{1}{2}\text{Id})\text{ or }(\Omega\text{lim}_{\xi_{k}}M_{\theta}+\frac{1}{2}\text{Id})
\]
is singular for some parameter $\xi_{k}$. Then we have
\[
\lim_{\xi_{k}}\mathcal{I}_{\bar{x}}(\theta)=\infty,
\]
which means the estimation of $\theta$ can be infinitely precise
if measured in a proper measurement scheme (e.g. the heterodyne measurement
as explained in the last section). Moreover, a similar conclusion
also holds for $\mathcal{I}_{V}\left(\theta\right)$, especially when
each mode of the input probes is in a thermal state.

It is not hard to conclude that the Fisher information obtained for
the symplectic dilation should be the upper bound of that for the
bosonic sensing scheme, since there are less noise involved in the
whole process and more modes to measure. Therefore, a bosonic sensing
scheme can drastically improve the sensing precision if its symplectic
dilation corresponds to a mutli-mode squeezing operation, of which
the degree of squeezing can be infinitely increased. From another
point of view, to devise an exotic bosonic sensing scheme, we may
starting with a suitable squeezing-like symplectric transformation
and then tracing out the inaccessible modes, which account for the
losses and gains. As squeezing is not just a phenomenon of bosonic
system--spin squeezing is also an important effect, this picture
may serve as a novel and universal perspective for various exotic
physical sensing schemes.

\section{Conclusion and outlook}

In this chapter, we show how a careful investigation of the quantum
noise in exceptional system can shed light on developing exotic sensing
schemes. The toolbox and statements can be extended to a larger family
of Gaussian sensing. We see that the sensing precision enhancement
in such schemes can also be studied as a phenomenon of mutli-mode
squeezing. This observation may allow us to extend these ideas to
other physical platforms beyond the bosonic system.

This chapter also concludes our investigation of applications symplectic
geometry in solving practical physical problems in continuous variable
system. Amazed by the simplicity and efficacy of symplectic geometry
in continuous variable systems, in the next chapter, we try to bring
this powerful tool to the seeming irrelevant physical systems--the
discrete variable systems. We will see the formulas, previous derived
for bosonic systems, can also be applied, with little change, to interpreting
important physical phenomenon in the discrete variable systems.

\chapter{Beyond the continuous variable systems\label{chap:Beyond-the-continuous}}

\section{Introduction}

We have seen the applications of symplectic geometry in continuous
variable systems, leading to novel and promising solutions to practical
physical problems. In this chapter, we try to extend our consideration
to the discrete variable systems. First, we would like to introduce
how to suit the essential components of symplectic geometry to the
seemingly irrelevant discrete variable systems, based on the interesting
correspondence between displacement operations and Pauli operators.
Then we will see extension allows effortless transplantation of the
results originally derived for continuous variable systems, which
will be demonstrated by examples concerning the well-know notion of
quantum teleportation. After that we will show our attempts of tackling
a theoretical issue about this extension, in order to overcome certain
mathematical shortcomings and broaden its generality.

We start with the establishment of symplectic geometry for qubit systems
in the succeeding section.

\section{Symplectic geometry for qubit systems}

Concepts in continuous variable systems such as phase space, symplectic
trnansformations, and Wigner functions have their analogues in certain
discrete variable systems\footnote{This section is based on the methods introduced in \cite{Gross2013}.},
allowing extension of symplectic geometry to discrete variable systems.
Here, we demonstrate the general idea for qubit systems. We will see
how this extension reveals the connection between Pauli operators
and displacement operations, and between symplectic transformations
and Clifford operations.

To see this, we start with the qubit systems. Consider an $N$-qubit
system with $\hat{X}_{i}$ and $\hat{Z}_{i}$, $i\in\{1,\dots,N\}$
being the Pauli-X and Pauli-Z operators for the $i$-th qubit. Now
for any $u\in\left(\mathbb{Z}/2\mathbb{Z}\right)^{2N}=:E$, we can
define the generalized Pauli operator as
\begin{equation}
\hat{D}_{u}=\otimes_{j=1}^{N}\hat{Z}_{j}^{u_{i}^{(q)}}\hat{X}_{j}^{u_{i}^{(p)}}
\end{equation}
where $u^{\left(q\right)},u^{\left(p\right)}\in\left(\mathbb{Z}/2\mathbb{Z}\right)^{N}$
are defined by
\begin{equation}
u_{i}^{\left(q\right)}:=u_{2i-1},\,u_{i}^{(p)}:=u_{2i}.
\end{equation}
for any $a,b\in\left(\mathbb{Z}/2\mathbb{Z}\right)^{N}$. This operator
resembles the continuous variable displacement operator insofar as
$\hat{D}_{-u}=\hat{D}_{u}^{\dagger}$ and
\begin{equation}
\hat{D}_{u}\hat{D}_{v}\propto\hat{D}_{u+v}.
\end{equation}
Moreover, same as in the continuous variable systems, when an additional
phase will appear when exchanging the order of two generalized Pauli
operators:
\[
\hat{D}_{u}\hat{D}_{v}=e^{i\pi\sigma\left(u,v\right)}\hat{D}_{v}\hat{D}_{u}.
\]
Here, given the same $E$ considered as a discrete variable symplectic
space and hence a finite phase space, we define the symplectic form
$\sigma\left(\cdot,\cdot\right):E\times E\rightarrow\mathbb{Z}/2\mathbb{Z}$
by
\begin{equation}
\sigma\left(u,v\right)\rightarrow\sum_{j=1}^{N}u_{2j-1}v_{2j}-u_{2j}v_{2j-1}.
\end{equation}
Therefore, the group of symplectic transformations\emph{ }is similarly
defined as the group of linear transformations $S:E\rightarrow E$
preserving the symplectic form in the following sense:
\[
\sigma\left(Su,Sv\right)=\sigma\left(u,v\right)
\]
for any $u,v\in E$. 

In addition, given a $N$-qubit quantum state $\hat{\rho}$, we can
calculate the discrete characteristic\emph{ }function using
\begin{equation}
\chi_{\hat{\rho}}\left(u\right)=\frac{1}{2}\text{Tr}\left[\hat{D}\left(-u\right)\hat{\rho}\right].
\end{equation}
Then the Wigner function can be calculated by
\begin{equation}
W_{\hat{\rho}}\left(u\right)=\frac{1}{2}\sum_{v\in E}e^{i\pi\sigma\left(u,v\right)}\chi_{\hat{\rho}}\left(v\right).
\end{equation}
As expected, for each symplectic transformation $S:E\rightarrow E$,
there exists a unitary operator $\hat{U}_{S}$ such that
\[
W_{\hat{\rho}}\left(S^{-1}u\right)=W_{\hat{U}_{S}\hat{\rho}\hat{U}_{S}^{\dagger}}\left(u\right),
\]
and for every discrete displacement operator $\hat{D}\left(v\right)$,
we have
\[
W_{\hat{\rho}}\left(u+v\right)=W_{\hat{D}\left(v\right)\hat{\rho}\hat{D}\left(v\right)^{\dagger}}\left(u\right).
\]
Actually, the group generated by $\hat{U}_{S}$ and $\hat{D}\left(v\right)$
with every possible $S$ and $v$ is the group of Clifford operators.
Note that $\hat{U}_{S}\hat{D}\left(v\right)\hat{U}_{S}=\hat{D}\left(S^{t}v\right)$.

One can easily extend these results to qu$d$its systems with $d$
being any prime number, just by letting $E=\left(\mathbb{Z}/d\mathbb{Z}\right)^{N}$,
\[
\hat{D}\left(u\right)=\prod_{i=1}^{N}e^{-i\frac{\pi}{d}u^{(q)t}u^{(p)}}\hat{Z}_{i}^{u_{i}^{(q)}}\hat{X}_{i}^{u_{i}^{(p)}},
\]
\[
\chi_{\hat{\rho}}\left(u\right)=\frac{1}{d}\text{Tr}\left[\hat{D}\left(-u\right)\hat{\rho}\right],
\]
and
\[
W_{\hat{\rho}}\left(u\right)=\frac{1}{d}\sum_{v\in E}e^{i\frac{2\pi}{d}\sigma\left(u,v\right)}\chi_{\hat{\rho}}\left(v\right).
\]

Although it remains unclear how to replicate the concepts of Gaussian
states in the discrete variable systems, we can still find the counterparts
of infinitely squeezed states quite easily. As we know, in continuous
variable systems, infinitely squeezed states are the eigenstates of
the quadrature operators and hence the displacement operators. Therefore,
the counterparts of the continuous variable infinitely squeezed states
in the discrete variable systems are nothing but the stabilizer states,
e.g, the eigenstates of the generalized Pauli operators (products
of the clock and shift operators) in the discrete variable system.
Moreover, the homodyne measurement can similarly be mapped to the
syndrome measurement in discrete variable systems.

\section{Understanding discrete variable quantum teleportation}

Now that all the key ingredients can be extended to discrete variable
systems, we show in examples how to extend our previous theorems of
generalized teleportation to discrete variable systems. Note that
the idea of simulating Clifford operations using linear transformations
is well known \cite{Gottesman1998}. Here we are focusing on the extending
our results in the continuous variable systems to the discrete variable
systems.

\begin{figure}[h]
\includegraphics[width=1\textwidth]{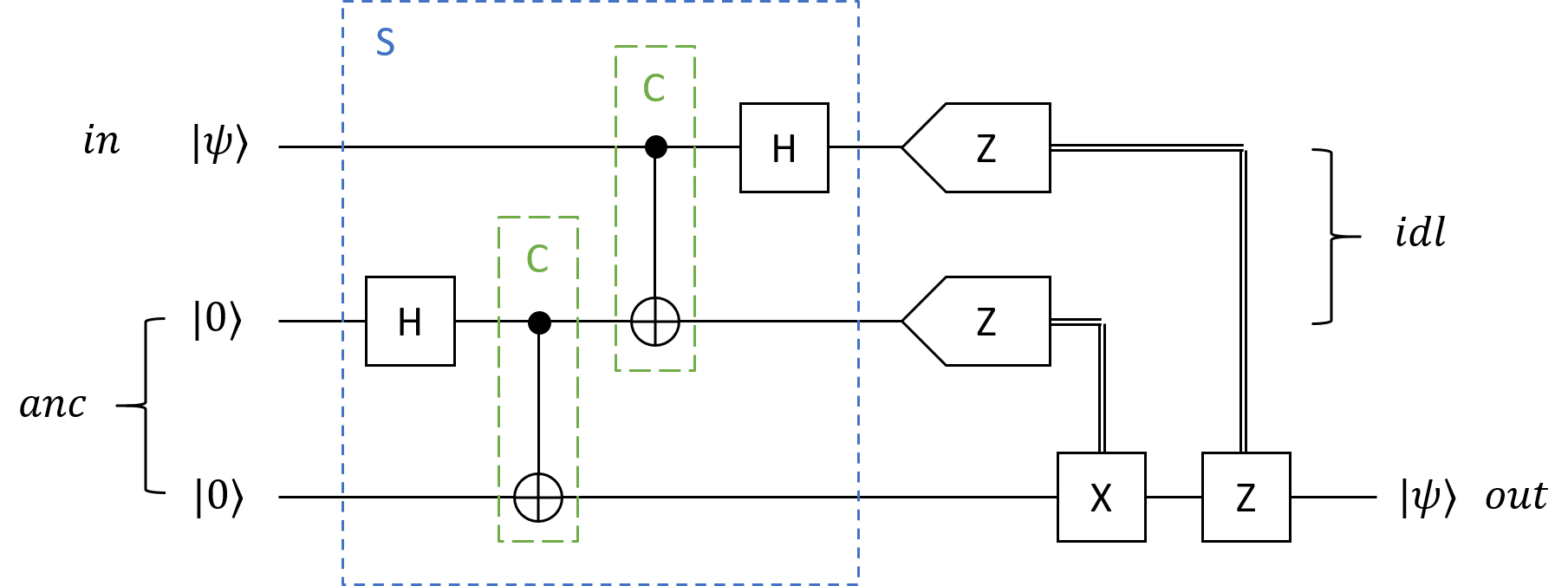}
\begin{doublespace}
\centering{}\caption{\emph{Quantum teleportation circuit.}}
\begin{minipage}[t]{0.9\textwidth}%
The boxes labeled by H are Hadamard gates, labled by X and Z are Pauli-X
and Pauli-Z operators. The two-qubit gates, encircled by the green
dashed boxes and labeled by C, are the CNOT gates. The solid double
lines represent classical communication channels. The classical communication
channels connects the syndrome measurement, i.e. the Pauli-Z measurements,
to the corresponding Pauli operators. As stated in the main text,
the Hadmard gates and the CNOT gates, encircled by the blue dashed
box, can be treated as a single symplectic transformation $S$. The
Pauli-operators are performed to modified the output mode based on
the measurement outcomes of the syndrome measurements.%
\end{minipage}
\end{doublespace}
\end{figure}

\begin{example}
(Quantum teleportation) We first consider the well-known quantum teleportation
circuit. The circuit is composed only of Clifford operations, stabilizer
states, classical transmission and syndrome measurement and thus can
be understood using our theorem. To see this, we first demonstrate
the symplectic transformation representation of each Clifford gate
in the circuit. The Hadmard gate 
\[
\hat{H}=\frac{1}{\sqrt{2}}\begin{pmatrix}1 & 1\\
1 & -1
\end{pmatrix},
\]
by definition, transforms Pauli-X operator to Pauli-Z operators and
vice versa. Therefore, it can be correspondingly represented by a
symplectic transformation
\[
S_{H}=\begin{pmatrix}0 & 1\\
1 & 0
\end{pmatrix}.
\]
The CNOT gate
\[
\text{CNOT}=\begin{pmatrix}1 & 0 & 0 & 0\\
0 & 1 & 0 & 0\\
0 & 0 & 0 & 1\\
0 & 0 & 1 & 0
\end{pmatrix}
\]
transforms
\[
\hat{D}(\begin{pmatrix}1 & 0 & 0 & 0\end{pmatrix}^{t})=\hat{Z}\otimes\text{Id}\text{ to }\hat{D}(\begin{pmatrix}1 & 0 & 0 & 0\end{pmatrix}^{t})=\hat{Z}\otimes\hat{Z},
\]
\[
\hat{D}(\begin{pmatrix}0 & 1 & 0 & 0\end{pmatrix}^{t})=\hat{X}\otimes\text{Id}\text{ to }\hat{D}(\begin{pmatrix}0 & 1 & 0 & 1\end{pmatrix}^{t})=\hat{X}\otimes\text{Id},
\]
\[
\hat{D}(\begin{pmatrix}0 & 0 & 1 & 0\end{pmatrix}^{t})=\text{Id}\otimes\hat{Z}\text{ to }\hat{D}(\begin{pmatrix}1 & 0 & 1 & 0\end{pmatrix}^{t})=\text{Id}\otimes\hat{Z},
\]
and
\[
\hat{D}(\begin{pmatrix}0 & 0 & 0 & 1\end{pmatrix}^{t})=\text{Id}\otimes\hat{X}\text{ to }\hat{D}(\begin{pmatrix}1 & 0 & 0 & 0\end{pmatrix}^{t})=\hat{X}\otimes\hat{X}.
\]
Therefore, the symplectic transformation of the CNOT gate is 
\[
S_{C}=\begin{pmatrix}1 & 0 & 0 & 0\\
0 & 1 & 0 & 1\\
1 & 0 & 1 & 0\\
0 & 0 & 0 & 1
\end{pmatrix}.
\]
Then, the composition of the Clifford gates in the circuit can be
represented by the symplectic transformation
\begin{align*}
S & =\left(S_{H}\oplus\text{Id}_{4\times4}\right)\left(S_{C}\oplus\text{Id}_{2\times2}\right)\left(\text{Id}_{2\times2}\oplus S_{C}\right)\left(\text{Id}_{2\times2}\oplus S_{H}\oplus\text{Id}_{2\times2}\right)\\
 & =\begin{pmatrix}0 & 1 & 0 & 0 & 0 & 0\\
1 & 0 & 0 & 0 & 0 & 0\\
0 & 0 & 1 & 0 & 0 & 0\\
0 & 0 & 0 & 1 & 0 & 0\\
0 & 0 & 0 & 0 & 1 & 0\\
0 & 0 & 0 & 0 & 0 & 1
\end{pmatrix}\begin{pmatrix}1 & 0 & 0 & 0 & 0 & 0\\
0 & 1 & 0 & 1 & 0 & 0\\
1 & 0 & 1 & 0 & 0 & 0\\
0 & 0 & 0 & 1 & 0 & 0\\
0 & 0 & 0 & 0 & 1 & 0\\
0 & 0 & 0 & 0 & 0 & 1
\end{pmatrix}\begin{pmatrix}1 & 0 & 0 & 0 & 0 & 0\\
0 & 1 & 0 & 0 & 0 & 0\\
0 & 0 & 1 & 0 & 0 & 0\\
0 & 0 & 0 & 1 & 0 & 1\\
0 & 0 & 1 & 0 & 1 & 0\\
0 & 0 & 0 & 0 & 0 & 1
\end{pmatrix}\begin{pmatrix}1 & 0 & 0 & 0 & 0 & 0\\
0 & 1 & 0 & 0 & 0 & 0\\
0 & 0 & 0 & 1 & 0 & 0\\
0 & 0 & 1 & 0 & 0 & 0\\
0 & 0 & 0 & 0 & 1 & 0\\
0 & 0 & 0 & 0 & 0 & 1
\end{pmatrix}\\
 & =\left(\begin{array}{cccccc}
0 & 1 & 1 & 0 & 0 & 1\\
1 & 0 & 0 & 0 & 0 & 0\\
1 & 0 & 0 & 1 & 0 & 0\\
0 & 0 & 1 & 0 & 0 & 1\\
0 & 0 & 0 & 1 & 1 & 0\\
0 & 0 & 0 & 0 & 0 & 1
\end{array}\right).
\end{align*}
Now we have
\[
S_{out,in}=\begin{pmatrix}0 & 0\\
0 & 0
\end{pmatrix},\,S_{out,z'}=\begin{pmatrix}1 & 0\\
0 & 1
\end{pmatrix},\,S_{h,z'}=\begin{pmatrix}0 & 1\\
1 & 0
\end{pmatrix},\,S_{h,in}=\begin{pmatrix}0 & 1\\
1 & 0
\end{pmatrix}.
\]
(the syndrome measurement is determined by projections $\Pi_{0}=\frac{1}{2}\left(\text{Id}-Z\right),\Pi_{1}=\frac{1}{2}\left(\text{Id}+Z\right)$.)
Then we obtain that
\begin{equation}
\tilde{S}=S_{out,in}-S_{out,z'}\left(S_{h,z'}\right)^{-1}S_{h,in}=\text{Id},
\end{equation}
and
\begin{equation}
F_{\star}=S_{out,z'}\left(S_{h,z'}\right)^{-1}=\begin{pmatrix}0 & 1\\
1 & 0
\end{pmatrix}.
\end{equation}
It is then easy to check this these results lead to the correct controlled
Pauli operations in the circuit. For example, if the syndrome measurement
returns $0$ for the first qubit (the $|1\rangle$ state) and $1$
for the second qubit (the $|0\rangle$ state), then we will apply
$\hat{D}\left(F_{\star}\begin{pmatrix}0\\
1
\end{pmatrix}\right)=\hat{D}\left(\begin{pmatrix}1\\
0
\end{pmatrix}\right)=\hat{Z}$.
\end{example}

\begin{figure}[h]
\includegraphics[width=1\textwidth]{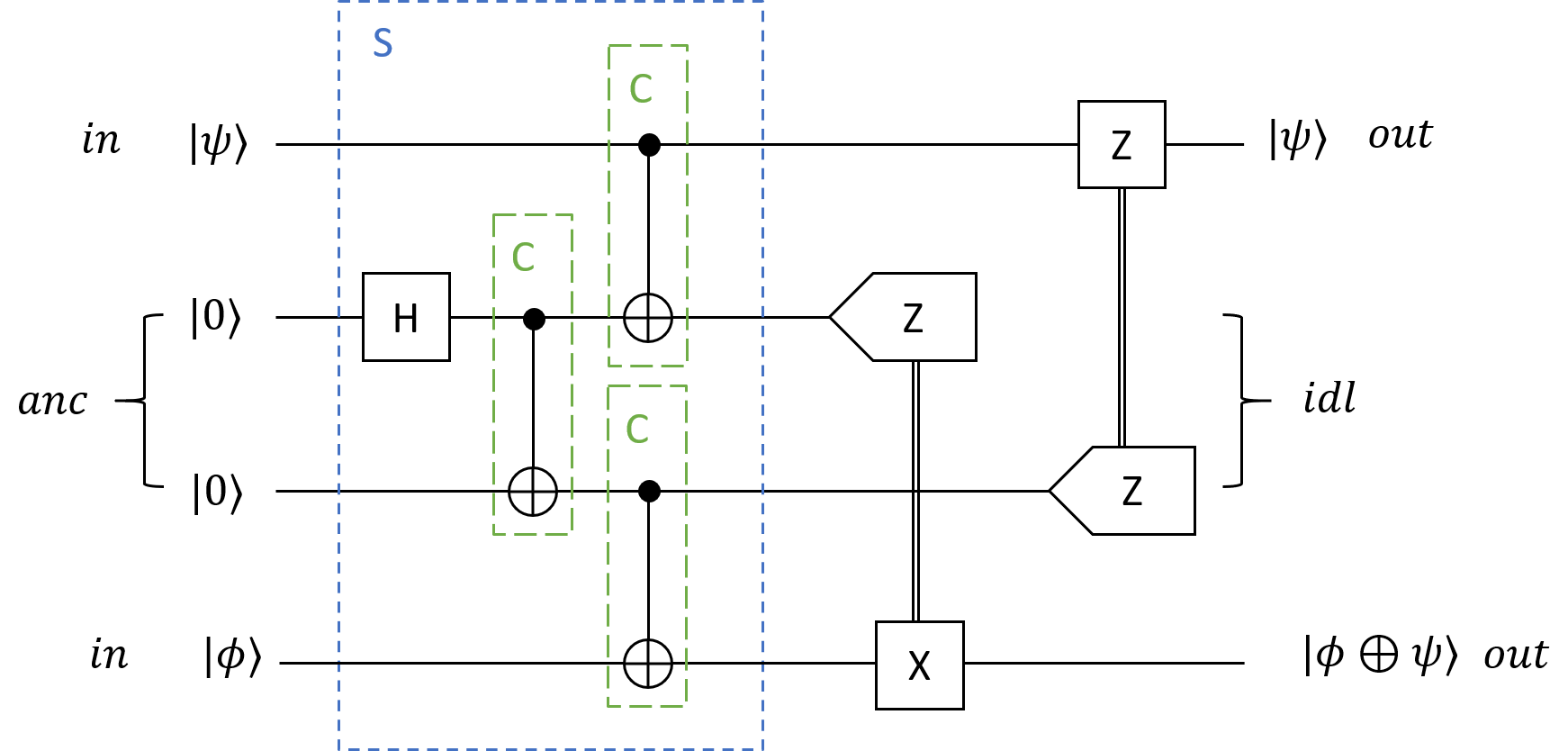}
\begin{doublespace}
\centering{}\caption{\emph{Quantum gate teleportation.}}
\begin{minipage}[t]{0.9\textwidth}%
The boxes labeled by H are Hadamard gates, labled by X and Z are Pauli-X
and Pauli-Z operators. The two-qubit gates, encircled by the green
dashed boxes and labeled by C, are the CNOT gates. The solid double
lines represent classical communication channels. The classical communication
channels connects the syndrome measurement, i.e. the Pauli-Z measurements,
to the corresponding Pauli operators. As stated in the main text,
the Hadmard gates and the CNOT gates, encircled by the blue dashed
box, can be treated as a single symplectic transformation $S$. The
Pauli-operators are performed to modified the output mode based on
the measurement outcomes of the syndrome measurements.%
\end{minipage}
\end{doublespace}
\end{figure}

\begin{example}
Then we test our theorem on a slightly more complicated circuit --
a gate teleportation circuit \cite{Chou2018}. The 3-qubit Clifford
gate composed of the three CNOTs and the Hadmard gate in this circuit
can be represented by a symplectic transformation
\[
S=\left(\begin{array}{cccccccc}
1 & 0 & 0 & 0 & 0 & 0 & 0 & 0\\
0 & 1 & 1 & 0 & 0 & 1 & 0 & 0\\
1 & 0 & 0 & 1 & 0 & 0 & 0 & 0\\
0 & 0 & 1 & 0 & 0 & 1 & 0 & 0\\
0 & 0 & 0 & 1 & 1 & 0 & 0 & 0\\
0 & 0 & 0 & 0 & 0 & 1 & 0 & 1\\
0 & 0 & 0 & 1 & 1 & 0 & 1 & 0\\
0 & 0 & 0 & 0 & 0 & 0 & 0 & 1
\end{array}\right).
\]
Then we have
\[
S_{out,in}=\left(\begin{array}{cccc}
1 & 0 & 0 & 0\\
0 & 1 & 0 & 0\\
0 & 0 & 1 & 0\\
0 & 0 & 0 & 1
\end{array}\right),
\]
\[
S_{out,z'}=\left(\begin{array}{cc}
0 & 0\\
0 & 1\\
1 & 0\\
0 & 0
\end{array}\right),
\]
\[
S_{h,z'}=\left(\begin{array}{cc}
1 & 0\\
0 & 1
\end{array}\right),
\]
and
\[
S_{h,in}=\left(\begin{array}{cccc}
1 & 0 & 0 & 0\\
0 & 0 & 0 & 1
\end{array}\right).
\]
Thus, it is easy to calculate
\begin{equation}
\tilde{S}=S_{out,in}-S_{out,z'}\left(S_{h,z'}\right)^{-1}S_{h,in}=\left(\begin{array}{cccc}
1 & 0 & 0 & 0\\
0 & 1 & 0 & 1\\
1 & 0 & 1 & 0\\
0 & 0 & 0 & 1
\end{array}\right)=S_{C}
\end{equation}
and
\begin{equation}
F_{\star}=S_{out,z'}\left(S_{h,z'}\right)^{-1}=\left(\begin{array}{cc}
0 & 0\\
0 & 1\\
1 & 0\\
0 & 0
\end{array}\right).
\end{equation}
This explains why the corresponding controlled Pauli operations adapted
to the syndrome measurement results can generated the CNOT gate between
the first and last qubits. 
\end{example}

\section{Symplectic geometry for general discrete variable systems}

We have so far restricted ourselves to working on qu$d$it systems
with $d$ being a prime number, or, in other words, with $\mathbb{Z}/d\mathbb{Z}$
being a field. The reason of this restriction is that a field has
no zero divisors and any element of a field has an inverse, which
allows us to effortlessly reproduce the constructions and proofs of
symplectic space and symplectic algebra in the continuous variable
systems. However, the integer $d$ is not a prime number in general,
which implies that the commutative ring $\mathbb{Z}/d\mathbb{Z}$
is not a field\footnote{Note that $\mathbb{Z}/d\mathbb{Z}$ is not a field when $d$ is a
power of a prime number. } and hence contains zero divisors for most discrete variable systems.
Therefore, a straightforward extension of the symplectic algebra to
general discrete variable systems seems not existing. Motivated by
the successful applications of symplectic algebra in continuous variable
systems, we attempt to give a solution to this problem here.

First, we briefly review the structure of generalized Pauli-operators
in a $N$-partite $d$-level system. The generalized Pauli-X and Pauli-Y
operators in a single $d$-level system are defined as the shift and
clock operators as follows
\[
\hat{X}=\begin{pmatrix}0 & 0 & \cdots & 0 & 1\\
1 & 0 & \cdots & 0 & 0\\
0 & 1 & \ddots & \vdots & \vdots\\
\vdots & \ddots & \ddots & 0 & 0\\
0 & \cdots & 0 & 1 & 0
\end{pmatrix},\,\hat{Z}=\begin{pmatrix}1 & 0 & 0 & \cdots & 0\\
0 & \omega & 0 & \ddots & \vdots\\
0 & 0 & \omega^{2} & \ddots & 0\\
\vdots & \ddots & \ddots & \ddots & 0\\
0 & \cdots & 0 & 0 & \omega^{d-1}
\end{pmatrix},
\]
with 
\[
\omega=e^{2\pi i/d}.
\]
It follows that
\[
\hat{Z}\hat{X}=\omega\hat{X}\hat{Z}.
\]
Now we define the generalized Pauli-operators for the $N$-partite
$d$-level system with $d$ being an arbitrary integer as
\[
\hat{D}(q,p)=\otimes_{j=1}^{n}\hat{Z}_{j}^{q_{j}}\hat{X}_{j}^{p_{j}},
\]
where each $q_{j}$ and $p_{j}$ are elements of $\mathbb{Z}/d\mathbb{Z}$,
and hence the ``vectors'' $q,p$ are elements of a free of rank
$N$ module over $\mathbb{Z}/d\mathbb{Z}$\footnote{A free module is the generalization of vector space when $\mathbb{Z}/d\mathbb{Z}$
is a ring. Both of a free module and a vector space are spanned by
subsets of them, known as the bases \cite{Hungerford1990}.} Then we have
\[
\hat{D}(q,p)\hat{D}(q',p')=\omega^{-(q^{t}p'-p^{t}q')}\hat{D}(q',p')\hat{D}(q,p),
\]
for any two generalized Pauli operators $\hat{D}(q,p)$ and $\hat{D}(q',p')$,
with $\omega=e^{i2\pi/d}$. The relation amounts to the following:
For any two elements
\[
\begin{pmatrix}q\\
p
\end{pmatrix}\text{ and }\begin{pmatrix}q'\\
p'
\end{pmatrix}
\]
of a free of rank $2N$ module over the ring $\mathbb{Z}/d\mathbb{Z}$,
denoted by $E$, we have
\[
\sigma\left(\begin{pmatrix}q\\
p
\end{pmatrix},\begin{pmatrix}q'\\
p'
\end{pmatrix}\right)=q^{t}p'-p^{t}q',
\]
with the bilinear map $\sigma:E\times E\rightarrow\mathbb{Z}/d\mathbb{Z}$.
Therefore,
\[
\left[\hat{D}(q,p),\hat{D}(q',p')\right]=0
\]
if and only if
\[
\sigma\left(\begin{pmatrix}q\\
p
\end{pmatrix},\begin{pmatrix}q'\\
p'
\end{pmatrix}\right)=0.
\]
It is tempting to claim that $E$ is the symplectic space and $\sigma$
is the symplectic form as we have for the continuous variable systems.
But there are important issues which must be identified and addressed.

Let us now focus on $p^{r}$-level systems with an integer $r>1$
and $p$ a prime number. It means that we will work on the free of
rank $2N$ module $E$ over the ring $\mathbb{Z}/p^{r}\mathbb{Z}$.
The first issue we have to solve here is how to identify or find a
symplectic basis
\[
\mathcal{B}=\left\{ e_{1},e_{2},\dots,e_{N},f_{1},f_{2},\dots,f_{N}\right\} \subset E
\]
satisfying
\[
\sigma\left(e_{j},e_{k}\right)=\sigma\left(f_{j},f_{k}\right)=0
\]
and
\[
\sigma\left(e_{j},f_{k}\right)=-\sigma\left(f_{j},e_{k}\right)=\delta_{j,k}
\]
for all $\left(j,k\right)\in\left\{ 1,2,\dots,N\right\} \times\left\{ 1,2,\dots,N\right\} $.
Let us consider the following example, and we will see this task is
nontrivial:
\begin{example}
Let $N=1$. Then
\[
\left\{ \begin{pmatrix}1\\
0
\end{pmatrix},\begin{pmatrix}0\\
1
\end{pmatrix}\right\} 
\]
is a symplectic basis while
\[
\left\{ \begin{pmatrix}p\\
0
\end{pmatrix},\begin{pmatrix}0\\
p^{r-1}
\end{pmatrix}\right\} 
\]
is not, since
\[
\sigma(\begin{pmatrix}p\\
0
\end{pmatrix},\begin{pmatrix}0\\
p^{r-1}
\end{pmatrix})=\left(p\cdot p^{r-1}-0\cdot0\right)\,\text{mod }p^{r}=0,
\]
even if the latter is obtained by multiplying each element in the
former with an element in the ring $\mathbb{Z}p^{r}\mathbb{Z}$.
\end{example}

\noindent To understand this issue, here we treat the problem in
an isomorphic ring of the $\mathbb{Z}/p^{r}\mathbb{Z}$, the quotient
polynomial ring (corresponding to the coordinate ring of a singular
point\cite{Vakil2017})
\[
\mathbb{F}_{p}\left[x\right]/\left(x^{r}\right)\cong\mathbb{Z}/p^{r}\mathbb{Z},
\]
where the isomorphism is given by
\[
\sum_{j=0}^{r-1}a_{j}p^{j}\mapsto\sum_{j=0}^{r-1}a_{j}x^{j},
\]
with $a_{j}\in\left\{ 0,1,\dots,p-1\right\} $, for all $j\in\{0,1,\dots,r-1\}$. 

Since the only prime ideal and hence the maximal ideal of $\mathbb{Z}/p^{r}\mathbb{Z}$
is $\mathfrak{p}=\left(p\right)$, $\mathbb{Z}/p^{r}\mathbb{Z}$ is
a local ring with its localization at $\mathfrak{p}$, the residue
field being \cite{Vakil2017}
\[
k\left(\mathfrak{p}\right)=\left(\mathbb{F}_{p}\left[x\right]/\left(x^{r}\right)\right)_{\mathfrak{p}}\cong\mathbb{F}_{p}.
\]
Note that the image of any element
\[
b=\sum_{j=0}^{r-1}a_{j}x^{j}\in\mathbb{F}_{p}\left[x\right]/\left(x^{r}\right)
\]
is equal to $a_{0}$, denoted by $b_{\mathfrak{p}}$, which corresponds
to the ``function value'' of $b$. We can also define the formal
differential $d:\mathbb{F}_{p}\left[x\right]/\left(x^{r}\right)\rightarrow\text{Hom}\left(\mathbb{F}_{p}\left[x\right]/\left(x^{r}\right),\mathbb{F}_{p}\right)$
as \cite{Hungerford1990}
\[
d\left(\sum_{j=0}^{r-1}a_{j}x^{j}\right)=\sum_{j=1}^{r-1}ja_{j}x^{j-1}dx.
\]
It is easy to check that $d$ satisfy the Leibniz rules.

In view of these properties of $\mathbb{F}_{p}\left[x\right]/\left(x^{r}\right)$,
we can now represent each of its elements by a $r$-tuple. Specifically,
this means the following map from $\mathbb{F}_{p}\left[x\right]/\left(x^{r}\right)$
to $k\left(\mathfrak{p}\right)^{\times r}\cong\mathbb{F}_{p}^{\times r}$:
\begin{align*}
b=\sum_{j=0}^{r-1}a_{j}x^{j} & \mapsto\left(b_{\mathfrak{p}},\left(d\left(b\right)\right)_{\mathfrak{p}},\left(d^{2}\left(b\right)\right)_{\mathfrak{p}},\dots,\left(d^{r-1}\left(b\right)\right)_{\mathfrak{p}}\right)\\
 & =\left(a_{0},a_{1}dx,a_{2}dx^{2},\dots,a_{r-1}dx^{r-1}\right).
\end{align*}
Now we can define the symplectic basis on the free module $E$ over
$\mathbb{F}_{p}\left[x\right]/\left(x^{r}\right)$ as a subset
\[
\mathcal{B}=\left\{ e_{1},e_{2},\dots,e_{N},f_{1},f_{2},\dots,f_{N}\right\} \subset E
\]
satisfying the following
\begin{equation}
\sigma\left(e_{j},e_{k}\right)_{\mathfrak{p}}=\sigma\left(f_{j},f_{k}\right)_{\mathfrak{p}}=0,
\end{equation}
\begin{equation}
\sigma\left(e_{j},f_{k}\right)_{\mathfrak{p}}=-\sigma\left(f_{j},e_{k}\right)_{\mathfrak{p}}=\delta_{j,k},
\end{equation}
and
\begin{equation}
\left(d^{l}\left(\sigma\left(e_{j},e_{k}\right)\right)\right)_{\mathfrak{p}}=0
\end{equation}
for all $\left(j,k\right)\in\left\{ 1,2,\dots,N\right\} \times\left\{ 1,2,\dots,N\right\} $
and all $l\in\left\{ 1,2,\dots,r-1\right\} $.\footnote{Note that the differential must be computed first in $\mathbb{F}_{p}\left[x\right]$
before we map the result to the quotient ring $\mathbb{F}_{p}\left[x\right]/\left(x^{r}\right)$.}
\begin{example}
Let $p=3$, $r=2$ and $N=1$. The two elements
\[
u=\begin{pmatrix}x\\
0
\end{pmatrix},\,v=\begin{pmatrix}0\\
x
\end{pmatrix}\in E
\]
satisfy
\[
\sigma\left(u,v\right)=0
\]
and therefore
\[
\sigma\left(u,v\right)_{\mathfrak{p}}=0.
\]
However, we have
\begin{align*}
\left(d\left(\sigma\left(e_{j},e_{k}\right)\right)\right)_{\mathfrak{p}} & =\left(dx\right)x+x\left(dx\right)-0\cdot0\\
 & =2x\,dx\\
 & \neq0.
\end{align*}
Therefore, $u,v$ cannot be contained in the same symplectic basis.

Accordingly, the symplectic homomorphism should be defined to preserve
all of the following constraints for any two elements $u,v\in E$:
\begin{equation}
\left(d^{l}\left(\sigma\left(Su,Sv\right)\right)\right)_{\mathfrak{p}}=\left(d^{l}\left(\sigma\left(u,v\right)\right)\right),
\end{equation}
for all $l\in\left\{ 0,1,2,\dots,r-1\right\} $. An immediate corollary
of this definition is the following: Given a symplectic homomorphism
$S$, we have
\begin{equation}
\Omega(S^{-1}dS)^{t}+\left(S^{-1}dS\right)\Omega=0,
\end{equation}
which resembles the definition of symplectic Lie algebra in the continuous
variable system.
\end{example}

Now we shift our focus to the case with $d$ being any positive integer.
The whole idea is premised on Chinese remainder theorem \cite{Hungerford1990}.
Therefore, before proceeding, we first state the theorem here:
\begin{thm}
Let $n_{1},n_{2},\dots,n_{k}$ be arbitrary pairwise coprime integers,
and $a_{1},a_{2},\dots,a_{k}$ be arbitrary integers. Then there always
exists an integer $x$ such that
\begin{align*}
x\equiv a_{1} & \left(\text{mod }n_{1}\right)\\
x\equiv a_{2} & \left(\text{mod }n_{2}\right)\\
\cdots\\
x\equiv a_{k} & \left(\text{mod }n_{k}\right).
\end{align*}
Moreover, the integer $x$ is unique modulo $n_{1}n_{2}\cdots n_{k}$.
\end{thm}

\begin{rem}
This theorem states a fact in the scheme theory \cite{Vakil2017}.
For example, when $n_{1},n_{2},\dots,n_{k}$ are distinct coprime
numbers, the spectrum of the ring $\mathbb{Z}/d\mathbb{Z}$ is given
by
\[
X:=\text{Spec}\mathbb{Z}/d\mathbb{Z}=\{\left(n_{1}\right),\left(n_{2}\right),\dots,\left(n_{k}\right)\}.
\]
Then we have the $\mathcal{O}_{X}=\mathbb{Z}/d\mathbb{Z}$ consisting
of the global sections on $X$, and the stalks $\mathcal{O}_{X,(n_{l})}\cong\mathbb{Z}/n_{l}\mathbb{Z}$
for all $l\in\{1,2,\dots,k\}$. Since all the integers $n_{l}$ are
coprime, the generic points in $X$ will not overlap with each other.
Therefore, we have $\mathcal{O}_{X}\cong\mathcal{O}_{X,(n_{1})}\times\mathcal{O}_{X,(n_{2})}\times\cdots\times\mathcal{O}_{X,(n_{k})}$. 
\end{rem}

\noindent In other words, for any $\mathbb{Z}/d\mathbb{Z}$ with
$d=n_{1}n_{2}\cdots n_{k}$ and $n_{1},n_{2},\dots,n_{k}$ are pairwise
coprime, we have the following isomorphism
\[
\mathbb{Z}/d\mathbb{Z}\cong\mathbb{Z}/n_{1}\mathbb{Z}\times\mathbb{Z}/n_{2}\mathbb{Z}\times\cdots\times\mathbb{Z}/n_{k}\mathbb{Z}.
\]
This means every element in $\mathbb{Z}/d\mathbb{Z}$ can be represented
by a $k$-tuple. And addition and multiplication in $\mathbb{Z}/d\mathbb{Z}$
can be carried out separately in each $\mathbb{Z}/n_{l}\mathbb{Z}$,
for all $l\in\{1,2,\dots,k\}$.
\begin{example}
Let $d=6=2\times3$. Then each element in $\mathbb{Z}/d\mathbb{Z}$
can be represented b a 2-tuple:
\[
0\mapsto\begin{pmatrix}0, & 0\end{pmatrix},
\]
\[
1\mapsto\begin{pmatrix}1, & 1\end{pmatrix},
\]
\[
2\mapsto\begin{pmatrix}0, & 2\end{pmatrix},
\]
\[
3\mapsto\begin{pmatrix}1, & 0\end{pmatrix},
\]
\[
4\mapsto\begin{pmatrix}0, & 1\end{pmatrix},
\]
and
\[
5\mapsto\begin{pmatrix}1, & 2\end{pmatrix}.
\]
Then the addition of two elements, e.g. 2 and 5 can be calculated
by
\[
2+5\equiv\begin{pmatrix}0, & 2\end{pmatrix}+\begin{pmatrix}1, & 2\end{pmatrix}\equiv\begin{pmatrix}1, & 1\end{pmatrix}\equiv1\left(\text{mod }6\right),
\]
as well as the multiplication

\[
2\cdot5\equiv\begin{pmatrix}0, & 2\end{pmatrix}\cdot\begin{pmatrix}1, & 2\end{pmatrix}\equiv\begin{pmatrix}0, & 1\end{pmatrix}\equiv4\left(\text{mod }6\right).
\]
\end{example}

Now it seems that we have all we need to carry out the symplectic
algebraic calculation on a module $E$ over $\mathbb{Z}/d\text{\ensuremath{\mathbb{Z}}}$
free of rank $2n$ (a generalization of the concept of vector space
when $\mathbb{Z}/d\mathbb{Z}$ is not a field). But just as discussed
in the previous scenario, we will encounter unsolvable problems as
shown in the following example:
\begin{example}
Consider a module $E$ over $\mathbb{Z}/6\mathbb{Z}$ free of rank
$2$. Let $u,v\in E$ be
\[
u=\begin{pmatrix}2\\
0
\end{pmatrix}=2\begin{pmatrix}1\\
0
\end{pmatrix},\,\text{and }v=\begin{pmatrix}0\\
3
\end{pmatrix}=3\begin{pmatrix}0\\
1
\end{pmatrix}.
\]
Then we have
\[
\sigma\left(u,v\right)=\left(2\cdot3\text{ mod }6\right)=0,
\]
while
\[
\sigma\left(\begin{pmatrix}1\\
0
\end{pmatrix},\begin{pmatrix}0\\
1
\end{pmatrix}\right)=1\neq0.
\]
 This problem can never be observed when $E$ is an actual vector
space and has to be fixed for otherwise we will not be able to identify
the symplectic bases.
\end{example}

\noindent To address this problem, we utilize the Chinese remainder
theorem to establish an isomoprhism
\[
E\cong E_{n_{1}}\times E_{n_{2}}\times\cdots\times E_{n_{k}}
\]
with $d=n_{1}n_{2}\cdots n_{k}$ and $n_{1},n_{2},\dots,n_{k}$ pairwise
coprime integers. Moreover, since each integer $n_{l}$ is either
a prime number or a power of a prime number, each $E_{n_{m}}$ is
either a $2n$-dimensional symplectic vector space over the finite
field $\mathbb{Z}/n_{m}\mathbb{Z}$ or a module which has already
been studied in the previous scenario. Naturally, any multiplication
and addition in $E$ will be conducted separately in each $E_{n_{m}}$.
\begin{example}
The same $u$ and $v$ in the lase examples are now represented by
2-tuples:
\[
u\mapsto(\begin{pmatrix}0\\
0
\end{pmatrix},\begin{pmatrix}2\\
0
\end{pmatrix})
\]
and
\[
v\mapsto(\begin{pmatrix}0\\
1
\end{pmatrix},\begin{pmatrix}0\\
0
\end{pmatrix}).
\]
Then we have
\begin{align*}
\sigma(u,v) & \mapsto(\sigma_{E_{2}}(\begin{pmatrix}0\\
0
\end{pmatrix},\begin{pmatrix}0\\
1
\end{pmatrix}),\sigma_{E_{3}}(\begin{pmatrix}2\\
0
\end{pmatrix},\begin{pmatrix}0\\
0
\end{pmatrix}))\\
 & =(0,0)\\
 & \mapsto0.
\end{align*}
At the same time, the original $\begin{pmatrix}1\\
0
\end{pmatrix}$ and $\begin{pmatrix}0\\
1
\end{pmatrix}$ in $E$ are now represented by 2-tuples as the following:
\[
\begin{pmatrix}1\\
0
\end{pmatrix}\mapsto(\begin{pmatrix}1\\
0
\end{pmatrix},\begin{pmatrix}1\\
0
\end{pmatrix})
\]
and
\[
\begin{pmatrix}0\\
1
\end{pmatrix}\mapsto(\begin{pmatrix}0\\
1
\end{pmatrix},\begin{pmatrix}0\\
1
\end{pmatrix}),
\]
with
\begin{align*}
\sigma(\begin{pmatrix}1\\
0
\end{pmatrix},\begin{pmatrix}0\\
1
\end{pmatrix}) & \mapsto(\sigma_{E_{2}}(\begin{pmatrix}1\\
0
\end{pmatrix},\begin{pmatrix}0\\
1
\end{pmatrix}),\sigma_{E_{3}}(\begin{pmatrix}1\\
0
\end{pmatrix},\begin{pmatrix}0\\
1
\end{pmatrix}))\\
 & =(1,1)\\
 & \mapsto1.
\end{align*}
However this time, the difference between the two skew-inner product
will not cause a problem, since there is no element $(c_{1},c_{2})\in\mathbb{Z}/6\mathbb{Z}$
with $c_{1}\neq0$ and $c_{2}\neq0$, such that
\[
(\begin{pmatrix}0\\
0
\end{pmatrix},\begin{pmatrix}2\\
0
\end{pmatrix})=(c_{1},c_{2})\cdot(\begin{pmatrix}1\\
0
\end{pmatrix},\begin{pmatrix}1\\
0
\end{pmatrix})
\]
or
\[
(\begin{pmatrix}0\\
1
\end{pmatrix},\begin{pmatrix}0\\
0
\end{pmatrix})=(c_{1},c_{2})\cdot(\begin{pmatrix}0\\
1
\end{pmatrix},\begin{pmatrix}0\\
1
\end{pmatrix}).
\]
\end{example}

Correspondingly, every homomorphism $S:E\rightarrow E$ should be
represented as a $k$-tuple:
\[
S\mapsto(S_{1},S_{2},\dots,S_{k})
\]
with each $S_{l}:E_{n_{l}}\rightarrow E_{n_{l}}$ being a module homomorphism
(or linear transfomration if $n_{l}$ is a prime number). Now we can
define the symplectic homorphism: The homomorphism $S$ is a symplectic
if each $S_{l}:E_{n_{l}}\rightarrow E_{n_{l}}$ is a symplectic homomorphism
(with $n_{l}$ not a prime number) or a symplectic transformation
(with $n_{l}$ a prime number).
\begin{example}
We consider the module $E$ over $\mathbb{Z}/6\mathbb{Z}$ free of
rank $2$. We see the following homomorphism is symplectic:
\[
\left(\begin{array}{cccc}
5 & 4 & 0 & 0\\
3 & 5 & 0 & 0\\
0 & 0 & 5 & 3\\
0 & 0 & 2 & 5
\end{array}\right)
\]
with the 2-tuple representation
\[
(\left(\begin{array}{cccc}
1 & 0 & 0 & 0\\
1 & 1 & 0 & 0\\
0 & 0 & 1 & 1\\
0 & 0 & 0 & 1
\end{array}\right),\left(\begin{array}{cccc}
2 & 1 & 0 & 0\\
0 & 2 & 0 & 0\\
0 & 0 & 2 & 0\\
0 & 0 & 2 & 2
\end{array}\right)),
\]
since each component of this 2-tuple is a symplectic transformation
on the corresponding symplectic space. Its symplecity can also be
checked directly as follows:
\[
\left(\begin{array}{cccc}
5 & 4 & 0 & 0\\
3 & 5 & 0 & 0\\
0 & 0 & 5 & 3\\
0 & 0 & 2 & 5
\end{array}\right)\left(\begin{array}{cccc}
0 & 0 & 1 & 0\\
0 & 0 & 0 & 1\\
5 & 0 & 0 & 0\\
0 & 5 & 0 & 0
\end{array}\right)\left(\begin{array}{cccc}
5 & 4 & 0 & 0\\
3 & 5 & 0 & 0\\
0 & 0 & 5 & 3\\
0 & 0 & 2 & 5
\end{array}\right)^{t}\equiv\left(\begin{array}{cccc}
0 & 0 & 1 & 0\\
0 & 0 & 0 & 1\\
5 & 0 & 0 & 0\\
0 & 5 & 0 & 0
\end{array}\right).
\]
Note that $5\equiv-1\left(\text{mod }6\right)$.
\end{example}

Now, using the tuple representation, the symplectic bases can be easily
constructed or identified.
\begin{example}
Let
\[
\mathcal{B}_{1}=\{\begin{pmatrix}1\\
1
\end{pmatrix},\begin{pmatrix}0\\
1
\end{pmatrix}\}
\]
be a symplectic basis of $E_{2}$, and
\[
\mathcal{B}_{2}=\{\begin{pmatrix}2\\
0
\end{pmatrix},\begin{pmatrix}2\\
2
\end{pmatrix}\}
\]
be a symplectic basis of $E_{3}$. Then the set
\[
\{(\begin{pmatrix}1\\
1
\end{pmatrix},\begin{pmatrix}2\\
0
\end{pmatrix}),(\begin{pmatrix}0\\
1
\end{pmatrix},\begin{pmatrix}2\\
2
\end{pmatrix})\}\mapsto\{\begin{pmatrix}5\\
3
\end{pmatrix},\begin{pmatrix}2\\
5
\end{pmatrix}\}
\]
is a symplectic basis of the module $E$ since
\[
\sigma(\begin{pmatrix}5\\
3
\end{pmatrix},\begin{pmatrix}2\\
5
\end{pmatrix})=1
\]
and this set spanned the whole module $E$ for obvious reason.
\end{example}

\noindent In summary, the symplectic basis for the most general situation
may be defined as such: The subset
\[
\mathcal{B}=\{e_{1},e_{2},\dots,e_{n},f_{1},f_{2},\dots,f_{n}\}
\]
of $E$ is a symplectic basis if the set
\[
\mathcal{B}_{l}=\{\left(e_{1}\right)_{l},\left(e_{2}\right)_{l},\dots,\left(e_{n}\right)_{l},\left(f_{1}\right)_{l},\left(f_{2}\right)_{l},\dots,\left(f_{n}\right)_{l}\}
\]
is a symplectic basis of $E_{n_{l}}$for all pair $(s,t)\in\{1,2,\dots,n\}\times\{1,2,\dots,n\}$
and all $l\in\{1,2,\dots,k\}$\@. Here the notation $(e_{s})_{k}$
denotes the $k$-th component of $e_{s}$'s tuple representation.
Note that this definition simply amounts to the following familiar
relations:
\[
\sigma(e_{s},e_{t})=\sigma(f_{s},f_{t})=0,
\]
and
\[
\sigma(e_{s},f_{t})=-\sigma(f_{s},e_{t})=\delta_{st},
\]
for all pair $(s,t)\in\{1,2,\dots,n\}\times\{1,2,\dots,n\}$ when
all of the $n_{l}$ are prime numbers. Although the defintion of symplectic
basis may remain the same, definitons of some important related concepts
require more visible adjustments. For instance, the conventioanl defintions
of the Lagrangian planes and the symplectic subspaces fail in the
example below:
\begin{example}
Let
\[
e=\begin{pmatrix}1\\
0
\end{pmatrix},\,f=\begin{pmatrix}0\\
1
\end{pmatrix}
\]
be a symplectic basis of the free module $E$ over $\mathbb{Z}/6\mathbb{Z}$
of rank 2. Let
\[
u=\begin{pmatrix}2\\
0
\end{pmatrix}=2e,\,v=\begin{pmatrix}0\\
3
\end{pmatrix}=3f
\]
be two elements of $E$. Conventionally, the subset of $E$ spanned
by $e$ is regarded as the Lagrangian plane. However, as we know $\sigma(u,v)=0$,
according to its traditional defintion, this Lagrangian plane must
also include $v$. This gives rise to a conflict since $\sigma(e,v)=3\neq0$
which implies that $v$ should not be included in this Lagrangian
plane. This contradiction cannot be resolved unless we look into the
2-tuple representation of $E$. Then $e$ and $f$ yields the following
symplectic bases
\[
e_{1}=\begin{pmatrix}1\\
0
\end{pmatrix},\,f_{1}=\begin{pmatrix}1\\
0
\end{pmatrix}
\]
for the vector space $E_{2}$ and
\[
e_{2}=\begin{pmatrix}1\\
0
\end{pmatrix},\,f_{2}=\begin{pmatrix}1\\
0
\end{pmatrix}
\]
for the vector space $E_{3}$. And we know
\[
u=(\begin{pmatrix}0\\
0
\end{pmatrix},\begin{pmatrix}2\\
0
\end{pmatrix}),\,v=(\begin{pmatrix}0\\
1
\end{pmatrix},\begin{pmatrix}0\\
0
\end{pmatrix}).
\]
It now follows that $v\in l_{1}\times l_{2}$ while $u\notin l_{1}\times l_{2}$,
where $l_{1}$ is the Lagrangian plane spanned by $e_{1}$ in $E_{2}$
and $l_{2}$ is the Lagrangian plane spanned by $e_{2}$ in $E_{3}$.

Thus, we define the module counterparts of the crucial subspaces of
a symplectic space as follows: A subset $\mathcal{E}\in E$ is a Lagrangian
plane (symplectic submodule, ...) if the projection of $\mathcal{E}$
to each $E_{n_{k}}$ is a Lagrangian plane (symplectic submodule,
...). Naturally, the symplectic conjugate $\mathcal{E}'$ of a subset
$\mathcal{E}$ in $E$ will be the following set
\[
\mathcal{E}'=\mathcal{E}_{1}'\times\mathcal{E}_{2}'\times\cdots\times\mathcal{E}_{k}'
\]
with each $\mathcal{E}_{l}'$ defined as
\[
\mathcal{E}_{l}'\cong E_{n_{l}}/\mathcal{E}_{l}
\]
where $\mathcal{E}_{l}$ is the projection of $\mathcal{E}$ to the
$E_{n_{l}}$. At last, we have all necessary ingredients to prove
the symplectic Gram-Schmidt theorem for every $d$-level systems,
which allows us to construct a symplectic basis given a subset of
it.  The proof of this theorem is trivial since the validity for
this theorem is already known when restricted to each $E_{n_{l}}$.
\end{example}

Due to the relation of this construction to the scheme theory we briefly
mentioned, what we obtained in this section can be further generalized.
For example, the ring $\mathbb{Z}/d\mathbb{Z}$ may be generalized
to a general scheme, and the module $E$ may be generalized to coherent
sheaves \cite{Vakil2017}. Hopefully, this picture can open the way
to describing more stabilizer-based quantum processes using the powerful
tools provided symplectic algebra.

\section{Conclusion and outlook}

We demonstrated a framework of defining symplectic geometry in discrete
variable systems and its immediate application in understanding the
quantum teleportation schemes. It is also possible to further generalize
this framework to other quantum systems, in particular the systems
consisting of stabilizer states. We are looking forward to seeing
and understanding more interesting applications of the symplectic
geometry and mathematical tools developed for classical mechanics
in quantum mechanics.

\appendix

\chapter{Formulas for bosonic scattering process \label{chap:Scattering-process-and}}

\section{Introduction}

Linear bosonic scattering processes are Gaussian processes that are
widely in theoretical models in this thesis. In this appendix, we
hope to demonstrate formulas suitable for analyzing these processes
in general situations. Specifically, a linear bsonic scattering process
is completely determined by a quadratic Hamiltonian, describing the
interactions between system modes, and the coupling between the system
modes and the external propagating modes (i.e. the modes that will
be scattered). By properly organizing these data as matrices, we systematically
develop formulas that can be applied to general linear scattering
processes (Eq.~\ref{eq:scattering_matrix}), no matter these processes
are passive or active, and which are meanwhile compatible with the
symplectic geometrical nature of Gaussian process. These formulas
are particularly friendly for computer programs, and have greatly
simplified our analysis in this thesis. We hope they will also be
useful for other studies concerning complex scattering processes.

\section{Quadratic bosonic Hamiltonian}

In Sec.~\ref{symplectic_transformation}, we mentioned that the elements
of the metaplectic Lie algebra $\text{mp}(2N,\mathbb{R})$, which
generates all the metaplectic operators (unitary operators determined
by symplectic transformations up to complex constants), are quadratic
Hamiltonians. Required by the later derivations in the following sections,
we describe the connection between the quadratic Hamiltonians and
the metaplectic operators in detail.

We consider an $N$-mode bosonic system. A quadratic bosonic Hamiltonian
for this system is a Hamiltonian of the following form
\begin{equation}
\hat{H}=\sum_{j,k=1}^{N}\left(\mathcal{Y}_{j,k}\hat{a}_{j}^{\dagger}\hat{a}_{k}+\frac{1}{2}\mathcal{W}_{j,k}\hat{a}_{j}^{\dagger}\hat{a}_{k}^{\dagger}+\frac{1}{2}\mathcal{W}_{j,k}^{*}\hat{a}_{j}\hat{a}_{k}\right)+\text{const.},
\end{equation}
with $\hat{a}_{j}$, $\hat{a}_{j}^{\dagger}$, $j\in\{1,2,\dots,N\}$
being the bosonic annihilation and creation operators for each mode,
and complex matrices
\[
\mathcal{Y}=\mathcal{Y}^{\dagger},
\]
and
\[
\mathcal{W}=\mathcal{W}^{t}.
\]
Then the dynamics of the system is given by the metaplectic unitary
operator
\[
\hat{U}=e^{-i\hat{H}t}
\]
with the real parameter $t$ representing time. Specifically, in the
Heisenberg picture, the equation of evolution of the system is given
as follows:
\begin{align*}
\frac{d\hat{a}_{j}}{dt} & =i[\hat{H},\hat{a}_{j}]\\
 & =-i\sum_{k=1}^{n}\mathcal{Y}_{j,k}\hat{a}_{k}-\frac{i}{2}\sum_{k=1}^{n}\mathcal{W}_{j,k}\hat{a}_{k}^{\dagger}-\frac{i}{2}\sum_{k=1}^{n}\mathcal{W}_{k,j}\hat{a}_{k}^{\dagger},
\end{align*}
for all $j\in\{1,2,\dots,N\}$, and equivalently
\begin{align*}
\frac{d\hat{a}_{j}^{\dagger}}{dt} & =i[\hat{H},\hat{a}_{j}^{\dagger}]\\
 & =i\sum_{k=1}^{n}\mathcal{Y}_{j,k}^{*}\hat{a}_{k}^{\dagger}+\frac{i}{2}\sum_{k=1}^{n}\mathcal{W}_{j,k}^{*}\hat{a}_{k}^{\dagger}+\frac{i}{2}\sum_{k=1}^{n}\mathcal{W}_{k,j}^{*}\hat{a}_{k}.
\end{align*}
With the usual definition of the quadrature operators, it turns out
that the
\begin{align*}
\hat{U}= & e^{-i\hat{H}t}\\
= & \hat{U}_{S}
\end{align*}
with the symplectic transformation
\begin{equation}
S=\exp\left(\begin{pmatrix}\Im[\mathcal{Y}+\mathcal{W}] & \Re[\mathcal{Y}-\mathcal{W}]\\
-\Re[\mathcal{Y}+\mathcal{W}] & \Im[\mathcal{Y}-\mathcal{W}]
\end{pmatrix}t\right)
\end{equation}
if we arrange the canonical symplectic basis as $\hat{q}_{1},\hat{q}_{2},\dots,\hat{q}_{N},\hat{p}_{1},\hat{p}_{2},\dots,\hat{p}_{N}$.
In fact, the matrix
\[
\begin{pmatrix}\Im[\mathcal{Y}+\mathcal{W}] & \Re[\mathcal{Y}-\mathcal{W}]\\
-\Re[\mathcal{Y}+\mathcal{W}] & \Im[\mathcal{Y}-\mathcal{W}]
\end{pmatrix}
\]
is a member of the symplectic Lie algebra $\text{sp}(2N,\mathbb{R})$
as it should be.

For simplicity, we denote
\[
\hat{a}=\begin{pmatrix}\hat{a}_{1}\\
\hat{a}_{2}\\
\vdots\\
\hat{a}_{N}
\end{pmatrix},\ \hat{a}^{\dagger}=\begin{pmatrix}\hat{a}_{1}^{\dagger}\\
\hat{a}_{2}^{\dagger}\\
\vdots\\
\hat{a}_{N}^{\dagger}
\end{pmatrix}.
\]
as well as
\[
\hat{q}=\begin{pmatrix}\hat{q}_{1}\\
\hat{q}_{2}\\
\vdots\\
\hat{q}_{N}
\end{pmatrix},\ \hat{p}=\begin{pmatrix}\hat{p}_{1}\\
\hat{p}_{2}\\
\vdots\\
\hat{p}_{N}
\end{pmatrix}.
\]
Therefore, the above can also be expressed as follows:
\[
\begin{pmatrix}\hat{a}\\
\hat{a}^{\dagger}
\end{pmatrix}\left(t\right)=\exp\left(i\begin{pmatrix}-\mathcal{Y} & -\mathcal{W}\\
\mathcal{W}^{*} & \mathcal{Y}^{*}
\end{pmatrix}t\right)\begin{pmatrix}\hat{a}\\
\hat{a}^{\dagger}
\end{pmatrix}
\]
and
\[
\begin{pmatrix}\hat{q}\\
\hat{p}
\end{pmatrix}(t)=\exp\left(\begin{pmatrix}\Im[\mathcal{Y}+\mathcal{W}] & \Re[\mathcal{Y}-\mathcal{W}]\\
-\Re[\mathcal{Y}+\mathcal{W}] & \Im[\mathcal{Y}-\mathcal{W}]
\end{pmatrix}t\right)\begin{pmatrix}\hat{q}\\
\hat{p}
\end{pmatrix}.
\]

\section{Passive scattering process \label{sec:Scattering-process-without}}

Apart from the correspondence between the quadratic Hamiltonians and
the symplectic transformations shown in the last section, we can also
obtain symplectic transformations using scattering processes. Let
$\hat{A}_{j}^{(in/out)},\hat{A}_{j}^{\dagger(in/out)}$ be the annihilation
and creation operators of each input/output propagating bosonic mode.
Then the dynamics of such a scattering process is determined by the
following quantum Langevin equation \cite{gardiner_quantum_2004}:
\[
\frac{d}{dt}\begin{pmatrix}\hat{a}\\
\hat{a}^{\dagger}
\end{pmatrix}(t)=i[\hat{H},\begin{pmatrix}\hat{a}\\
\hat{a}^{\dagger}
\end{pmatrix}]-\frac{\mathcal{B}}{2}\begin{pmatrix}\hat{a}\\
\hat{a}^{\dagger}
\end{pmatrix}+\mathcal{C}\begin{pmatrix}\hat{A}^{(in)}\\
\hat{A}^{(in)}{}^{\dagger}
\end{pmatrix},
\]
and the input-output relation
\[
\begin{pmatrix}\hat{A}^{(out)}\\
\hat{A}^{(out)}{}^{\dagger}
\end{pmatrix}=\mathcal{D}\begin{pmatrix}\hat{a}\\
\hat{a}^{\dagger}
\end{pmatrix}+\begin{pmatrix}\hat{A}^{(in)}\\
\hat{A}^{(in)}{}^{\dagger}
\end{pmatrix}.
\]
In the above,
\[
\hat{H}=\sum_{j,k=1}^{N}\left(\mathcal{Y}_{j,k}\hat{a}_{j}^{\dagger}\hat{a}_{k}+\frac{1}{2}\mathcal{W}_{j,k}\hat{a}_{j}^{\dagger}\hat{a}_{k}^{\dagger}+\frac{1}{2}\mathcal{W}_{j,k}^{*}\hat{a}_{j}\hat{a}_{k}\right)
\]
is a quadratic Hamiltonian, the complex matrices $\mathcal{B}$, $\mathcal{C}$,
and $\mathcal{D}$ are introduced to describe the effects of the coupling
between the system and the propagating modes.

With a fixed frequency $\omega$, the Fourier transform of these equations
leads to
\begin{equation}
\left(i\begin{pmatrix}\mathcal{Y}-\omega\text{Id} & \mathcal{W}\\
-\mathcal{W}^{*} & -\mathcal{Y}^{*}-\omega\text{Id}
\end{pmatrix}+\frac{\mathcal{B}}{2}\right)\begin{pmatrix}\hat{a}[\omega]\\
\hat{a}[-\omega]^{\dagger}
\end{pmatrix}=\mathcal{C}\begin{pmatrix}\hat{A}^{(in)}[\omega]\\
\hat{A}^{(in)}[-\omega]^{\dagger}
\end{pmatrix},
\end{equation}
and
\begin{equation}
\begin{pmatrix}\hat{A}^{(out)}[\omega]\\
\hat{A}^{(out)}[-\omega]^{\dagger}
\end{pmatrix}=\mathcal{D}\begin{pmatrix}\hat{a}[\omega]\\
\hat{a}[-\omega]^{\dagger}
\end{pmatrix}+\begin{pmatrix}\hat{A}^{(in)}[\omega]\\
\hat{A}^{(in)}[-\omega]^{\dagger}
\end{pmatrix},
\end{equation}
with
\[
\hat{O}[\omega]:=\frac{1}{\sqrt{2\pi}}\int\hat{O}(t)e^{-i\omega t}dt
\]
for an operator $\hat{O}$. Then we have the following equation:
\begin{equation}
\begin{pmatrix}\hat{A}^{(out)}[\omega]\\
\hat{A}^{(out)}[-\omega]^{\dagger}
\end{pmatrix}=\left[\text{Id}+\mathcal{D}\left(i\begin{pmatrix}\mathcal{Y}-\omega\text{Id} & \mathcal{W}\\
-\mathcal{W}^{*} & -\mathcal{Y}^{*}-\omega\text{Id}
\end{pmatrix}+\frac{\mathcal{B}}{2}\right)^{-1}\mathcal{C}\right]\begin{pmatrix}\hat{A}^{(in)}[\omega]\\
\hat{A}^{(in)}[-\omega]^{\dagger}
\end{pmatrix}.
\end{equation}
Supposing there is no intrinsic losses or gains in the system and
letting $\omega=0$
\[
O=\frac{1}{\sqrt{2}}\begin{pmatrix}\text{Id} & i\text{Id}\\
\text{Id} & -i\text{Id}
\end{pmatrix},
\]
\[
B=O^{-1}\mathcal{B}O,\quad C=O^{-1}\mathcal{B}O,\quad D=O^{-1}\mathcal{B}O,
\]
the matrix
\begin{equation}
S=\text{Id}+D\left(\begin{pmatrix}\Im[\mathcal{Y}+\mathcal{W}] & \Re[\mathcal{Y}-\mathcal{W}]\\
-\Re[\mathcal{Y}+\mathcal{W}] & \Im[\mathcal{Y}-\mathcal{W}]
\end{pmatrix}+\frac{B}{2}\right)^{-1}C
\end{equation}
is symplectic, since the transformation from the input propagating
modes to the output porpagating modes is purely unitary and $O$ is
the matrix transforming the quadrature operators to the annihilation
and creation operators.

In the above, the matrices $B,C$, and $D$ must be real. Let $C\in\mathbb{R}^{2N\times2M}$,
$D\in\mathbb{R}^{2M\times2N}$ , with $N\leq M$, are full-ranked,
where $N$ is the number of the system modes, and $N$ is the number
of the propagating modes. Therefore, according to the symplectic singular-value
decomposition which will be introduced later, we can conclude that
there exists a symplectic transformation $\mathcal{S}$, such that
\[
C=-\Omega C'\Omega\begin{pmatrix}\text{Id} & 0\end{pmatrix}\mathcal{S},\quad D=\mathcal{S}^{-1}\begin{pmatrix}\text{Id}\\
0
\end{pmatrix}C'^{t}
\]
with $C',D'\in\mathbb{R}^{2N\times2M}$. and if there are no intrinsic
losses or gains,
\[
-\Omega\left(C'\right)^{-1}\Omega BC'^{t}+C'B^{t}\Omega\left(C'^{t}\right)^{-1}\Omega=2\Omega.
\]
For the situation where intrinsic losses or gains are significant,
the right-hand-side of the last equation should be modified to some
arbitrary skew-symmetric matrix. Because most of times the symplectic
spaces where the input and output modes are defined artificially,
we will assume $M=N$ from now on.
\begin{example}
Consider the quadratic Hamiltonian
\[
\hat{H}=g(\hat{a}_{1}^{\dagger}\hat{a}_{2}^{\dagger}+\hat{a}_{1}\hat{a}_{2}),
\]
and the system-environment interaction determined by
\[
\mathcal{B}=\begin{pmatrix}\kappa_{1} & 0\\
0 & \kappa_{2}
\end{pmatrix},
\]
and
\[
\mathcal{C}=\mathcal{D}=\begin{pmatrix}\sqrt{\kappa_{1}} & 0\\
0 & \sqrt{\kappa_{2}}
\end{pmatrix}.
\]
We thus have
\[
\mathcal{Y}=0,
\]
and
\[
\mathcal{W}=\begin{pmatrix}0 & g\\
g & 0
\end{pmatrix}.
\]
It follows that
\[
S=\begin{pmatrix}0 & t' & r' & 0\\
t' & 0 & 0 & r'\\
r' & 0 & 0 & t'\\
0 & r' & t' & 0
\end{pmatrix}
\]
with
\[
r'=\frac{\chi+1}{\chi-1},\quad t'=\frac{2\sqrt{\chi}}{1-\chi},
\]
satisfying $r'^{2}-t'^{2}=1$, and
\[
\chi=\frac{g^{2}}{\kappa_{1}\kappa_{2}}.
\]
\end{example}

We call a scattering process passive if there is no coupling between
any two annihilation operators of the system. Then the symplectic
transformations will only transform $\hat{Q}[\omega]^{(in)}$ and
$\hat{P}[\omega]^{(in)}$ to $\hat{Q}[\omega]^{(out)}$ and $\hat{P}[\omega]^{(out)}$,
and will be in the form of
\[
S=\text{Id}+D\left(\begin{pmatrix}\Im[\mathcal{Y}] & \Re[\mathcal{Y}]\\
-\Re[\mathcal{Y}] & \Im[\mathcal{Y}]
\end{pmatrix}+\omega\Omega+\frac{B}{2}\right)^{-1}C.
\]

\begin{example}
Consider the quadratic Hamiltonian
\[
\hat{H}=g\left(\hat{a}_{1}^{\dagger}\hat{a}_{2}+\hat{a}_{2}^{\dagger}\hat{a}_{1}\right),
\]
and the system-environment interaction given by
\[
\mathcal{B}=\begin{pmatrix}\kappa_{1} & 0\\
0 & \kappa_{2}
\end{pmatrix},
\]
and
\[
\mathcal{C}=\mathcal{D}=\begin{pmatrix}\sqrt{\kappa_{1}} & 0\\
0 & \sqrt{\kappa_{2}}
\end{pmatrix}.
\]
Then the above equation leads to
\[
S=\begin{pmatrix}0 & -t & r & 0\\
t & 0 & 0 & r\\
r & 0 & 0 & -t\\
0 & r & t & 0
\end{pmatrix}
\]
with
\[
r=\frac{\chi-1}{\chi+1},\quad t=\frac{2\sqrt{\chi}}{\chi+1}
\]
satisfying the condition that $r^{2}+t^{2}=1$, and
\[
\chi=\frac{g^{2}}{\kappa_{1}\kappa_{2}}.
\]
\end{example}

\section{A matrix representation of Clifford algebra $Cl_{3,0}$}

Let $V$ be a 3-dimensional real vector space. Let $Q:V\rightarrow\mathbb{R}$
be a quadratic form (e.g. the norm $\left\Vert \cdot\right\Vert $
induced by the standard Euclidean inner product $\langle\cdot,\cdot\rangle:V\times V\rightarrow\mathbb{R}$).
Let $A$ be an unital assocaitve algebra over $\mathbb{R}$. Then
there exists a unital algebra over $\mathbb{R}$, denoted by $B$,
with a linear map $i:V\rightarrow Cl_{3,0}$ satisfying $i\left(v\right)^{2}=Q\left(v\right)\text{Id}$
for all $v\in V$, such that for any linear map $j:V\rightarrow A$
satisfying $j\left(v\right)^{2}=Q\left(v\right)\text{Id}$ for all
$v\in V$, there exists an unique algebra homomorphism satisfying
$f\circ i=j$. Thus, the pair $\left(B,i\right)$ is a universal object,
meaning it is uniquely determined up to an algebra isomorphism, which
is called a Clifford algebra \cite{Karoubi1978}.

For the case of $Q=\left\Vert \right\Vert $, we denote $B$ as $Cl_{3,0}$.
Let $v_{1},v_{2},v_{3}$ be the orthonormal basis of $V$. The $Cl_{3,0}$
is generated by $i\left(v_{1}\right),i\left(v_{2}\right),i\left(v_{3}\right)$.
These elements satisfy
\[
i\left(v_{k}\right)i\left(v_{l}\right)+i\left(v_{l}\right)i\left(v_{k}\right)=2\langle v_{k},v_{l}\rangle\text{Id},\,\text{for all }v_{k},v_{l}\in V.
\]
As an real vector space, $Cl_{3,0}$ has the following set of elements
as a basis
\[
\left\{ \text{Id},i\left(v_{1}\right),i\left(v_{2}\right),i\left(v_{3}\right),i\left(v_{1}\right)i\left(v_{2}\right),i\left(v_{2}\right)i\left(v_{3}\right),i\left(v_{1}\right)i\left(v_{3}\right),i\left(v_{1}\right)i\left(v_{2}\right)i\left(v_{3}\right)\right\} .
\]

Now we introduce two matrix representations of $Cl_{3,0}$. First,
we can represent any element of $Cl_{3,0}$ as a $4\times4$ complex
matrix by letting
\begin{equation}
\epsilon_{1}=i\begin{pmatrix}0 & 1 & 0 & 0\\
-1 & 0 & 0 & 0\\
0 & 0 & 0 & 1\\
0 & 0 & -1 & 0
\end{pmatrix},\quad\epsilon_{2}=\begin{pmatrix}0 & 1 & 0 & 0\\
1 & 0 & 0 & 0\\
0 & 0 & 0 & 1\\
0 & 0 & 1 & 0
\end{pmatrix},\quad\epsilon_{3}=\begin{pmatrix}1 & 0 & 0 & 0\\
0 & -1 & 0 & 0\\
0 & 0 & -1 & 0\\
0 & 0 & 0 & 1
\end{pmatrix},
\end{equation}
be the representations of $i\left(v_{1}\right),i\left(v_{2}\right)$,
and $i\left(v_{3}\right)$ correspondingly. We can also let the representaions
of $i\left(v_{1}\right),i\left(v_{2}\right)$, and $i\left(v_{3}\right)$
be the $4\times4$ real matrices
\begin{equation}
e_{1}=\begin{pmatrix}0 & 0 & 0 & 1\\
0 & 0 & 1 & 0\\
0 & 1 & 0 & 0\\
1 & 0 & 0 & 0
\end{pmatrix},\quad e_{2}=\begin{pmatrix}0 & 0 & 1 & 0\\
0 & 0 & 0 & -1\\
1 & 0 & 0 & 0\\
0 & -1 & 0 & 0
\end{pmatrix},\quad e_{3}=\begin{pmatrix}1 & 0 & 0 & 0\\
0 & 1 & 0 & 0\\
0 & 0 & -1 & 0\\
0 & 0 & 0 & -1
\end{pmatrix},
\end{equation}
so that each element of $Cl_{3,0}$ can be represented by a $4\times4$
real matrix.

Let
\begin{equation}
U=\frac{1}{\sqrt{2}}\begin{pmatrix}\text{Id} & i\text{Id} & 0 & 0\\
0 & 0 & \text{Id} & -i\text{Id}\\
0 & 0 & \text{Id} & i\text{Id}\\
\text{Id} & -i\text{Id} & 0 & 0
\end{pmatrix},
\end{equation}
it is easy to check that
\[
e_{k}=U^{-1}\epsilon_{k}U,\quad\text{for }k=1,2,3.
\]

\section{Active scattering process}

We now resume our discussion about the scattering processes to include
the situations where the couplings between creation operators are
present. Such scattering processes are regarded as active scattering
processes. Throughout this section, we assume $\omega>0$ and the
complex matrices $\mathcal{B}$, $\mathcal{C}$, $\mathcal{D}$ are
diagonal and full-ranked. We define
\[
\Theta=\omega\mathcal{B}^{-1},
\]
\[
Y=\mathcal{C}^{-1}\mathcal{Y}\mathcal{D}^{-1},
\]
\[
W=\mathcal{C}^{-1}\mathcal{W}\mathcal{D}^{-1},
\]
and
\[
\Gamma=\frac{\mathcal{C}^{-1}\mathcal{B}\mathcal{D}^{-1}}{2}.
\]
Then we have
\[
\begin{pmatrix}\hat{A}^{(out)}[\omega]\\
\hat{A}^{(out)}[-\omega]^{\dagger}\\
\hat{A}^{(out)}[-\omega]\\
\hat{A}^{(out)}[\omega]^{\dagger}
\end{pmatrix}=\mathcal{S}\begin{pmatrix}\hat{A}^{(in)}[\omega]\\
\hat{A}^{(in)}[-\omega]^{\dagger}\\
\hat{A}^{(in)}[-\omega]\\
\hat{A}^{(in)}[\omega]^{\dagger}
\end{pmatrix},
\]
where
\begin{equation}
\mathcal{S}=\text{Id}-\left(\text{Id}\otimes\left(\Im Y+\frac{1}{2}\text{Id}\right)-\epsilon_{1}\otimes\Re W+\epsilon_{2}\otimes\Im W-\epsilon_{1}\epsilon_{2}\otimes\Re Y+\epsilon_{1}\epsilon_{2}\epsilon_{3}\otimes\Theta\right)^{-1}.
\end{equation}
Since
\[
\begin{pmatrix}\hat{a}[\omega]\\
\hat{a}[-\omega]^{\dagger}\\
\hat{a}[-\omega]\\
\hat{a}[\omega]^{\dagger}
\end{pmatrix}=U\begin{pmatrix}\hat{q}[\omega]\\
\hat{p}[\omega]\\
\hat{q}[-\omega]\\
\hat{p}[-\omega]
\end{pmatrix}
\]
with
\[
U=\frac{1}{\sqrt{2}}\begin{pmatrix}\text{Id} & i\text{Id} & 0 & 0\\
0 & 0 & \text{Id} & -i\text{Id}\\
0 & 0 & \text{Id} & i\text{Id}\\
\text{Id} & -i\text{Id} & 0 & 0
\end{pmatrix},
\]
we also have
\[
\begin{pmatrix}\hat{Q}^{(out)}[\omega]\\
\hat{P}^{(out)}[\omega]\\
\hat{Q}^{(out)}[-\omega]\\
\hat{P}^{(out)}[-\omega]
\end{pmatrix}=S\begin{pmatrix}\hat{Q}^{(in)}[\omega]\\
\hat{P}^{(in)}[\omega]\\
\hat{Q}^{(in)}[-\omega]\\
\hat{P}^{(in)}[-\omega]
\end{pmatrix}
\]
with
\begin{align}
S= & U^{-1}\mathcal{S}U\nonumber \\
= & \text{Id}-\left(\text{Id}\otimes\left(\Im Y+\Gamma\right)-e_{1}\otimes\Re W+e_{2}\otimes\Im W-e_{1}e_{2}\otimes\Re Y+e_{1}e_{2}e_{3}\otimes\Theta\right)^{-1}\label{eq:scattering_matrix}
\end{align}
a symplectic transformation when there are no intrinsic losses or
gains, i.e., $\Gamma=\text{Id}/2$. Note that he symplectic form can
be represented by
\[
\Omega=e_{1}e_{2}\otimes\text{Id}.
\]

Furthermore, we also note that the symplectic transformation $\mathcal{S}$
of a scattering process can be written as
\begin{align}
\mathcal{S}= & \text{Id}-(\Omega\mathcal{M}+\frac{1}{2}\text{Id})^{-1}\nonumber \\
= & (\Omega\mathcal{M}+\frac{1}{2}\text{Id})^{-1}(\Omega\mathcal{M}-\frac{1}{2}\text{Id})
\end{align}
with $\mathcal{M}$ being a symmetric matrix, which is called the
symplectic Cayley transform \cite{DeGosson2006}. The matrix $\mathcal{M}$
can be understood as the dimensionless quadratic Hamiltonian, since
it corresponds to the actual quadratic Hamiltonian normalized by the
system-environment coupling rates.

\section{Symplectic singular-value decomposition}

Let $E$ be a symplectic vector space. Let $\Omega$ be the symplectic
form. Then for any $H$ being a real symmetric positive semi-definite
linear transformation on $E$, there exists a symplectic transformation
$S$, such that $H=S\Lambda S^{t}$, where $\Lambda$ is real, positive-semi-definite,
and diagonal\footnote{This is a straightforward result of Appendix 6 of \cite{Arnold2013}.}.
Now consider an arbitray $M\in\mathbb{R}^{2m\times n}$, with $\text{rank}\left(M\right)=n$.
Since $MM^{t}$ is positve-semi-definite, there exists a symplectic
matrix $S$, such that
\[
SMM^{t}S^{t}=\begin{pmatrix}\Lambda\\
 & 0
\end{pmatrix},
\]
where $\Lambda$ is positive-definite and diagonal. We thus have
\[
\begin{pmatrix}\Lambda^{-1/2}\\
 & \text{Id}
\end{pmatrix}SMM^{t}S^{t}\begin{pmatrix}\Lambda^{-1/2}\\
 & \text{Id}
\end{pmatrix}=\begin{pmatrix}\text{Id}\\
 & 0
\end{pmatrix}.
\]
We now let
\[
\begin{pmatrix}\Lambda^{-1/2}\\
 & \text{Id}
\end{pmatrix}SM=\begin{pmatrix}Q\\
T
\end{pmatrix},
\]
with $Q\in\mathbb{R}^{n\times n}$ and $T\in\mathbb{R}^{\left(2m-n\right)\times n}$.
Then we have
\[
QQ^{t}=\text{Id},
\]
and
\[
TT^{t}=0,
\]
which imply that $Q$ is orthogonal and $T=0$. Therefore,
\begin{prop}
For any $M\in\mathbb{R}^{2m\times n}$ with $\text{rank}\left(M\right)=n$,
there exist a symplectic matrix $S$ on $\mathbb{R}^{2m}$ and an
orthognal matrix $Q$ on $\mathbb{R}^{n}$, such that
\begin{equation}
M=S\begin{pmatrix}\Lambda\\
 & 0
\end{pmatrix}Q,
\end{equation}
where $\Lambda$ is positive-definite and diagonal.
\end{prop}

This result has also been proved in \cite{Xu2003} using a different
approach.

\printindex{}

\bibliographystyle{elsarticle-num}
\bibliography{dissertation}

\end{document}